\documentclass[graybox]{svmult}
\usepackage{mathptmx}       
\usepackage{helvet}         
\usepackage{courier}        
\usepackage{type1cm}        
\usepackage{makeidx}         
\usepackage{graphicx}        
\usepackage{multicol}        
\usepackage{amsmath}        
\usepackage{amssymb}        
\usepackage[bottom]{footmisc}
\usepackage{hyperref}        
\usepackage[utf8]{inputenc}

\usepackage{journals}        

\newcommand{\beq}{\begin{equation}}
\newcommand{\eeq}{\end{equation}}
\newcommand{\bea}{\begin{eqnarray}}
\newcommand{\eea}{\end{eqnarray}}
\newcommand{\subscr}[1]{_{\rm #1}}

\newcommand{\doverd}[2]{\frac{d #1}{d #2}}
\newcommand{\doverdt}[1]{\frac{d #1}{d t}}
\newcommand{\pdoverdt}[1]{\frac{\partial #1}{\partial t}}
\newcommand{\pdoverdr}[1]{\frac{\partial #1}{\partial r}}
\newcommand{\pdoverdphi}[1]{\frac{\partial #1}{\partial {\varphi}}}
\newcommand{\pdoverd}[2]{\frac{\partial #1}{\partial #2}}
\newcommand{\RHS}{right-hand side}
\newcommand{\sub}[1]{_{\rm #1}}

\def\Op{\Omega\sub{p}}
\def\rp{r\sub{p}}

\newcommand{\AU}{\mbox{AU}}

\makeindex             

\begin{document}

%
%

\title*{Planet formation and disk-planet interactions}
\author{Wilhelm Kley}
\institute{Wilhelm Kley \at Institute of Astronomy \& Astrophysics, 
Universit\"at T\"ubingen, Morgenstelle 10, D-72070 T\"ubingen, Germany, \email{wilhelm.kley@uni-tuebingen.de}
}
%
%

\maketitle
%
%
\abstract*{
This review is based on lectures given at the 45th Saas-Fee Advanced Course “From Protoplanetary Disks to Planet Formation”
held in March 2015 in Les Diablerets, Switzerland. Starting with an overview of the main characterictics of the Solar System and the
extrasolar planets, we describe the planet formation process in terms of the sequential accretion scenario.
First the growth processes of dust particles to planetesimals and subsequently to terrestrial planets or planetary cores
are presented. This is followed by the formation process of the giant planets either by core accretion or gravitational instability.
Finally, the dynamical evolution of the orbital elements as driven by disk-planet interaction and the overall evolution of multi-object systems
is presented.
}

\abstract{
This review is based on lectures given at the 45th Saas-Fee Advanced Course “From Protoplanetary Disks to Planet Formation”
held in March 2015 in Les Diablerets, Switzerland. Starting with an overview of the main characterictics of the Solar System and the
extrasolar planets, we describe the planet formation process in terms of the sequential accretion scenario.
First the growth processes of dust particles to planetesimals and subsequently to terrestrial planets or planetary cores
are presented. This is followed by the formation process of the giant planets either by core accretion or gravitational instability.
Finally, the dynamical evolution of the orbital elements as driven by disk-planet interaction and the overall evolution of multi-object systems
is presented.
}

%
%
\section{Introduction}
\label{lect0:intro}
The problem of the formation of the Earth and the Solar System has a very long
tradition in the human scientific exploration, and has caught the attention of many
philosophers and astronomers. 
Often it is referred to as {\it one of the
most fundamental problems of science. Together with the origin of the Universe, galaxy 
formation, and the origin and evolution of life, it forms a crucial piece in understanding, were
we, as a species, come from}. This statement was made in 1993 by J. Lissauer in his excellent 
review about the planet formation process \cite{1993ARA&A..31..129L}, just before the discovery of
the first extrasolar planet orbiting a solar type star.
Today, as about 20 years have passed since the discovery of the first extrasolar planet orbiting a
solar type star in 1995, the understanding of the origin of planets and planetary
systems has indeed become a major focus of research in modern astrophysics.

Applying different observational strategies the number of confirmed detections of exoplanets
has nearly reached 2000 as of today. While already the very first discoveries of hot Jupiter planets
such as 51~Peg \cite{1995Natur.378..355M} and very eccentric planets, such as 16~Cyg~B \cite{1997ApJ...483..457C},
have hinted at the differences to our own
Solar System, later the numerous detections by the Kepler Space Telescope and others have given us full
insight as to the extraordinary diversity of the exoplanetary systems in our Milkyway.
Planets come in very different masses and sizes and show interesting dynamics in their orbits.
Full planetary systems with up to 7 planets have been found as well as planets in binary stars systems,
making science fiction become a reality.
 
At the same time, it has become possible to study  so called
protoplanetary disks in unprecedented detail. These are flattened, disk-like structures that orbit young stars,
as seen for example clearly in the famous silhouette disks in the Orion nebula,
observed by the Hubble Space telescope, also in 1995 \cite{1996AJ....111.1977M}.
Being composed by a mixture of about 99\% gas and 1\% dust, these disks hold the reservoir of
material from which planets may form. Indeed protoplanetary disks are considered to be the birthplaces of
planets as anticipated already long time ago by Kant and Laplace \cite{1755anth.book.....K,1776Laplace}
in their thoughts about the origin of the Solar System.
Following the close connection between planets and disks, particular structures
(gaps, rings, and non-axisymmetries) observed in these disks are often connected to the
possible presence of young protoplanets. The most famous recent example is the ALMA-observation
of the disk around the star HL~Tau that shows a systems of ring-like structures
which may have been carved by a planetary system forming in this relatively young disk \cite{2015ApJ...808L...3A}.

As it is well established that planets form in protoplanetary disks, many aspects of this formation process
are still uncertain and depend on details of the gas disk structure and the embedded solid (dust) particles.
At the same time the evolution of planets and planetary systems is driven by the evolving disk, and we can
understand the exoplanet sample and its architecture only by studying both topics (disks and planets) simultaneously.
In this lecture we will summarize the current understanding of the planet formation and evolution process, 
while aspects of the physics of disks have been presented in the chapter by P.~Armitage in this volume.
The presentation will focus more on the basic physical concepts while for the specific aspects we will refer to the
recent review articles and other literature. 

\subsection{The Solar System}
\label{lect0:solar}
Any theory on planet formation has to start by analyzing the physical properties of 
the observed planetary systems. Here, we start out with a brief summary of the most relevant facts
of the Solar System, with respect to its formation process, for a more detailed list see the review by J.~Lissauer \cite{1993ARA&A..31..129L}.
The Solar System is composed of 8 planets, 5 dwarf planets, probably thousands of minor bodies such as trans-Neptunian objects (TNOs),
asteroids, and comets, and finally millions of small dust particles.
The planets come in two basic flavors, terrestrial and giant planets. The first group (Mercury, Venus, Earth and Mars) are
very compact, low mass planets that occupy the inner region of the Solar System, from 0.4 to 2.5 AU. 
Separated by the asteroid-belt the larger outer planets (Jupiter, Saturn, Uranus and Neptune) occupy the region from about
5 to 30 AU. Sometimes the giant planets are sub-divided into the {\it gas-giants} (Jupiter, Saturn) that have a mean density
similar or even below that of water (1 g/cm$^3$) and are composed primarily of Hydrogen and Helium, and the {\it ice-giants} (Uranus, Neptune)
with a mean density of about 1.5 g/cm$^3$. All giants are believed to have a solid core (rocks) in their centers, while
the atmospheres of the ice-giants are much less massive than those of the gas-giants and contain more ices of water,
ammonia and methane. The dwarf planets, asteroids and TNOs are primarily composed of solid material.

The most important {\it dynamical} property of the Solar System is its {\it flatness}. The maximum inclination (i.e. the
angle of the planetary orbit with the ecliptic plane) a planet has is about $7^\circ$ for Mercury, while all the other, larger
planets have $i < 3.4^\circ$. All planets orbit the Sun in the same direction (prograde),
their angular momentum vectors are roughly aligned with that of the Sun, and
their orbits are nearly circular. The giant planets have an eccentricity $e < 0.055$, 
only the smallest planets have a significant eccentricity, $e=0.2$ for Mercury and in particular $e=0.1$ for Mars, which allowed
Johannes Kepler to infer the elliptic nature of the planetary orbits.
The spin-axis of the planets is also approximately aligned with the orbital angular momentum, only Venus and Uranus 
(and Pluto, even though not a planet anymore) represent exceptions. 
From meteoric dating the age of the planets and asteroids in the Solar System has been estimated
to be about $4.56$ billion years, i.e. the planets have about the same age as the Sun itself which implies a coeval origin
of the Solar System as a whole.
The orbital spacings of the planets are such that their mutual separation increases with semi-major axis, in a way that 
they can be ordered into a geometric series, the Titius-Bode law. It has often been suggested that this orbital sequence must be
a direct consequence of the formation process, which has become very doubtful after the discovery of the extrasolar planets
and noticing the importance of physical processes like migration and scattering.
Instead the simple requirement of long-term stable planetary orbits implies a sort of geometric spacing between planets
\cite{1998Icar..135..549H}.

\begin{figure}
    \begin{center}
        \includegraphics[width=0.90\textwidth]{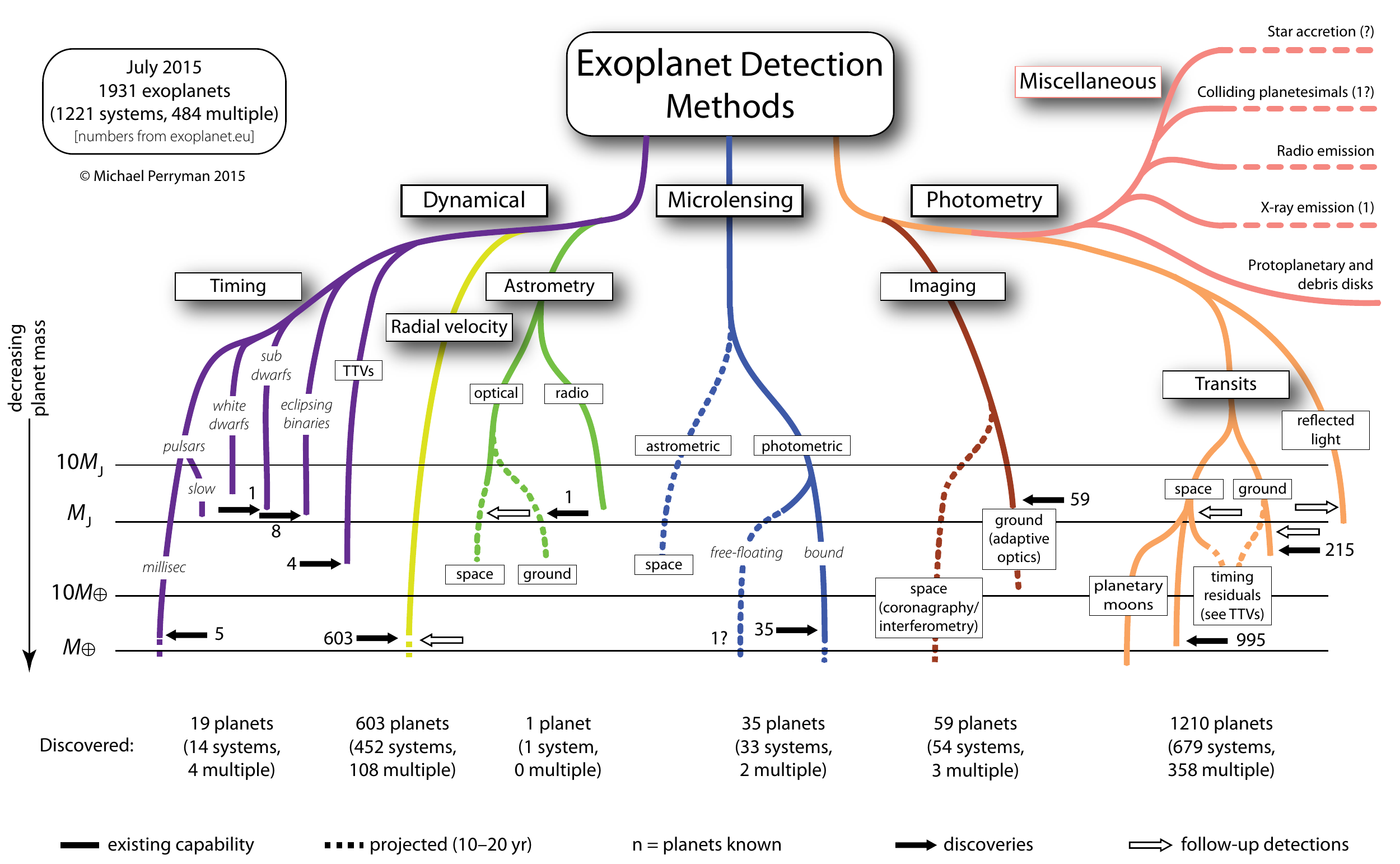} \\
    \end{center}
    \caption{Graphics to indicate the number of exoplanet detections using a particular detection method.
       Courtesy: M. Perryman}
    \label{fig:intro-detections}
\end{figure}
The prevalence of solid material suggests that the main formation process has started
from the accumulation of small bodies via {\it sequential accretion} where bodies grow through a sequence of trillions
of collisions from small dust particles to full fledged planet.
Additionally, the flat structure of the Solar System indicates that this process has taken place within a protoplanetary disk,
the {\it Solar Nebula}, that orbited the early Sun. These findings are supported by the observational fact that many protostars are surrounded by a
flat disk consisting of gas and dust, with extensions similar to that of the Solar System.
This nebular hypothesis of the Solar System's origin was already the basis of the formation theories of Kant and Laplace. While
Kant focused on the evolution of the solid dust material in the Solar System \cite{1755anth.book.....K}, Laplace focused more on the
hydrodynamical aspects \cite{1776Laplace}. Much later, V.~Weizs\"acker has taken up these ideas to develop his hydrodynamical
theory of planet formation in a disk containing several vortices \cite{1943ZA.....22..319W}.
An overview of these and many subsequent ideas are contained in \cite{2000oess.book.....W} or \cite{1998Fahr}.
A modern (pre-extrasolar planet) review of the formation of the Solar System from an astrophysicists perspective
is given by \cite{1993ARA&A..31..129L}, while the chronological aspect is emphasized by \cite{2006EM&P...98...39M}.

\subsection{Properties of the extrasolar planets}
\label{lect0:exoplanet}
During the past 20 years numerous extrasolar planets orbiting Sun-like stars have been detected,
and their physical and dynamical properties provide us with the opportunity to examine
the generality of our ideas about the planet formation process, and modify those
chapters of the whole story that are not compliant with the new observational data.
We will not go in any detail into the detection methods that are used to discover new planets, an overview is given
in \cite{2011exha.book.....P}, and see also the lecture notes by A.~Quirrenbach in the 
Saas-Fee Advanced Course 31 \cite{2006expl.conf....1Q}.
In Fig.~\ref{fig:intro-detections} we display graphically the number of planets discovered using a particular detection
method as of July, 2015.
As shown, in absolute numbers the transit method has been by far the most successful method since over
half of all detections have been achieved using transits, most of them with the Kepler space telescope.
From the ground the radial velocity method is still the most successful. As shown, there are nearly
500 multiple systems. After the end of the regular Kepler-mission the rate of detections has somewhat slowed,
but new space missions such as TESS \cite{2015JATIS...1a4003R} or PLATO \cite{2014ExA....38..249R} will surely enhance it again.
A status of the actual numbers can be found in several online catalogues such as
\url{http://exoplanet.eu/}, \url{http://exoplanets.org}, or \url{http://www.openexoplanetcatalogue.com/}. 

\begin{figure}
    \begin{center}
        \includegraphics[width=0.70\textwidth]{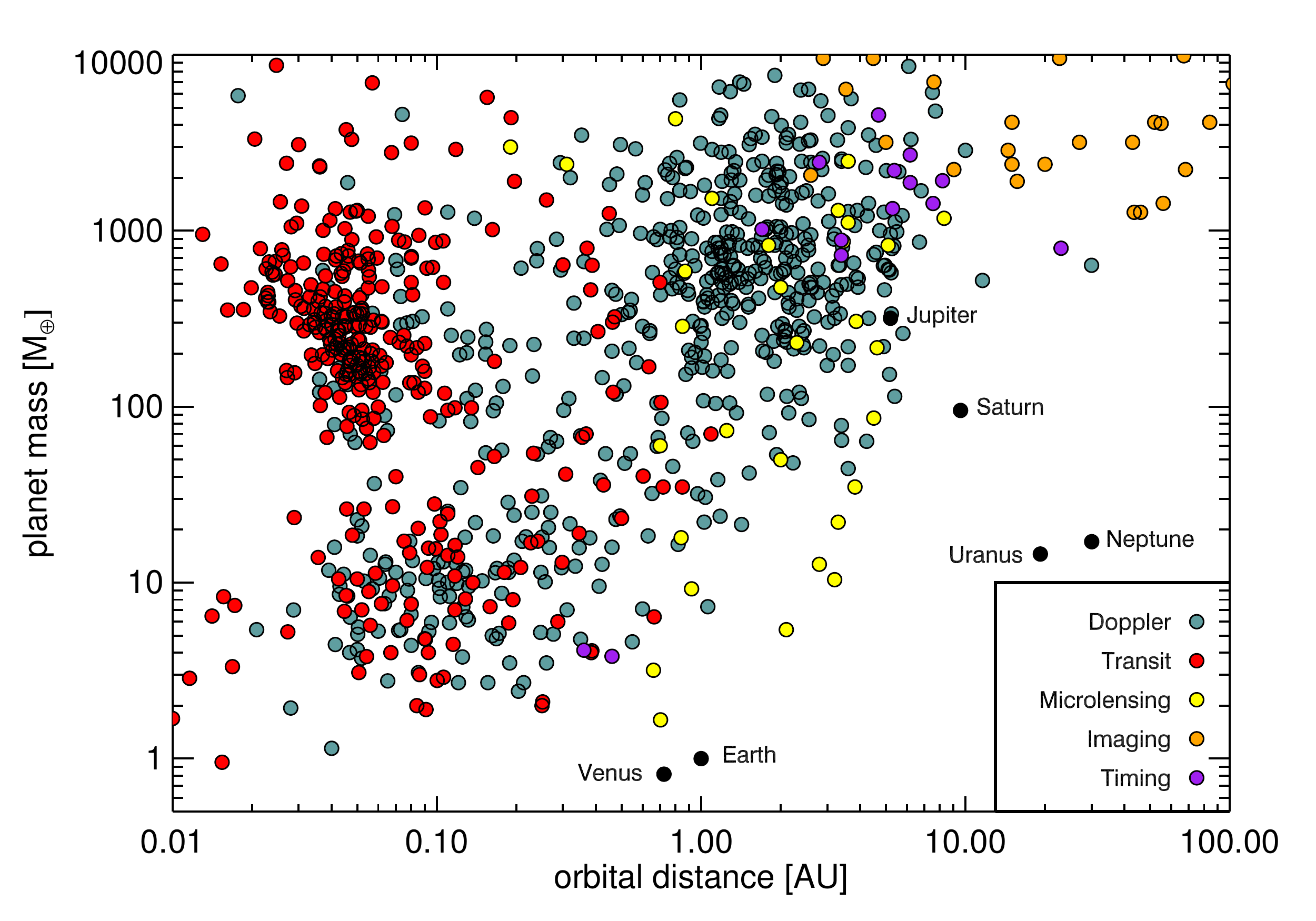} \\
    \end{center}
    \caption{Mass of the extrasolar planets with respect to the distance from their host star.
       From Winn and Fabrycky \cite{2015ARA&A..53..409W}.
    }
    \label{fig:intro-mass-distance}
\end{figure}
The amount of data collected so far allows for extensive statistical analyses of the exoplanet properties,
which serves in particular to understand the similarities, as well as the differences with respect to our Solar System.
A very important diagram is shown in Fig.~\ref{fig:intro-mass-distance} where we display the mass of the planet 
against semi-major axis. Unfortunately, for the Kepler planets their mass is known only for a small fraction of all discovered
planets because the low apparent magnitude of the host stars does not allow for an easy radial velocity follow-up, but
see \cite{2014ApJS..210...20M}, where this has been successful. 
Hence, the plotted data in Fig.~\ref{fig:intro-mass-distance} refer primarily to the radial velocity measurements.
Here, one has to keep in mind that for this detection method the quoted masses refer to the minimum
masses because of the unknown inclination to the systems.
In the plot the different detection techniques have been indicated and it is clear that they are sensitive to
different regions in the $M-a$ diagramme, but we will not discuss here any detection limits or biases.
The planets populate different regions in the plot which have been used to separate the whole planet population
into various planetary species.
The top left corner, for semi-major axis smaller below about 0.1~AU and masses above 100 $M_{\oplus}$, is inhabited
by the so-called {\it Hot Jupiters}, as these planets are comparable in mass to Jupiter in the Solar System but are 
located very close to the star such that the stellar irradiation leads to surface temperatures often well above
1000 K. Already the very first exoplanet discovered, 51 Peg \cite{1995Natur.378..355M}, 
with a 4 day period and a separation from the Sun 
about 1/20 AU, belongs to this category. It is very difficult to form such a planet at that location
because the high disk temperatures do not allow for simple condensation of material.
Hence, already this first discovery of an exoplanet required a modification to the standard planet formation 
theory by allowing some migration process \cite{1996Natur.380..606L}. Within the same mass range, or
even a bit higher, up to 10 $M\sub{Jup}$, and with a larger distance between 1 and 10 AU we find the {\it classical giant}
planets, i.e. planets of Jupiter mass or higher within a similar distance from the star.
Below these, there is a whole group of planets with masses between 2 and about 30 $M_\oplus$, and distances from 0.03 up to about
1 AU. These are the so-called {\it Super Earths}\footnote{Even though some of the planets lie more in the size range of Neptune
($4 R_\oplus$) and are sometimes termed Sub-Neptunes, we refer in this text to the whole group as Super-Earths.},
as they are located in the same distance regime as our home planet
but with a somewhat larger mass. Even though not directly apparent in this limited sample plot, from the Kepler data it is clear
that the smaller planets are by far the largest population of the detected exoplanets.
Only less than 10\% of all Kepler planets are larger than Jupiter while nearly half are in the range of Neptune, with
(2 -- 6) $R_\oplus$, and the rest smaller (for the most up to date statistics on the distribution of planetary radii of Kepler-planets
we refer to the NASA-Homepage). 
Observational limitations do presently
not allow for the detection of the very low mass and small objects in this regime,
but from statistical analyses it is estimated that at least {\it 11\% of Sun-like stars
harbor an Earth-size planet receiving between one and four times the stellar intensity as Earth} \cite{2013PNAS..11019273P},
a statement with strong implication with respect to the possibility of life in the Milkyway.

Concerning the shape of the orbits it became clear very early on that the average eccentricity of
exoplanets is with $\bar{e} \sim 0.3$ much higher than for the Solar System. 
This value applies for the radial velocity planets, which show a weak trend 
of an increase in $e$ for larger planet masses.
The Kepler systems show, on average, a smaller eccentricity and have lower planet masses.
The main orbital and dynamical properties of exoplanets have recently been summarized in \cite{2015ARA&A..53..409W}.
From a formation point of view it is noteworthy that the multi-planet systems show only a very small mutual
inclinations between the planetary orbits, i.e. planetary systems are born flat.
Interestingly, this applies to the circumbinary planets as well which indicates that the protoplanetary disks from
which the planets formed were closely aligned with the central binary star.

\subsection{Pathways to planets}
\label{lect0:planetform}
After having summarized briefly the observational aspects of planetary systems including our own,
we will in the following sections outline the process of formation, based on the modern day view.  Historically, the Greek philosophers have already theorized a long time ago about the origin of the Earth and possible
other {\it Worlds}. In the 5th century BCE Leukippos suggested that {\it the worlds form in such a way, that
the bodies sink into the empty space
and connect to each other}, as translated in \cite{1998gdav.book.....H}.
In a very broad sense this is still the view today, because with {\it World} the whole Solar System
is implied, Sun and all planets. As we know today, stars form within a collapsing molecular cloud that turns into a highly
flatted configuration due to angular momentum conservation. In the center the proto-sun forms and in the disk the planets.
The problem we are facing is, that we need to illuminate a process that for the Solar System took place 4.5 billion years ago.
Nevertheless, the observational data from the Solar System, the data from protoplanetary disks and last not least the 
large and growing sample of extrasolar planetary systems allow
us to draw a coherent framework in which many details are left to be worked out but the main processes have probably 
been understood.

Two main pathways to make planets have been discussed in the literature. In the first scenario, planets are believed to have formed
directly from the protoplanetary disk by gravitational instability. In this {\it top-down} process spiral arms
become gravitationally unstable and fragment directly to form large protoplanets. The advantage of such a process may be the possibly very fast
formation timescale ($\sim 1000$ years) but for typical protoplanetary disks the required fast cooling of the disk material is
probably not satisfied expect possibly for very large distances (several 10s of AU) from the star.
In addition, the Solar System contains a multitude of small solid objects and the giant planets have most likely massive solid cores in their
centers.
These problems of the GI-scenario, and the properties of the Solar System have led to the current view that the planet formation
has occurred predominantly via a {\it bottom-up} process in which small dust particles have grown through millions and millions of sticking
collisions to form eventually large protoplanets, terrestrial planets, and the cores of the giant planets. 
The giants collect by their strong gravitational force in a final step a large amount gas.

Hence, in the following we describe the planet formation process on the basis of the sequential accretion scenario,
where we start out in Sect.~\ref{lect:01} with the growth from small dust to km-sized planetesimals, study the formation
of terrestrial planets in Sect.~\ref{lect:02}. The growth to massive planets via core-accretion will be described in 
Sect.~\ref{lect:03} followed a description of the GI pathway in Sect.~\ref{lect:04}. 
An important part of the planet formation process is the occurrence of dynamical evolution due to disk-planet interaction.
We will describe the most important results in Sect.~\ref{lect:05} of this lecture followed by a study of the dynamical behaviour
of multi-object planetary systems in Sect.~\ref{lect:06}.

\begin{svgraybox}
The physical and dynamical properties of the Solar System indicate that the planets formed in
a flat disk orbiting the young proto-sun. The information drawn from the solid bodies implies that the growth occurred via a
bottom-up process where planets were formed in a sequential accretion process starting from tiny interstellar dust grains,
all the way to full grown planets. The additional information drawn from the large sample of extrasolar
planetary systems supports this basic scenario but requires to take into account dynamical effects in multi-planet systems
and the gravitational interaction between the disk and the growing planets.
\end{svgraybox}

%
%
%

\section{From Dust to Planetesimals}
\label{lect:01}

Protoplanetary disks typically consist of a mixture of about 99\% gas and a small amount ($\sim 1\%$)
of solid, dust particles \cite{2011ARA&A..49...67W}.
As mentioned above in the introduction, in the Solar System and many of the observed extrasolar planetary systems,
the formation of planets is believed to be accomplished primarily through
a sequential growth process starting from small interstellar dust particles. 
Hence, to study this early planet formation phase we need two ingredients, an initial ensemble of dust
particles and a suitable disk model.

Concerning the dust, the now classical measurements of interstellar extinction \cite{1977ApJ...217..425M}
have shown that interstellar dust grains have a size range of about $0.1-1.0\mu$m with a size
distribution of $n(a) \propto a^{-3.5}$, where $n(a)$ denotes the number of particles with a given size $a$. 
New results indicate some deviation from this MRN-profile (after Mathis, Rumpl and Nordsieck, \cite{1977ApJ...217..425M}),
but the typical expected initial sizes
of interstellar dust particles is probably within the indicated size range \cite{2003ARA&A..41..241D}.
Through a sequence of trillions and trillions of collisions, starting from these tiny dust grains
full grown planets are eventually assembled.
In this section we deal with the first phase of this process and consider the initial growth up to about
km-sized planetesimals.

Concerning the disk structure, one often refers to simple disk models where the density and temperature distribution
can be described by suitable power laws, in order to simplify the analyses. Starting from the observed locations
and masses of the planets in the Solar System, Hayashi \& Weidenschilling \cite{1981PThPS..70...35H,1977Ap&SS..51..153W}
constructed a simple relation
for the variation of the surface density of the material with distance from the Sun. To this purpose
they spread out the observed mass of the planets in a number of radial bins and added the amount of gas
to match the observed dust/gas ratio in observed protoplanetary disk, typically 1/100 \cite{2011ARA&A..49...67W}.
This mass distribution is now referred to as the {\it Minimum Mass Solar Nebula} (MMSN) as it contains just the right
total amount of matter and density slope to make the Solar System, as we observe it today. 
The resulting gas density distribution is then given by \cite{1981PThPS..70...35H}
\beq
\label{eq:MMSN}
   \Sigma\sub{g} (r) \approx  1700 \, \left(\frac{r}{1 \mathrm{AU}}\right)^{-3/2} \,  \mbox{g/cm}^2 \,.
\eeq
Such a model yields a typical total disk mass of about $0.01-0.05$ M$_\odot$.
Of course, initially the mass could have been substantially larger and the density distribution different, because
planets tend to move (migrate) in disks and do not remain at their birth locations. Consequently, the simple 
distribution (\ref{eq:MMSN}) has been criticized frequently and different slopes have been suggested
\cite{2007ApJ...659..705A,2007ApJ...671..878D}, but is nevertheless still frequently used,
let it be only as a suitable reference model.
Observed disks typically tend to have a flatter profile and show at some radius a cutoff,
as seen also in time-dependent viscous disks model \cite{1974MNRAS.168..603L}.
Such distributions can be described in parameterized form as an exponentially tapered power law
\beq
\label{eq:dust01-sigmadisk}
   \Sigma(r)  \propto \left(\frac{r}{R\sub{c}}\right)^{-\gamma} \, \exp \left[ - \left(\frac{r}{R\sub{c}}\right)^{2-\gamma}\right] \,.
\eeq
Here the exponent $\gamma$ and the characteristic radius, $R\sub{c}$, can be chosen to match the observations.
For $\gamma$ mean values of about 0.9 have been determined while the value of $R\sub{c}$ ranges from 
$20 - 200$ AU \cite{2011ARA&A..49...67W}.
For the temperature in the disk one can assume as a first approximation a passive disk, where
internal heat is solely generated by the illumination of the central star. At larger radii (beyond a few AU)
this is a good approximation because there the internally generated heat is lower than stellar irradiation.
In this case the radial temperature of the disk is given by \cite{1981PThPS..70...35H}
\beq
\label{eq:dust01-tdisk}
   T(r) \approx  280 \,  \left(\frac{r}{1 \mathrm{AU}}\right)^{-1/2} \,  \mbox{K}  \,.
\eeq
This distribution assumes a vertically isothermal stratification, hence eq.~(\ref{eq:dust01-tdisk}) describes
the midplane as well as the surface temperature of the disk. Assuming now a constant temperature in the $z$-direction,
the vertical hydrostatic equation can be solved for the density, and a Gaussian distribution is obtained, with
\beq
\rho(r,z) = \rho_0(r) \, \exp{(- z^2/H^2)} \,. 
\eeq
Here $\rho_0(r)$ is the midplane density,
$H(r) = c\sub{s}/\Omega\sub{K}$ the vertical scale height (often termed the {\it disk half-thickness}),
$\Omega\sub{K}(r)$ is the Keplerian rotational velocity in the disk, and $c\sub{s}(r)$ the local sound speed.
By integrating $\rho(r,z)$ over $z$
one obtains the radial surface density distribution 
\beq
\Sigma(r) = \int_{-\infty}^\infty \rho(r,z) dz \,.
\eeq
By varying the exponent $\gamma$
in eq.~(\ref{eq:dust01-sigmadisk}) different disk models can be constructed, that can be used as a basis for analyzing
the dust motion within the disk.
More detailed analytical disk models including internal as well as external heating 
have been calculated by \cite{1997ApJ...490..368C}. 

These disk models can be used as a basis to study the motion and growth of embedded dust particles. 
As shown in the accompanying chapter by P.~Armitage in this volume the particles have a different
velocity as the gas because they do not feel the effects of gas pressure. As a consequence they experience a drag
force which is proportional to the velocity difference between the gas and dust particles and depends on the size 
of the particle and the gas density. This frictional force leads to a drift of the particles which is directed
towards the pressure maximum of the gas. As can easily be inferred from the gas and temperature distribution the
pressure, that is $\sim \rho T$, drops rapidly with radius in a protoplanetary nebula. Consequently, the particles
experience a rapid inward radial drift which is for typical disk parameter about $v\sub{drift} \sim 50$m/s. 
This leads to a rapid loss of particles into the central star, and any successful growth process has to be sufficiently fast to
overcome this {\it drift-barrier}.
For more details on the motion of particles in gas disks see \cite{1977MNRAS.180...57W},
or the review article by \cite{2010EAS....41..187Y}. The overall growth phase from dust to planetesimals has been reviewed
more recently in \cite{2010AREPS..38..493C} or \cite{2014prpl.conf..547J}. 
 
\subsection{Study the initial growth phase}
\label{subsec:dust02}
Starting from the initially $\mu$m sized solid particles that are embedded in the accretion disk the growth process
can only proceed through a sequence of collisions where two partners hit each other and stick together due
to some adhesive forces. The outcome of these mutual collisions and the important sticking probability depend on the  
friction coefficients, the compactification of the material (i.e. how much kinetic energy can be dissipated upon collisions)
and the relative velocity of the components in the collision process. 

The actual growth of small particles has been studied in the laboratory and via numerical simulations, where
for the latter the experimental results have been used to calibrate the simulations.
On the experimental side the most complete set of studies have been performed using spherical,
mono-disperse SiO$_2$ particles (silica)
that have the additional advantage of an easy direct usage in numerical studies \cite{2008ARA&A..46...21B}.
Although these particles seem to be rather special, experiments have indicated that the variations 
in material properties between silica and silicates (that are often detected in protoplanetary disks)
do not seem to play a major role in comparison to morphological and size differences \cite{2000ApJ...533..454P}. 
Hence, the usage of this material in the laboratory studies seems to be well justified.
In the following we describe the basic results of the laboratory work and the numerical studies.

\subsubsection{Experiments}
\label{subsubsec:dust02-experiments}
The experimental studies have been performed under a variety of different laboratory setups,
partly under vacuum and zero gravity conditions.
For the latter the following facilities have been utilized: the fall-tower in Bremen, parabola flights as offered for example
by ESA, and the international space station. The outcome of these series of experimental studies 
have been summarized exhaustively in the review article \cite{2008ARA&A..46...21B},
and we will focus here on the main results.

For the very small $\mu$m sized particles the initial growth is very clearly fractal, i.e. the mass-size relation is
given by
\beq
        m \propto s^{D\sub{f}}
\eeq
with a fractal dimension $D\sub{f}$ smaller than 3. Here $s$ denotes the size of the particle (effective radius)
with mass $m$.
The fractal growth can be understood by the
fact that small particles are well coupled to the gas and do show only very small relative velocities with respect to each other.
For the experiments relative velocities $v\sub{rel} \approx 10^{-4} - 10^{-2}$ m/s between the individual collision partners have been used.
Hence, upon collisions the direct sticking of individual particles/aggregates dominates over restructuring effects,
and the aggregates display a strongly elongated shape, with a fractal dimension typically
such that $1.4 \leq D\sub{f} \leq 1.8$. Images of the fractal agglomerates and their mass growth in time are shown
in \cite{2000PhRvL..85.2426B,2004PhRvL..93b1103K}. 

The subsequent growth after the initial fractal phase has been studied through experiments using multiple collisions
between the aggregates. In the experimental setup of \cite{2009ApJ...696.2036W} 
particles were enclosed in a plexiglass box whose floor could vibrate with a 
frequency of about 100 Hz. Using a high speed camera snapshots of the aggregates were taken and their mutual velocities
and changes in mass/volume during collisions were measured. 
The observed typical collision velocities for the particles have been in the range of $0.1 - 0.3$m/s.
An important quantity to describe the compactness of an agglomerate is the filling factor $\phi$ 
that gives the mass (density) ratio of a porous aggregate
to a solid object of the same base material and same volume 
(i.e. $\phi$ is also referred to as the volume fraction of the material)
\beq
\label{eq:filling}
       \phi = \rho\sub{agg}/\rho\sub{mat} \,.
\eeq
Initially, the prepared 'dust-cakes' are extremely fluffy and have a small filling factor of $\phi \approx 15\%$ which refers to a mean density
of only $\rho\sub{agg} = 0.3$g/cm$^3$ given that the solid matter density of SiO$_2$ is about $2.0$g/cm$^3$.
Often the {\it porosity} is quoted which is in a sense complementary to $\phi$ and given by $1 - \phi$.
The main outcome of these experiments (using mm sized particles) is a restructuring and compactification of the aggregates
such that the fractal dimension $D\sub{f}$ is increasing \cite{2009ApJ...696.2036W}.  
In terms of filling factor, starting from $\phi_0 = 0.15$ it increases with the number of collisions, $n$, according to
\beq
\phi(n) = \phi\sub{max}  - \Delta \phi \cdot e^{-n/\nu} \,,
\eeq
where $\phi\sub{max}$ is the final value of the filling factor,
$\Delta \phi = \phi\sub{max} - \phi_0$, and $\nu$ a constant determined to about $700$.
The final filling factor reached is about $\phi\sub{max} = 0.365$.
This compaction of the aggregates changes the surface to mass ratio, which is crucial for the dynamical behaviour
of the particles in the protosolar nebula \cite{1993prpl.conf.1031W}\footnote{This is known very well from daily experience where a feather
with the same weight as a small pebble falls much slower to the ground than the pebble due to the air resistance.}.
At the same time this restructuring modifies the macroscopic material
parameter that determine the tensile, shear and compressive strength of the material (see below), which is turn alters
their behaviour in subsequent collisions.

\begin{figure}
    \begin{center}
        \includegraphics[width=0.70\textwidth]{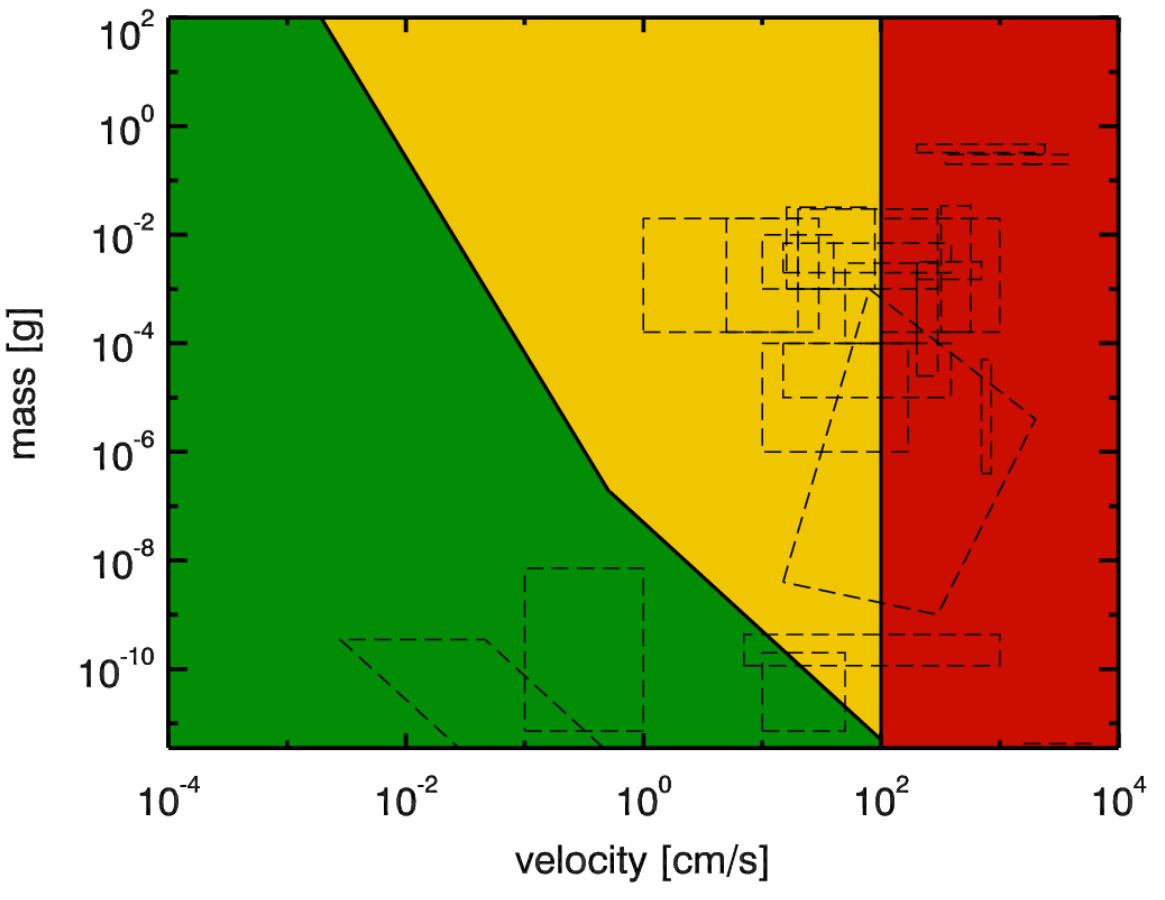} \\
    \end{center}
    \caption{The results of laboratory experiments with respect to the outcome of mutual
       collisions of equal sized dust aggregates. Green areas indicate sticking collisions, yellow bouncing ones,
       and red corresponds to fragmentation of the aggregates. The collision experiments correspond to the areas
       within the dashed lines.
       Courtesy: J\"urgen Blum}
    \label{fig:dust-collisions}
\end{figure}

In addition to the described results, a multitude of additional collision experiments have been performed and analyzed,
see e.g. \cite{2008ARA&A..46...21B} for an overview.
In \cite{2010A&A...513A..56G} the physical outcomes are classified with respect to different collision channels
corresponding to sticking, bouncing, fragmentation, or mass transfer from one aggregate to the other, see
also \cite{2011A&A...531A.166G} for an alternative classification scheme. 
A summary of the experimental results is presented in Fig.~\ref{fig:dust-collisions} where on the $y$-axis the mass
of collision partners is plotted and on the $x$-axis the relative collision velocity.
In all analyzed collisions the two colliding objects have the same size.
The areas enclosed by the dashed lines have been covered by the experiments of the research groups in Braunschweig
(around J\"urgen Blum) and Duisburg (around Gerhard Wurm). The colored regions indicate different growth or destruction regimes.
In the green area the two aggregates stick together (growth regime), in the yellow one they bounce off each other (neutral regime),
and in the red areas they are shattered into pieces and fragment (destruction regime).
For net growth, one clearly has to be in the green regime, i.e. sufficiently small relative velocities.
The diagram makes it very clear that there are different obstacles
to the successful growth from small sized grains to planetesimals, these are basically the {\it bouncing barrier} \cite{2010A&A...513A..57Z}
(the transition from green to yellow) and the {\it fragmentation barrier} (the transition from yellow to red).
Obviously, above $v\sub{rel} \approx 1$m/s growth is extremely difficult to achieve.
Together with the {\it drift barrier}, mentioned above,
these obstacles to successful growth are sometimes referred to as the {\it meter-sized barrier} to 
planetesimal growth \cite{2008ARA&A..46...21B}, as it is difficult to achieve growth beyond 1 m in size.
We will comment on possible pathways to overcome these obstacles to further growth in Sect.~\ref{subsec:dust02-barrier} below.

\subsubsection{Numerical studies}
\label{subsubsec:dust02-numerics}
\begin{figure}
    \begin{center}
        \includegraphics[width=0.70\textwidth]{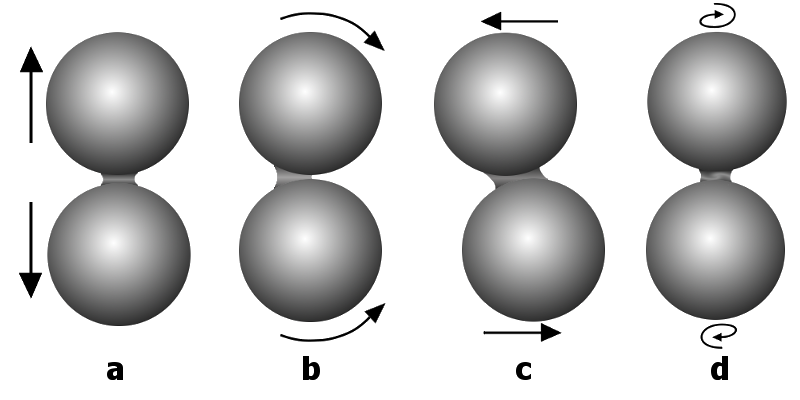} \\
    \end{center}
    \caption{Schematic view of the different type of interactions between two individual spherical monomers.
       These correspond to a) Compression/Adhesion, b) Rolling, c) Sliding, and d) Twisting.
       Courtesy: Alexander Seizinger}
    \label{fig:MD-forces}
\end{figure}

A very useful alternative to real laboratory experiments is performing {\it numerical experiments} and
study collisions in the computer using appropriate models to simulate reality. In the case of planetesimal
growth two approaches have proven to be useful. In the first approach the behaviour of aggregate collisions is
simulated on a microscopic level using a direct molecular dynamics (MD) approach with suitable forces between the individual
$\mu m$-sized monomers that the whole aggregate is composed off. In a second approach, suitable for larger particles,
a continuum model is constructed using averaged material parameters (such a elasticity, strengths, sound speed, etc.).
We will discuss both approaches in this section.

The MD approach relies on modeling in detail the microscopic interaction between two individual monomers, using
nanoscale molecular forces, such as the van der Waals force or other electrostatic forces. 
In the case of dust aggregates the forces can be divided into 4 categories as depicted in Fig.~\ref{fig:MD-forces},
see \cite{2007ApJ...661..320W} and references therein.
The formulation of the forces is based on a microscopic approach, and the implementation and application to 
the physics of dust coagulation under astrophysical conditions has been described in \cite{1997ApJ...480..647D}.
For the standard normal relative motions between individual monomers (panel $\bf a$ in Fig.~\ref{fig:MD-forces}) the force is based on the 
JKR-theory (after Johnson, Kendall and Roberts, \cite{1971RSPSA.324..301J}),
that constitutes an improvement to the original model which is due to to Heinrich Hertz \cite{1882Hertz}.
The rolling ($\bf b$), sliding ($\bf c$) and twisting ($\bf d$) frictional forces have been calculated by
Dominik \& Tielens, for example in \cite{1995PMagA..72..783D}, see summary in \cite{1997ApJ...480..647D}.
As hinted in Fig.~\ref{fig:MD-forces} by the 'necks' between the two spheres,
the intermolecular forces show a hysteretic behaviour. The forces
between two monomers do not only depend on the actual distance but on the past history. 
The forces between approaching particles, that have not been in contact yet, is different from the force (at the same separation)
if they just had been in contact. The way to imagine this is to assume that the particles are surrounded
by a thin layer of honey or glue. Before coming into contact there is no attractive fore, but after separation a
neck is formed that pulls the particles back together. 
The physical model by Dominik \& Tielens does take account for these hysteretic, dissipative effects. 

Later the model was improved by deriving the forces (torques) from appropriate potentials
which allows for detailed tracking of the different energy channels during the collisions \cite{2007ApJ...661..320W}.
Before applying the numerical model to physical collisions several numerical parameter have to be adjusted properly.
This is done by comparing the results of numerical simulations directly with laboratory experiments. 
In these well constructed experiments a small dust aggregate 
({\it dust-cake}) is prepared by simple ballistic aggregation procedure which leads to a sample with an
initial filling factor, $\phi = 0.15$. The dust-cake is then contained 
between two walls and compressed from above \cite{2009ApJ...701..130G}. 
Upon the compression the filling factor $\phi$ increases and the pressure on the adjacent walls as well.
The important quantity to calibrate is the pressure-porosity relation, that shows a characteristic behaviour, which
depends on the geometry of the particular setup.
Calibrations of this type have been done by \cite{2008A&A...484..859P} and later by \cite{2012A&A...541A..59S} who found for example a
modification of the rolling and sliding coefficients.

After a successful calibration process the numerical simulations allow to determine continuum properties of the aggregate, such
as sound speed or the shear modulus. These are sometimes difficult or impossible to determine experimentally. In numerical simulations
special treatments, such as adding a sort of glue between the particles and the wall can be introduced to make these
measurements numerically possible. Having successfully determined these parameter a new type of simulations can be performed where the
aggregates are treated as a smooth continuum average over the local details. In the astrophysical community a very important,
successful and widespread method is {\it Smooth Particle Hydrodynamics} (SPH). Here, the continuum is modeled again by individual
'particles' but now they do not represent individual monomers or real physical particles but they can rather 
be considered as a sort of Lagrangian nodes that move in space and time. Quantities such as density
or pressure are obtained from a suitable averaging procedure over neighboring particles.
The method was initially developed for simulating purely hydrodynamic flow (see review by Monaghan \cite{2005RPPh...68.1703M})
but was later extended to model the dynamics of solid objects, where time dependent equations to follow the stresses within the body
are added and solved simultaneously. In addition, the method has been augmented by an elasto-plastic model,
and special treatments for handling cracks and fragmentation have been implemented \cite{1994Icar..107...98B}.
Hence, these type of simulations allow to model the collisions of larger objects, starting from dm-size up to very large objects
where internal gravity may play a role. Indeed, using the SPH method collisions between objects that range in size from m-sized
bodies up to the Moon forming impact between the proto-Earth and a Mars-sized object have been simulated 
\cite{1999Icar..142....5B,1986Icar...66..515B}.

Similar to the collisions between tiny particles those between macroscopic objects will lead to a fragmentation and
destruction of the (larger) target if the relative collision speed becomes too large.
To analyse the outcome of these object-object 'encounters' the specific collision energy between target and projectile is 
a useful quantity. It is defined as
\beq
         Q\sub{D} = \frac{1}{2} \frac{m\sub{proj} v^2\sub{rel}}{M\sub{target} + m\sub{proj}} \,,
\eeq
where $m\sub{proj}$ is the mass of the projectile, 
$M\sub{target}$ the mass of the target and $v\sub{rel}$ the relative velocity between the two objects.

\begin{figure}
    \begin{center}
        \includegraphics[width=0.70\textwidth]{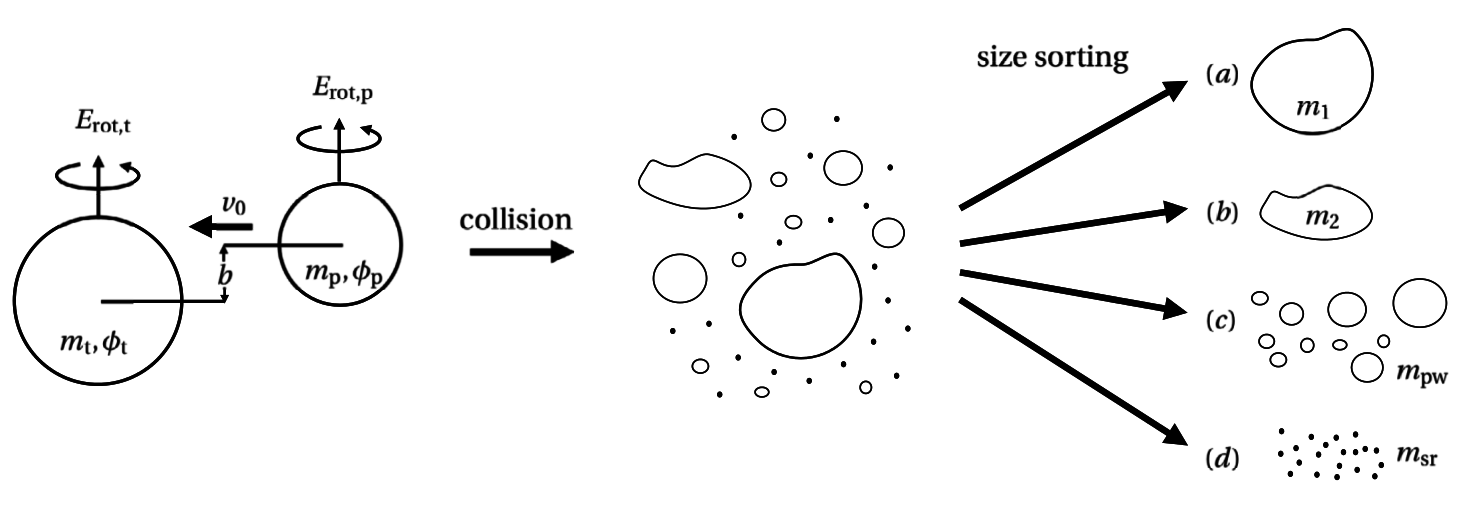} \\
    \end{center}
    \caption{Schematic representation of the outcome of a collision between two objects.
      The left side displays the situation before the collision: the target (t) and projectile (p)
      collide with impact velocity $v_0$, and impact parameter $b$. Before the collision they are described by their
      mass ($m$), filling factor ($\phi$) and rotational energy ($E\sub{rot}$). The outcome consists of different groups:
      a) the largest fragment; b) the second largest fragment; c) the power law population; and d) the sub-resolution population. 
       This classification scheme is based on the four-population model by \cite{2011A&A...531A.166G}.
      }
    \label{fig:four-pop}
\end{figure}

The general outcome of collisions is well described in Fig.~\ref{fig:four-pop}. It consists typically of a limited number
of bigger objects (here cases $\bf a,b$ on the right hand side), a large number of smaller particles ($\bf c$) that follow 
approximately a power law size distribution \cite{2009Icar..199..542l},
and a 'sea' of very small particles ($\bf d$) that are too small to be resolved (at least numerically). The four-population model
of \cite{2011A&A...531A.166G} assumes that there are only two major objects after the collision.
The catastrophic destruction threshold, $Q^*\sub{D}$, is now defined as that
specific collision energy at which the largest remaining fragment has 1/2 of the target mass.
As a function of target size $Q^*\sub{D}$ has a $V$-shaped behaviour,
in the strength dominated regime (for target radii smaller than about 1 km) $Q^*\sub{D}$ is decreasing with increasing target size.
For larger objects gravitational re-accumulation becomes important and the mass of the largest post-collision object increases, hence
$Q^*\sub{D}$ increases again with target size in this gravity dominated regime. The minimum occurs at the transition between the two regimes.
Using the SPH-method numerical experiments of collisions between two basalt spheres
with radii between 100m to 10km using different relative velocities have been performed \cite{1999Icar..142....5B}.
The results show that for a typical velocity, $v\sub{rel} = 20$km/s, the weakest bodies (with minimum $Q^*\sub{D}$) 
are those with radii of about 300m, here $Q^*\sub{D}$ has values of about $10^2$[J/kg].
Later \cite{2009Icar..199..542l} constructed fit formulae to calculate $Q^*\sub{D}$ as a function of target radii
which can be used in statistical simulations for an ensemble of objects (e.g. asteroids or Kuiper belt objects)
to follow their time evolution \cite{2013AJ....146...36S}.

\subsection{How to overcome growth barriers}
\label{subsec:dust02-barrier}
As sketched out for example in Fig.~\ref{fig:dust-collisions}, from laboratory experiments it is known that the mutual
collisions of mm to cm-sized aggregates result frequently in bouncing, e.g.
\cite{2009ApJ...696.2036W, 2010Icar..206..424H, 2012Icar..218..688W, 2012A&A...542A..80J}.
Incorporating the data from the various experimental setups, \cite{2010A&A...513A..56G} constructed an algorithm to
to describe analytically the results of mutual collisions of aggregates as a function of their relative collision speed,
their mass and their porosity. This model was then used to simulate the evolution of a swarm of dust aggregates 
in a protoplanetary disk in a statistical manner \cite{2010A&A...513A..57Z}.
Initially the aggregates grow by the above described process, but for larger sizes the relative velocities increase.
Due to the higher kinetic impact energy the aggregates become more and more compacted during successive collisions.
If the aggregates get too compact, i.e. the filling factor reaches values up to about $\phi \sim 0.4$,
then mutual collisions do not result in sticking anymore, rather they bounce off each other and the growth process is terminated
\cite{2010A&A...513A..56G}.
In agreement with the experiments this occurs within a size regime of centimeters to decimeters, and was termed
the {\it bouncing barrier} \cite{2010A&A...513A..57Z}. 

In early numerical studies to understand the origin of the bouncing it was found that for the aggregate parameters
used in the experiments typically sticking occurs, and bouncing only for much larger filling factors above $\phi=0.5$
\cite{2009ApJ...702.1490W}.
In subsequent numerical simulations the condition for rebound (bouncing)
has been studied in more detail by Wada et al. \cite{2011ApJ...737...36W} who showed that it depends on the {\it coordination number}, $n\sub{c}$,
of the individual monomers, which describes the number of contact points (neighbors) an individual monomer has. Clearly, the more
contacts an aggregate has the stiffer it reacts to collisions promoting bouncing rather than sticking, and
a value of $n\sub{c} = 0.6$ above which bouncing occurs has been found in the simulations \cite{2011ApJ...737...36W}.
Obviously, the higher the filling factor the larger the coordination number has to be. However, as shown recently 
there is not a one-to-one relation between $\phi$ and $n\sub{c}$ \cite{2013A&A...551A..65S}. The number of contacts that a specific aggregate with a given porosity has,
depends on the process by which it has been constructed. 
In laboratory experiments using aggregates composed of micron-sized dust grains, 
it is usually only possible to determine the global filling factor (via the mean density) but not the local coordination number
which is a microscopic quantity.
Thus, one has to be very careful when comparing results from numerical simulations for a given
filling factor directly to results from laboratory experiments that use the same $\phi$.

Even though the numerical studies are still not in full agreement with the experiments in the bouncing regime,
there are indications that non head-on collisions lead to increased bouncing \cite{2013A&A...551A..65S}.
Additionally, it was shown that for larger aggregate sizes the fragmentation velocity becomes higher, about 10 m/s.
Despite these purely geometric characteristics of the collision the sticking probability could be enhanced 
by special material properties,
such as sticky organic materials, magnetic or charged particles. However, as discussed in \cite{2008ARA&A..46...21B} 
the effects probably do not change the growth efficiency significantly. Another option is the aerodynamic re-accretion
as suggested by \cite{2001Icar..151..318W}. Here, the idea is that after a nearly destructive collision the small
fragments that surround the largest fragment fall back it due to aerodynamic drag exerted on them.
By this mechanism it is possible that the majority of particles become re-accreted leading to a net growth.
The process depends on the properties of the growing body in particular its porosity with respect to the gas flow through
it and on the speed of the ejecta, and its importance for a successful growth is not fully settled \cite{2008ARA&A..46...21B}.

In a process related to aerodynamic re-accretion the bouncing barrier can be overcome by a sweep-up process.
Using the standard results from the collision experiments in a coagulation code one finds for 
the general dust population that bouncing collisions prevent any growth above millimeter-sizes.
However, adding in the models a few cm-sized particles to a sea of smaller ones, 
which could happen in a real disk for example by vertical mixing or radial drift, these can
act as a catalyst by starting to sweep up the smaller particles, which results in very rapid growth.
As shown in \cite{2012A&A...540A..73W} using this mechanism, 100-m-sized bodies can be formed on a timescale of 1 Myr
at a distance of 3~AU from the central star.
In this process the existence of the bouncing barrier is highly beneficial for promoting growth, 
as it prevents the formation of too many larger particles that would otherwise destroy each other via fragmenting
collisions. Hence, a reservoir of small particles is maintained that can be swept up by larger bodies. 
The single requirement for this greatly enhanced growth process is the creation of a few {\it lucky} particles 
of cm-size or larger that can then sweep-up the smaller ones \cite{2012A&A...540A..73W}. 

A possibility to overcome the fragmentation barrier is through collisions of particle with very different sizes.
As shown in some experimental studies \cite{2005Icar..178..253W},
collisions between millimeter-sized dust projectiles and centimeter-sized dust targets lead to net mass growth of the
target up to collision velocities of 25~m/s. The authors suggest that for even higher velocities 
growth can be achieved which supports the idea that planetesimal formation via collisional growth is a
viable mechanism at higher impact velocities. 
Following up on this this idea of high velocities and different collision partners,
\cite{2012A&A...544L..16W} and \cite{2013ApJ...764..146G} demonstrated that it is the combination of a statistical 
relative velocity distribution function in the coagulation models and high-mass-ratio collisions that
allows for successful growth of larger bodies. The velocity distribution allows to overcome the bouncing barrier,
and the different masses to cross the fragmentation barrier \cite{2005Icar..178..253W}.
\cite{2013ApJ...764..146G} even suggest that via this mechanism the problem of planetesimal formation close to the central star,
the presence of millimeter- to centimeter-sized particles far out in the disk,
and the persistence of $\mu$-sized grains for millions of years, can be solved.

In the outer regions of the protoplanetary-disk the growth of particles is made easier by two effects. First, 
water has condensed to ice beyond the so-called {\it snow line} at temperatures below about 170~K, which increases the
local surface density of solids by about a factor upto 4 \cite{1981PThPS..70...35H}. 
In addition to the enhanced surface density of the solid particles the icy particles stick together up to much higher
relative velocities of about 50 km/s \cite{2011ApJ...737...36W}. The higher density and improved stickiness 
clearly will enhance the growth of small bodies. For disk parameter according to the MMSN
the snow line lies around 2.7~AU in good agreement with the location of the planets in the Solar System \cite{1981PThPS..70...35H}.
In addition, the sudden change in the opacity of the disk at the snow line will lead to a reduction in turbulent activity in the disk
and subsequently to a pressure maximum of the gas, which in turn will
further increase the local surface density of solid particles because they drift towards the local pressure maxima
\cite{1977MNRAS.180...57W}.
Hence, the inward drift of particle is prevented at the snow line and the growth of planetesimals enhanced
\cite{2007ApJ...664L..55K}.
Finally, considering the low catastrophic destruction threshold of 10\,m sized objects made of basalt, recent SPH simulations
have indicated that the inclusion of a porous equation of state allow non-fragmenting collisions to higher
velocities \cite{2011A&A...531A.166G}.


\subsection{Dust concentration}
\label{subsec:dust04}
As mentioned above there are several new suggestions available now how to overcome the growth barriers 
on the way from dust to planetesimals. The suggested mechanism rely on a better understanding of individual 
collisions and a statistical treatment of a whole ensemble of growing objects.
Another, often discussed mechanism to cross the growth barriers is the pre-concentration of particles
in inhomogeneities of the protoplanetary disk.
Often discussed has been the collection of particles in turbulent eddies or in vortices.
When the dust concentration has reached a critical value a streaming instability can set in \cite{2005ApJ...620..459Y},
and gravity can lead to rapid growth to larger objects \cite{2007Natur.448.1022J}.
Similarly gravitational instability in the condensed dust layer in the midplane of the disk dust layer 
could enhance planetesimal growth \cite{1973ApJ...183.1051G}.
These mechanisms have been presented in detail in the contribution by P.~Armitage and will not be discussed here further.


\begin{svgraybox}
The initial growth from dust to planetesimals suffers a growth crisis when particles reach a typical size of about decimeter to
meter. In this regime, the particles experience the fastest radial drift and possible loss into the star and
collisions often lead either to destruction or bouncing and no net growth. While this obstacle is presently not fully mastered,
several possible new paths have been investigated recently to solve the problem. These include a better understanding of individual
collisions, improvements in the statistical treatment as well as large scale collective effects by interaction with the ambient
disk.
\end{svgraybox}

%
%
%

\section{Terrestrial planet formation}
\label{lect:02}

In this chapter we describe the growth from planetesimals to full terrestrial type planets.
Following \cite{2006expl.conf..369C} we define a planetesimal to be an object large enough that 
the motion of nearby particles are significantly perturbed through gravitational interaction.
This implies that the velocity perturbations induced by planetesimals 
are larger than the typical drift velocity of particles which is about $10$ m/s. 
Hence, the term planetesimal refers then to objects of a km in size or larger, while below this
size the term pre-planetesimal is sometimes used \cite{2011A&A...531A.166G}.
As we shall see, this growth from km-sized planetesimals to a full grown terrestrial planet
several 1000 km in size proceeds in two major steps that are quite different in duration. 
In the first phase the planetesimals grow in a fast, runaway process to produce
a relatively small number of Moon to Mars sized {\it planetary embryos} (also called {\it protoplanets}).
After this phase, the inner parts of the protoplanetary disk contains very little
amounts of gas and the embryos grow via a much longer collisional phase to a set of
terrestrial type planets. 
Excellent overviews on this era in the history of the planet formation process 
have been presented elsewhere \cite{2004E&PSL.223..241C,2012AREPS..40..251M}. 

\subsection{Concepts}
\label{subsec:terr01-concepts}
Before going into the actual growth scenario from planetesimals to protoplanets we explain a few
important physical concepts or processes that are of relevance with respect to this phase in the planet formation process.
The problem we are facing when growing in size from few km-sized planetesimals to Moon-sized
planetary embryos, is the fact that for this mass range the aerodynamic drag forces become negligible and
cannot serve anymore to provide for any change in the relative velocity between particles necessary
to have collisions. Additionally, the induced inhomogeneities in the disk by the growing object
are not strong enough yet to provide significant drag and subsequent migration.
Finally, with initial 1 km-sized planetesimals
over $10^{11}$ particles are required to make an embryo. This is numerically very demanding as it requires
lots of particles and very long evolution times (a few 10 million years), see \cite{2006expl.conf..369C}.
Consequently, this part of planetary growth is described best by a combination of statistical and numerical methods.
%
\begin{figure}
    \begin{center}
        \includegraphics[width=0.90\textwidth]{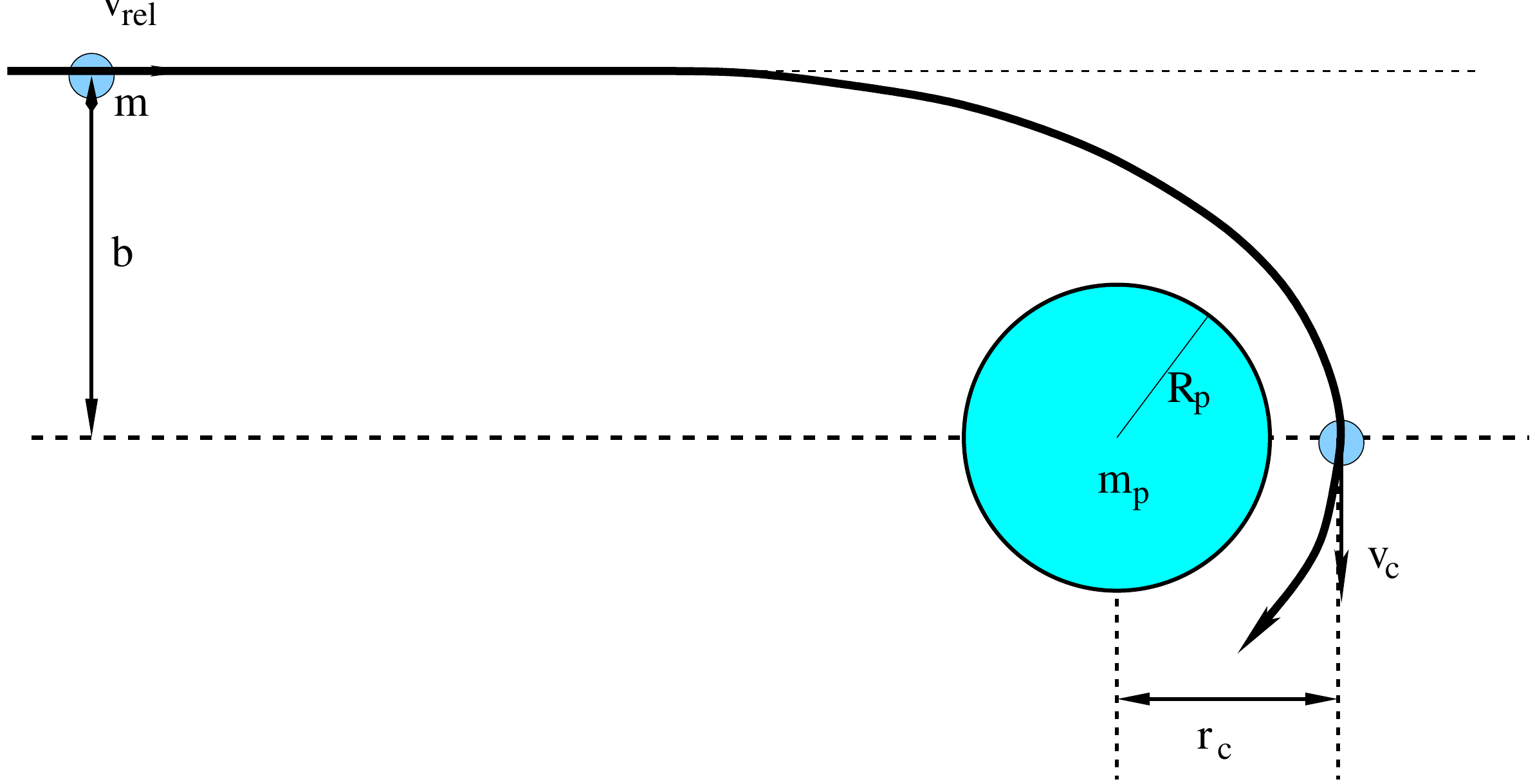} \\
    \end{center}
    \caption{Principle of the gravitational focusing process. A small body with mass $m$ approaches a larger
    planetesimal with mass $m\sub{p}$ with impact parameter $b$ and relative velocity $v\sub{rel}$.
    Gravitational attraction results in a close encounter with smallest distance $r\sub{c}$, and velocity $v\sub{c}$
    at the time of closest approach.
       }
    \label{fig:terr-focus}
\end{figure}
\subsubsection{Gravitational focusing} 
\label{subsubsec:terr01-focus}
The solid bodies can only growth via physical collisions, i.e. the shortest approach of two objects must
be smaller than the sum of their radii. Here, we assume that the growing 
planetesimal has a radius $R\sub{p}$ and that the accreted particle with radius $R$ is much smaller than this,
$R \ll R\sub{p}$.
The geometrical cross section for such a collision is then given simply by $\sigma_0 = \pi R\sub{p}^2$. 
If there were no other means of increasing $\sigma_0$ then planetary growth would finish soon.  
Fortunately, when the objects become a few km in size the situation becomes much more
favorable because the gravitational interaction between the two bodies will start to play an
important role in bringing the particles closer together, a process called {\it gravitational focusing}.
Let us consider the situation of two objects that have masses $m\sub{p}$ (the large body) and $m$ (the small body 
to be accreted) which have initially the relative velocity $v\sub{rel}$ with the impact parameter $b$, as shown 
in Fig.~\ref{fig:terr-focus}. 
To have a physical collision between the two objects the distance of closest approach, $r\sub{c}$, must be smaller than
$R\sub{p}$, i.e. $r\sub{c} < R\sub{p}$. 
As the sketch in Fig.~\ref{fig:terr-focus} implies the number of particles that possibly can directly hit the growing planetesimal
is greatly enhanced over the purely geometrical cross section $\sigma_0$.
The enhancement of the cross section through gravitational focusing 
can be obtained from the conservation of angular momentum
and energy, by considering the initial state when the objects are very far away from each other and the situation
of closest approach. We neglect the presence of the very distant central star and
considering only these two objects. To have a physical collision we require that $r\sub{c} = R\sub{p}$, and angular momentum conservation yields
\beq
\label{eq:terr-angmom}
                 b \, v\sub{rel}  = R\sub{p} \, v\sub{c} \,,
\eeq
where $v\sub{c}$ is the velocity at the closest approach (Fig.~\ref{fig:terr-focus}). 
Energy conservation (in the center of mass frame) gives 
\beq
\label{eq:terr-energy}
            \frac{1}{2} \mu \, v\sub{rel}^2 = \frac{1}{2} \mu \, v\sub{c}^2 
             - \frac{G (\mu M\sub{tot}) }{R\sub{p}} \,,
\eeq
with the reduced mass $\mu = m m\sub{p} / (m +m\sub{p})$ and total mass $M\sub{tot} = m +m\sub{p}$.
Here we assumed that initially the two particles are
sufficiently far apart such hat we can neglect the potential energy on the left side of eq.~(\ref{eq:terr-energy}),
and on the \RHS\ we evaluated the energy again at the point of closest approach.
Combining these two equations results in the following expression for the {\it effective cross section}, $\sigma$,
of the interaction
\beq
\label{eq:terr-grav-focus}
     \sigma \equiv  \pi b^2  \,  = \,
   \pi R\sub{p}^2  \, F\sub{grav} \, = \,
   \sigma_0
     \left[ 1 + \left(\frac{v\sub{esc}}{v\sub{rel}} \right)^2 \right] \,,
\eeq
where we introduced the joint {\it escape velocity}
\beq
      v_{\rm esc}  = \left( \frac{2 G  M\sub{tot} }{R\sub{p}} \right)^{1/2} \,,
\eeq
and the gravitative {\it focussing factor} over the purely geometrical cross section 
\beq
\label{eq:terr-enhance}
    F\sub{grav} =  \left[ 1 + \left(\frac{v\sub{esc}}{v\sub{rel}} \right)^2 \right]  \,.
\eeq 
From eq.~(\ref{eq:terr-grav-focus}) and Fig.~\ref{fig:terr-focus} we can see that for large relative velocities
only particles that arrive directly from the front of the object will collide with it, i.e.
the effective cross section is just the geometric one.
On the other hand for small relative speed, i.e. in a cold disk of planetesimals with $v_{\rm rel} \ll v_{\rm esc}$,
particles feel their mutual gravitational attraction for a longer time and hence particles from further away will be the drawn in.
The effective cross section is then {\it much} higher than without gravity, 
i.e. $\sigma \gg \sigma_0$ or $F_{\rm grav} \gg 1$. 
The quantity $\theta = (v\sub{esc}/v\sub{rel})^2$ 
in eqs.~(\ref{eq:terr-grav-focus}) or (\ref{eq:terr-enhance}) is sometimes called the {\it Safronov number}
after an early pioneer in studying the formation of the Solar System \cite{1972epcf.book.....S}.
If the approaching body is of size similar to the accreting planetesimal, then we can set $R\sub{p} \rightarrow  R\sub{p} + R$
in the above formula for $v_{\rm esc}$.
%
\begin{figure}
    \begin{center}
        \includegraphics[width=0.70\textwidth]{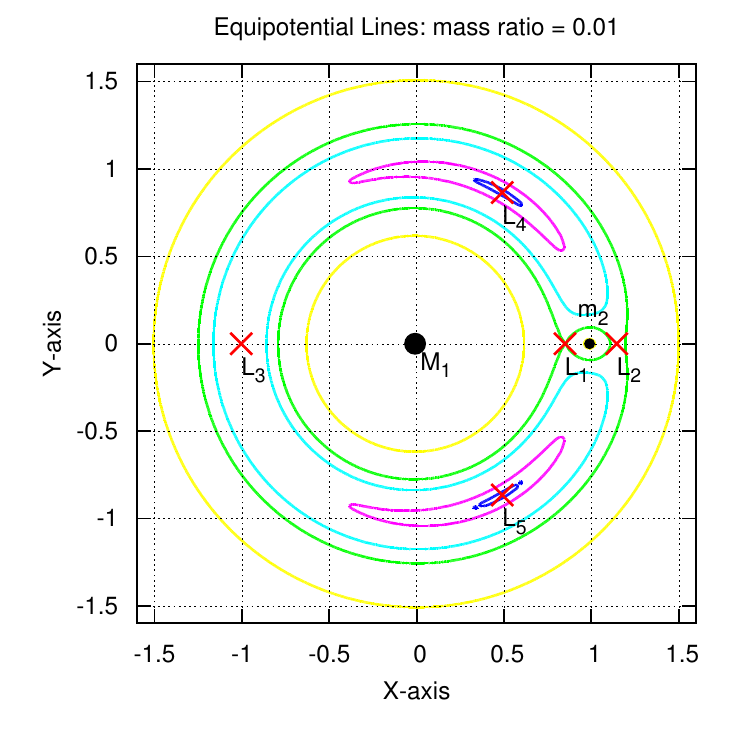} \\
    \end{center}
    \caption{The dynamical structure of the restricted circular three-body problem.
      Shown are equipotential lines of the
      corotating potential (eq.~\ref{eq:terr-potential-corot}) of a central star ($M_1$) which is orbited
      by secondary object ($m_2$). The classical 5 Lagrange points, $L_1$ to $L_5$ (marked by the red crosses),
      correspond to extrema of the potential.
     }
    \label{fig:terr-lagrange}
\end{figure}
\subsubsection{The Hill-radius}
\label{subsubsec:terr01-hill}
Another important concept in this context is the definition of the Hill-radius.
Here one considers the motion of a small (massless) particle in the presence of two larger 
gravitating bodies, that orbit each other on circular Keplerian motion. This is the so-called
circular {\it restricted three-body problem}.
In today's Solar System the primary (largest) object is the Sun, the secondary for example Jupiter and the third 
massless particle a small asteroid.
In our planet formation context the following objects are considered:
a protostar with mass $M_*$, the growing planetesimal (mass $m\sub{p}$) and a small particle
(much smaller mass) to be accreted onto the planetesimal.
As the motion of the star and the planetesimal is not affected by the small particle, these orbit each other with
Keplerian motion, i.e. with an orbital speed 
\beq
    \Omega_p = \sqrt{ \frac{G (M_*+m\sub{p})}{a\sub{p}^3} } \, \approx \, \sqrt{ \frac{G M_*}{a\sub{p}^3} }  \,,
\eeq
where $a\sub{p}$ is the semi-major axis of the planetesimal.

The type of motion of a massless (low mass) particle in the presence the other two can be most easily derived
from the effective gravitational potential written in a coordinate frame that rotates with the orbital speed,
$\Omega_p$, of the growing planetesimal see \cite{1999ssd..book.....M}.
In this corotating frame it reads
\beq
\label{eq:terr-potential-corot}
   \Phi (\vec{r} ) = - \frac{G M_*}{| \vec{r} - \vec{r_*} | } 
                     - \frac{G m_p}{| \vec{r} - \vec{r_p} | }
                     - \frac{1}{2} \, \Omega_p^2 \, r^2 \,, 
\eeq
where $\vec{r_*}$ and $\vec{r_p}$ denote the positions of the star and planetesimal, respectively.
The first two terms of the right hand side refer to the individual potentials of the star and planetesimal while
the last term denotes the {\it centrifugal potential} as we consider the motion in the corotating frame.
The equations of motion of the massless particle follow, as usual, from the gradient of the potential.
As can be inferred from eq.~(\ref{eq:terr-potential-corot}) that shows selected equipotential lines for a secondary to primary
mass ratio of $m_1/M_* = 0.01$.  The extrema of this potential are given
as the roots of a 5th order polynomial, and one obtains the classical 5 Lagrange points, $L_1$ to $L_5$,
as depicted in Fig.~\ref{fig:terr-lagrange}.
The Lagrange points mark possible equilibrium positions of the particle but only two of those are stable, $L_4$ and $L_5$,
which lie $60^\circ$ in front and behind the secondary.
Typically one finds that the orbits are around the primary object, either inside the distance of
the secondary (e.g. the main belt asteroids in the Solar System), or outside of it as indicated by the inner and outer
lines in Fig.~\ref{fig:terr-lagrange}.
For particles having the same semi-major axis as the secondary (here the planetesimal)  the motion is either around
one of the stable Lagrange points ($L_4$ or $L_5$) in so-called tadpole,
or around both in horse shoe orbits \cite{1999ssd..book.....M}.
The prime example from the Solar System are the Trojan asteroids that have the
same semi-major axis and hence period as Jupiter and orbit around $L_4$ and/or $L_5$.
For stability reasons these particles come never too close to
the secondary object as can be seen by the equipotential lines in
Fig.~\ref{eq:terr-potential-corot}. If the object is in the very close vicinity of the secondary
it is physically bound to it and orbits the secondary (e.g. the Galilean moons orbiting Jupiter).
For our planetesimal this is the case for particle distances within a sphere of radius
\beq
\label{eq:terr-Hill}
    R\sub{H} = \left( \frac{m\sub{p}}{3 M_*} \right)^{1/3}  \, a\sub{p} \,
\eeq
from the planetesimal. Here the index $p$ refers to the planetesimal.
This sphere is enclosed between the Lagrange point $L_1$ and $L_2$ of the corotating potential,
see Fig.~\ref{fig:terr-lagrange}.
Using direct $3$-body simulations of a star, a planetesimal and small particles it was shown \cite{1990Icar...87...40G}
that the motion of the particle to be accreted onto the planetesimal is highly chaotic in the vicinity
of the planetesimal, in particular inside of the Hill-radius $R\sub{H}$.
These 3-body effects lead eventually to a limitation of the gravitational
focusing factor $F\sub{grav}$, that would otherwise diverge for small $v\sub{ref}$, 
to maximum values of about $10^4$ \cite{1993ARA&A..31..129L}. 

\subsubsection{Modes of growth}
\label{subsubsec:terr01-mass}
\begin{figure}
    \begin{center}
        \includegraphics[width=0.60\textwidth]{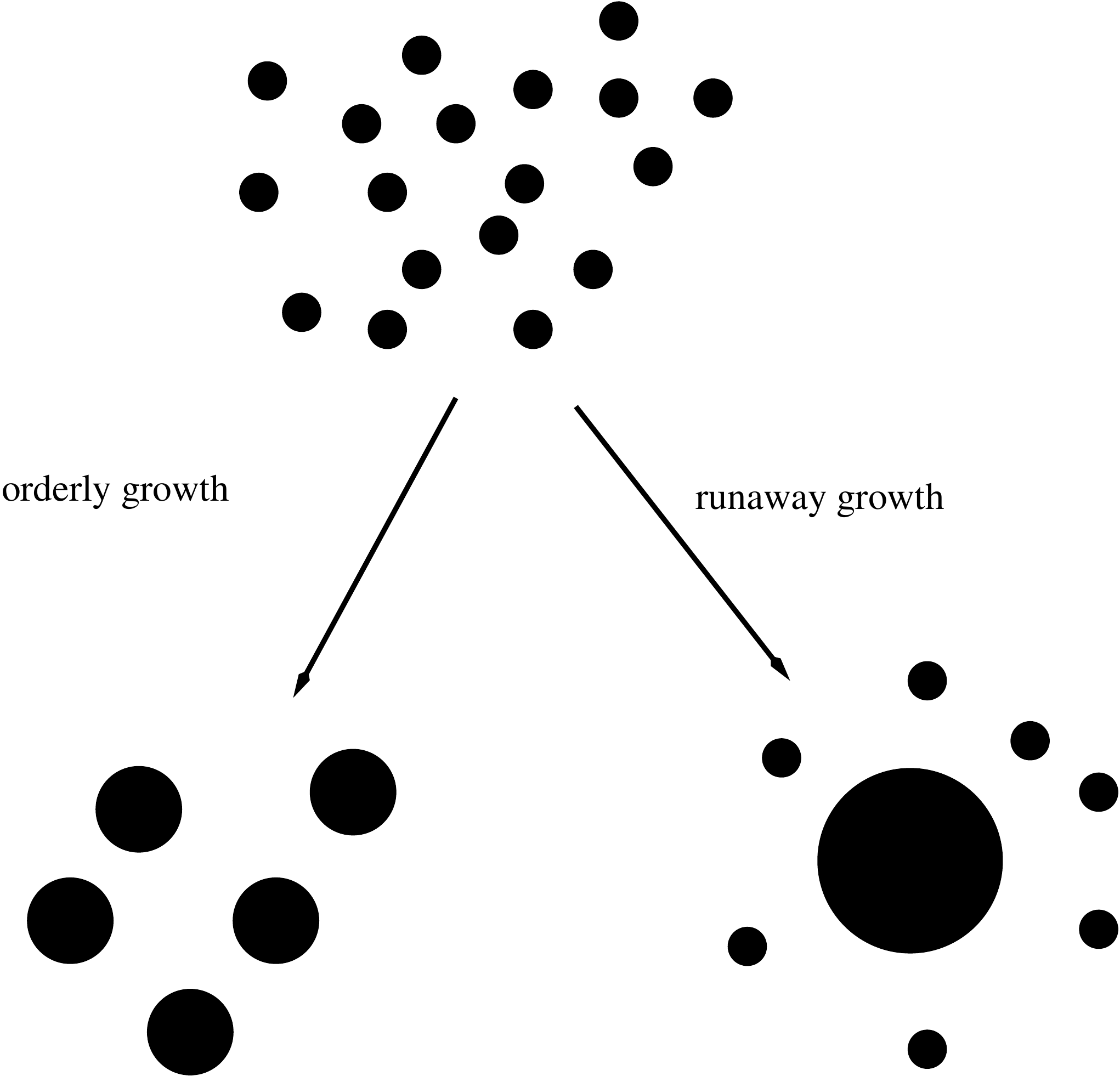} \\
    \end{center}
    \caption{The two modes of mass growth. In the case of orderly growth the whole ensemble has always
      similar particle sizes, while in the case of runaway growth one particle grows rapidly in a swarm of smaller ones. 
       (After E.~Kokubo \cite{2001RvMA...14..117K})
     }
    \label{fig:terr-modes}
\end{figure}

The accretion of small particles leads to an increase in mass of the growing planetesimal. In general one can distinguish
two modes of mass growth, ordered or runaway, as is schematically displayed in Fig.~\ref{fig:terr-modes}. 
In the ordered growth phase all objects grow at roughly the same rate and all the planetesimals in the whole ensemble
have approximately the same size, hence this mode is sometimes called {\it oligarchic} growth \cite{1998Icar..131..171K}.
On the other hand in the runaway case, one (or few) objects grow very rapidly at the expense of the smaller ones.
What type of growth mode operates can be analysed for example by considering the 
relative growth of two particles with mass $m_1$ and $m_2$ which can be obtained by expanding the time derivative
of the mass ratio
\beq
\label{eq:terr-mass}
    \doverdt{} \left( \frac{m_1}{m_2} \right) = \frac{m_1}{m_2}
     \left( \frac{1}{m_1} \doverdt{m_1} -  \frac{1}{m_2} \doverdt{m_2} \right)  \,.
\eeq
From eq.~(\ref{eq:terr-mass}) we can infer that the 
{\it relative mass growth}, $1/m (dm/dt)$, for each object is important.
Let us assume we start out with $m_1 > m_2$, then we see that this ratio is increasing if the 
relative growth increases with $m$,
because then the \RHS\ in (\ref{eq:terr-mass}) is positive.
This is exactly the situation for runaway-growth.
On the other hand, if relative growth decreases with $m$ we have an ordered growth because then the
mass ratio $m_1/m_2$ tends to unity.
As we shall see in the following both growth modes occur during the assembly of protoplanets, the first 
growth is via a runaway process which is followed later by an orderly growth.

Using the cross section from eq.~(\ref{eq:terr-grav-focus}) we find for the mass growth 
of a planetesimal with mass $m\sub{p}$ 
\beq
\label{eq:terr-massgrowth}
    \doverdt{m\sub{p}} = 
      \rho_{\rm part} \, v_{\rm rel} \, \sigma
    =  \rho_{\rm part} \, v_{\rm rel}  \, \pi R\sub{p}^2 \, F_{\rm grav}
\eeq
where $\rho_{\rm part}$ is the density of the incoming particles. 
In deriving eq.~(\ref{eq:terr-massgrowth}) we have assumed that 
each collision will results in growth (100\% sticking efficiency).
The outcome of collisions of km-sized objects can obviously not be studied in the
laboratory (where the maximum size is a fraction of a meter) and one has to rely on numerical simulations.
Here, the results indicate that typically the collisions lead to net accretion \cite{1999Icar..142....5B,2000Icar..146..133L}
unless the relative speeds are very high or the collisions are only grazing, but details depend on the internal strength of
the colliding objects \cite{2009ApJ...691L.133S}.

Before we evaluate (\ref{eq:terr-massgrowth}) for specific particle densities 
let us look at two illustrative examples (see \cite{2010apf..book.....A}) that illustrate the different growth modes.
Assuming a constant focusing factor, $F_{\rm grav} = const.$, 
we have for the mass growth of the planetesimal 
\beq
\label{eq:terr-linear}
      \frac{1}{m\sub{p}} \, \doverdt{m\sub{p}}  \propto m\sub{p}^{-1/3} \,.
\eeq
For objects with approximately constant density during the growth the mass scales as $m\sub{p} \propto R\sub{p}^3$,
and substituting this relation into eq.~(\ref{eq:terr-linear}) one finds $\dot{R\sub{p}} = const.$, which   
implies a linear growth of the particle with radius, $R\sub{p} \propto t$.
Assuming now constant relative velocities between growing planetesimal and
incoming particles, $v\sub{rel} = const.$, in eq.~(\ref{eq:terr-massgrowth}), and using the definition for 
the escape velocity then one obtains
\beq
\label{eq:terr-runaway}
      \frac{1}{m\sub{p}} \, \doverdt{m\sub{p}}  \propto R\sub{p} \, \propto m\sub{p}^{1/3} \,,
\eeq
which implies a growth of the particle to infinite mass, $m\sub{p} \rightarrow \infty$, in a finite time, corresponding
to strong runaway.
Of course, before this happens, the dynamics and space density of the ambient swarm of planetesimals will be changed 
which leads to a modification of relation (\ref{eq:terr-runaway}).

To obtain estimates of the actual growthrates of planetesimals within the protoplanetary disk
we assume that the incoming particle density is given by   
\[
        \rho_{\rm part} \approx \frac{\Sigma_{\rm part}}{2 H_{\rm part}}
                 = \frac{\Sigma_{\rm part} \Omega\sub{K}}{2 v_{\rm rel}} \,.
\]
Here $\Sigma_{\rm part}$ is the surface density of the particles, obtained by vertical integration over $\rho_{\rm part}$. 
To obtain the vertical thickness of the particle layer, $H_{\rm part}$, we assume that it is comparable to
the thickness of the gas density in the accretion disk, i.e. $H = c\sub{s}/\Omega\sub{K}$, where $c\sub{s}$ is the local sound speed and
$\Omega\sub{K}$ the Keplerian rotational angular velocity {(see Chapter by P.~Armitage)}.
In the case of the particle disk, we replace $c\sub{s}$
by the 'velocity dispersion', which is given here by the relative velocity $v_{\rm rel}$.
Using this in eq.~(\ref{eq:terr-massgrowth}) we obtain for the mass growth
\beq
\label{eq:terr-dmdt}
     \doverdt{m\sub{p}} = \frac{1}{2} \Sigma_{\rm part} \Omega\sub{K} \pi R\sub{p}^2
     \, \left[ 1 + \left(\frac{v_{\rm esc}}{v_{\rm rel}} \right)^2 \right]  \,.
\eeq
As can be noticed, for the mass increase, $dm\sub{p}/dt$, of the planetesimal the following conditions hold:
\begin{itemize}
\item growth is proportional to $\Sigma_{\rm part}$  
\item growth is proportional to $\Omega\sub{K}$, i.e. slower at larger distances 
\item $v_{\rm rel}$ enters only through the focusing factor 
\end{itemize}
During its growth the planetary embryo begins to influence and eventually alter its environment through its
increasing gravitational force which increases the velocity dispersion $v_{\rm rel}$.  
At the same time the particle density in its environment will be depleted, either due to accretion or scattering.
This will eventually lead to slow down of the runaway and the growth terminates.
For the relative velocity one can use here $v_{\rm rel} \approx \sqrt{e^2 + i^2} \, v\sub{K}$
where $e$ and $i$ are the (mean) eccentricity and inclination of the particle distribution and $v\sub{K}$ the Keplerian
velocity.
\subsection{Growth to protoplanets}
\label{subsec:terr02-proto}
Using the above estimate for the mass growth, eq.~(\ref{eq:terr-dmdt}), one can construct models that simulate the
growth of a whole ensemble of planetesimals to larger objects.
Numerically, this  growth process to protoplanets can be described by different methods
which combine statistical and direct methods, a topic that has been nicely reviewed by J.~Lissauer \cite{1993ARA&A..31..129L}. 
Here, we present briefly two types of approaches, the direct $N$-body method and a statistical method,
based on solving a Boltzmann type of equation.\\

\noindent
{\bf a) Direct N-Body methods} \\
In this method the equations of motion for $N$ planetesimals are solved by direct integration
of Newton's equations of motion. For the $i$-th planetesimal, which has the position $\vec{x}_i$,
the velocity $\vec{v}_i$ and mass $m_i$ the equation then reads
\beq
\label{eq:terr-nbody}
   \doverdt{\vec{v}_i} \, = \, - G M_* \, \frac{\vec{x}_i}{|\vec{x}_i|^3}
      \,  - \, \sum_{j\neq i}^N G m_j \frac{\vec{x}_i - \vec{x}_j}{|\vec{x}_i - \vec{x}_j|^3}
      \,  + \,  \vec{f}_{\rm gas} \,  + \, \vec{f}_{\rm coll} \,.
\eeq
In addition the positions need to be updated via 
\beq
    \doverdt{\vec{x}_i}  = \vec{v}_i  \,.
\eeq
In eq.~(\ref{eq:terr-nbody}) the first term on the \RHS\ is the gravitational force of the central star, the second refers
to the gravitational attraction of the $N-1$ other planetesimals, $\vec{f}_{\rm gas}$ is the frictional force 
exerted on the planetesimals by the gas in the protoplanetary disk, and
$\vec{f}\sub{coll}$ is the velocity change upon collisions between the individual planetesimals.
For details how to model these forces see for example \cite{2001RvMA...14..117K} and references therein.
The velocity dispersion of the growing planetesimals, $v_{\rm disp}$, is damped by the gas drag which enhances
their growth because of the reduced relative velocity between them, see eq.~(\ref{eq:terr-massgrowth}).
The advantage of this direct method is its accuracy because the growth of each individual particle
is modeled, and this is also its disadvantage because it requires to follow the evolution of very many particles.
Since it is impossible to include all planetesimals in such a simulation, the numerical computations follows
the evolution of so called super-particles that represent a sample of many planetesimals \cite{1998Icar..131..171K}.
In treating the outcome of physical collisions, the total momentum has to be conserved, while energy will have to
be dissipated in the growth processes.
Special numerical methods have been developed to integrate the $N$ gravitationally interacting bodies accurately over long times
\cite{1999PASP..111.1333A}.
\\

\noindent
{\bf b) Statistical method}  \\
In this method the mean density of particles in phase space, i.e. the 
probability distribution function $f(\vec{r}, \vec{v})$ is evolved in time.
The function $f$ gives the density of particles per space and velocity interval, such
that the particle density is given by integration over all velocities, $ n(\vec{r})  = \int f d^3 v $.
The evolution of the distribution is described by the collisional Boltzmann-equation
\beq
   \pdoverdt{f} \, + \, \dot{\vec{r}} \pdoverd{f}{\vec{r}}
  \,  + \, \dot{\vec{v}} \pdoverd{f}{\vec{v}} \, =
  \,  \left.\pdoverdt{f}\right|\sub{coll}
  \,  + \left.\pdoverdt{f}\right|\sub{grav}  \,,
\eeq
where $f_{\rm coll}$ describes the changes by individual collisions
and $f_{\rm grav}$ the gravitational scattering by the other particles.
It is typically assumed that the motion of individual particles is given approximately by Keplerian
orbits with eccentricity $e$ and inclination $i$ with randomly oriented orbits. 
Then, the distribution function $f$ can be simplified to follow a Rayleigh distribution
\beq
       f\sub{R} (x) \propto  \frac{x}{\sigma^2} \, \exp{\left[- \frac{x^2}{2 \sigma^2}\right]} \,,
\eeq
where $\sigma$ is related to the mean value of the distribution.
Here $f\sub{R}$ has two arguments, $e$ and $i$, given by $f\sub{R}(e,i)$ (see \cite{1993ARA&A..31..129L} for details).

The actual growth of the particles in this method is described by the 
coagulation equation
\beq
     \doverdt{n_k} \, = \, \frac{1}{2} \sum_{i+j=k} A_{ij} n_i n_j
  \, - \, n_k  \sum_{i=1}^\infty A_{ik} n_i  \,,
\eeq
where $n_k$ is proportional to the number of particles with a given mass $m_k$, and $A_{ij}$ represent the outcome
of physical collisions between two planetesimals. The first term on \RHS\ describes a gain in the number of objects
with mass $m_k$ and the second one a loss.
The outcome of individual physical collisions depends in a complicated way on the velocities and masses 
of the collision partners and requires extensive parameter studies \cite{2002Icar..159..306L,2009ApJ...691L.133S}.
The advantage of the Boltzmann method is that the complete ensemble is modeled, the disadvantage is its statistical nature.
How these kind of equations are actually solved numerically and the application to
planetesimal growth have been described in detail elsewhere, see \cite{1993Icar..106..190W,1997Icar..128..429W} and references therein.

\begin{figure}
    \begin{center}
        \includegraphics[width=0.45\textwidth]{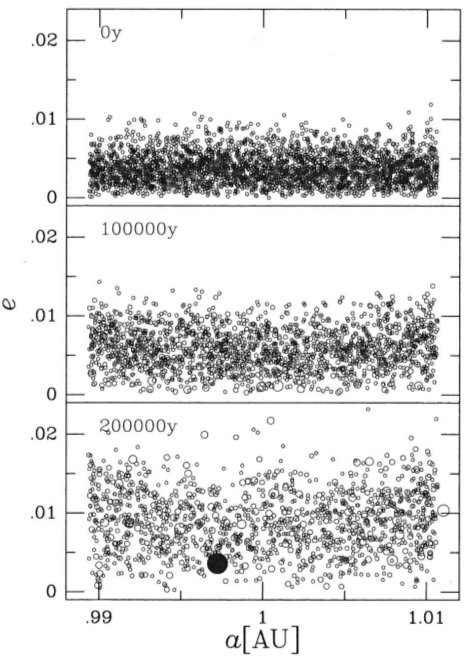} 
        \hfill
        \includegraphics[width=0.52\textwidth]{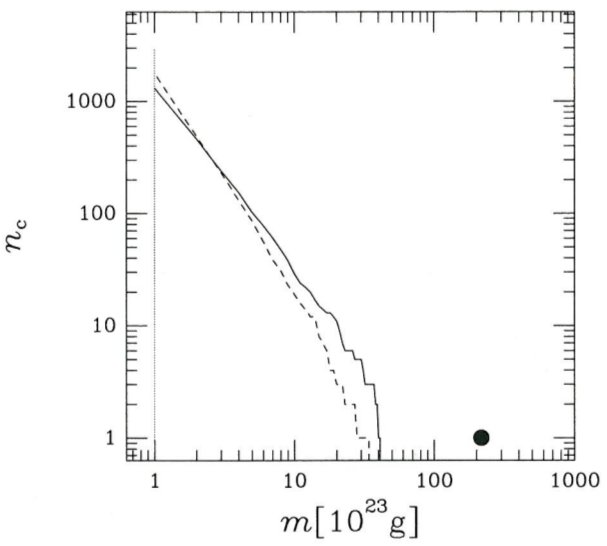} \\
    \end{center}
    \caption{Example of the runaway growth process from planetesimals to planetary embryos using the $N$-body method
    as described by Kokubo \cite{2001RvMA...14..117K}. The basic setup and initial conditions are given in the main text
    in section \ref{subsubsec:illustrative}.
    The left panel shows the eccentricity and distance distribution of the bodies for the initial setup (top panel) and
     at times 100,000 and 200,000 yrs (middle and bottom). 
   The right panel shows the cumulative mass distribution of the formed bodies 
     where $n\sub{c}(m)$ = number of particles with mass larger than $m$.
   The results are shown at the same times as in the left panel: 
    $t=0$ (vertical thin line), $t=100,000$ (dashed) and $t=200,000$ (solid).
    The large bullet in both panels denotes the single outstanding large object that has formed through a runaway process.
      From E.~Kokubo \cite{2001RvMA...14..117K}}
    \label{fig:terr-nbody}
\end{figure}

\subsubsection{An illustrative example}
\label{subsubsec:illustrative}
To be specific, we present in the following an example for the typical outcome of the planetesimal growth process. 
In \cite{2006expl.conf..369C} representative results of the second (statistical) type of approach
have been described in more detail \cite{1993Icar..106..190W,1997Icar..128..429W}
and we refer the reader to that excellent summary. 
For a complementary point of view, we summarize here results of the $N$-body approach, that
has been used for example by E.~Kokubo \cite{1998Icar..131..171K,2002ApJ...581..666K} to describe planetesimal growth.
In \cite{2001RvMA...14..117K} the main results of such simulations are presented, and we give here a short summary.
Initially a number of planetesimals are spread over certain region around the central star, here the Sun, 
typically centered at $1$AU with certain radial width. 
In \cite{2001RvMA...14..117K} a radial extent of 0.02 AU, centered a $a= 1$ AU, was chosen.
The simulations used 3000 bodies, each with an initial mass of $m=10^{23}$g. The mean material density of the growing planetesimals was
assumed to be $\rho_{\rm p} = 2$g\,cm$^{-3}$. The whole ensemble was evolved in time for several 100,000 yrs.

As shown in Fig.~\ref{fig:terr-nbody}, starting from the set of equal mass particles, at time $t=200,000$yrs the
distribution has evolved towards the situation where one large body (the $\bullet$) has formed
which has about 200 times the initial mass. 
It is embedded in a sea of smaller particles that have a continuous
mass distribution, see right panel in Fig.~\ref{fig:terr-nbody}.
The low eccentricity (and inclination) of ($\bullet$) comes through dynamical friction with the small objects.
Through (distant) gravitational interactions the smaller particles are dynamically excited,
and their mean $e$ and $i$ increase in time. 
Using equipartition of energy between $e$ and $i$ yields on average the following relation
$<e^2> = 4 <i^2> $ where $<x>$ denote mean values averaged over the ensemble of small particles.
The one large object orbits the star on a nearly circular orbit.

For the same $N$-body simulation, the right panel in Fig.~\ref{fig:terr-nbody} shows the cumulative 
particle distribution, after $10^5$yrs (dashed), and after $2\times 10^5$yrs (solid).
The objects between $10^{23}$-$10^{24}$g contain the majority of mass of the whole sample.
The distribution follows a power-law
\beq
     \doverd{n\sub{c}}{m} \propto  m^q \,.
\eeq
Here, $q$ describes the exponent of the power-law mass distribution, where a
value of $q=-2$ is equivalent to equal mass in each logarithmic mass bin.
A steeper distribution, $q < -2$, is characteristic for a runaway process \cite{1993prpl.conf.1061L},
where only very few particle reach larger masses.
Indeed, in the simulation shown in Fig.~\ref{fig:terr-nbody} the slope is about $q \approx -2.7$ and
one very massive particle ({$\bullet$}) 
is separated from the continuous distribution, i.e. it serves as a sink of particles.
These results indicate very clearly that in the early phase the planetesimal growth proceeds through a runaway phase.

Very similar results to those shown in Fig.~\ref{fig:terr-nbody} are obtained
for example by \cite{1993Icar..106..190W} and \cite{1997Icar..128..429W} using the statistical method, see
summary in \cite{2006expl.conf..369C}.
Their distribution of particle sizes follows a very similar slopes to that shown in the
right panel of Fig.~\ref{fig:terr-nbody}, again indicative of runaway growth.

\subsubsection{The end of the growth}
Obviously a runaway process, as just described, cannot continue forever. It is slowed down and eventually stopped mainly by
two processes. Upon growing to larger objects the planetesimals stir up their environment such that the
velocity dispersion is increasing and hence the relative velocity between them which leads, according to
eq.~(\ref{eq:terr-grav-focus}) to a reduction in the collisional cross section. Secondly, the accretion process
reduces the local density of particles, $\rho\sub{part}$, and the body isolates itself from further growth, because
fewer and fewer collision partners are available within the {\it feeding zone}.
One can get an estimate on the final size an embryo can reach, the {\it isolation mass} $M\sub{iso}$,
by assuming that the volume of accretion, the feeding zone, has a radial extend given by the width of the horse shoe region, 
which is approximately given by the Hill-radius (\ref{eq:terr-Hill}).
The total mass of all the particles within a region $\Delta a$ inside and outside of semi-major axis $a$ is given by
$ m =  2 \pi \, 2 a \, \Delta a \Sigma\sub{part}$. Using now $\Delta a = C R\sub{H}$  
we obtain with $m = M\sub{iso}$
\beq
\label{eq:lect02-iso}
    M\sub{iso} = 4 \pi a \, C \, \left(\frac{M\sub{iso}}{3 M_\odot}\right)^{1/3} \,
            a \Sigma\sub{part} \,,
\eeq
where $C$ is a factor of order unity \cite{2010apf..book.....A}.

We consider now for example the growth of the Earth in the Solar System, and assume that there were initially 
$2 M_{\oplus}$ located between $0.5$ und $1.5$AU. Using the standard condition of the MMSN 
with $\Sigma\sub{part} \sim a^{-3/2}$ and $\Sigma\sub{part} = 8$ gcm$^{-3}$ at 1AU and $C = 1$, then
we obtain for the isolation mass
\beq
     M\sub{iso} \approx 0.05 M_{\oplus} \,.
\eeq
This runaway process is essentially a local phenomenon, because the embryos accrete primarily from their immediate neighborhood,
their feeding zone.
The limited extend of each embryo's feeding zone implies that several objects in the protoplanetary nebula will 
experience runaway growth and grow at a similar rate. This is the {\it oligarchic phase} of terrestrial planet formation which
results eventually in about 40 planetary embryos which have a mean spatial separation of about
$\approx 0.01 - 0.025$AU \cite{1987Icar...69..249L,2004E&PSL.223..241C}.
The timescale for this growth of the oligarchs is about 0.1-1.0 Myr \cite{1997Icar..128..429W,2003Icar..161..431T}.
Due to the locality of the runaway growth, any radial compositional gradient present in the protoplanetary
disk should be reflected in the embryos' chemical compositions \cite{2012AREPS..40..251M}. 
Starting from these protoplanets the final assembly of the terrestrial planets can ensue.

\subsection{Assembly of the terrestrial planets}
\label{subsec:terr03-terr}
After the oligarchic phase there are only a few objects, the {\it embryos}, left over 
with masses of about Moon to Mars size.
The sea of planetesimals has mostly been depleted and only the gravitational 
interaction between these planetary embryos remains, i.e. in contrast to the previous growth phases
the problem is physically relatively clean. 
To model this final assemblage, classical $N$-body simulations are the standard choice.
In principle this is a straight-forward exercise because there are only very few particles ($\approx 100$) left over
whose motion needs to be integrated, but this process occurs over a very long time scale, about $10^7 - 10^8$ yrs.
Hence, the longterm integration of the equations of motion (similar to eq.~\ref{eq:terr-nbody} with vanishing gas and collision terms)
requires good symplectic integrators that conserve automatically the total energy of the system.
A well known, and often used example is the MERCURY-code developed by J.\,Chambers \cite{1999MNRAS.304..793C} which
is publicly available. Other codes are for example the SWIFT-package \cite{1998AJ....116.2067D} 
or the REBOUND-code \cite{2012A&A...537A.128R}, a modern $N$-body code, with the capability to treat collisions.

As an example we discuss briefly the results of Chambers et al.~\cite{1998Icar..136..304C,2001Icar..152..205C}.
The authors performed a series of $N$-body simulations, where as starting conditions they used
about 50 embryos in the first paper \cite{1998Icar..136..304C} and about 155 embryos in 
the second paper \cite{2001Icar..152..205C}.
In total about 2 $M_\oplus$ were distributed between 0.3 and 2.0 AU with different types
of initial mass distributions: all equal, bimodal or with a radial mass profile.
In all simulations Jupiter and Saturn were included on their present day orbits.
The collisions were treated as 
100\% sticking (perfectly inelastic) and the angular momentum of the coalesced bodies went into their spin.
The presence of Jupiter and Saturn may be surprising at this still early phase during the growth
of the terrestrial planets but, as we shall see in the following 
chapter, the formation timescale for these massive planets is indeed shorter than the time necessary for
the final assembly of the terrestrial planets. 

The results of those type of simulations show that indeed planetary system containing several terrestrial
planets are produced. The formation timescale is a few $10^7$ years, which is very long compared to the
previous phases. The reason is the fact that in this final growth phase only very few objects are remaining
which reduces the frequency of mutual collisions considerably and it takes a long time to 
produce full grown terrestrial planets. 
The presence of the massive planets
Jupiter and Saturn is required as they dynamically stir up the sample of embryos and prevent the formation of
an additional planet within the region of the main belt asteroids. The whole evolution is a highly {\it chaotic process}
because objects from different regions are scattered around and lead to a variety of collisions starting from
head-on to near misses. Many objects may be lost as they fall into the Sun, and 
it is estimated \cite{2001Icar..152..205C} that about 1/3 of the initial objects within 2\,AU may have to fear this fate. 
The typical outcome of these $N$-body simulations is a system with 3-4 planets on stable well separated orbits,
with the tendency for more planets in those runs that have more initial embryos.
Hence, the final systems in these simulations resemble roughly the situation in the
Solar System where the most massive planets are in the Venus-Earth region, while the innermost planets and those
in the Mars region are on average much smaller. The smallness of the Mercury type objects can be understood in terms
of the high collision speeds for this innermost orbits that often lead to fragmentation rather than growth.
Indeed the smallness and high density of
Mercury can be attributed to a high speed impact during this chaotic phase of terrestrial planet formation
\cite{2007SSRv..132..189B}. Most of the objects initially residing within the main asteroid region are scattered
out due to resonant action by Jupiter and Saturn \cite{2006Icar..184...39O}.

Nevertheless there are important differences when compared directly to the Solar System:
the mass concentration in the planets is not as high as for Venus and Earth, the planets have on average too high $e$
and $i$, and the spin-orientations are arbitrary. Specifically, the typical mass of a 'Mars'-object turns out to be
too large when compared to the Solar System, by a factor of about 5.
The presence of residual gas from the protoplanetary disk will reduce the eccentricities of the 
growing objects and shorten the formation time but too much gas will lower the collision rates such that
massive planets like Venus and Earth will not form at all \cite{2002Icar..157...43K}.
More recently, new evolutionary $N$-body simulations have been performed, some using a much larger number of initial bodies, 
a few thousand, that were spread over a wider radial domain ranging from 0.5-5.0 AU 
\cite{2004Icar..168....1R,2006Icar..183..265R,2006Icar..184...39O}. 
The influence of several input parameters to the simulations, such as the disk mass and radial density profile, 
the particle distribution in space and in mass, the orbits of the giant planets,
and the treatment of collisions have been analysed in detail in more elaborate simulations, see overview in \cite{2012AREPS..40..251M}.
From these one can infer that for the reduction of the orbital
excitement of the terrestrial planets the dynamical friction of the remaining population of planetesimals plays
an important role. Concerning the orbital architecture of the giant planets it was shown that for more eccentric giants
the terrestrial planets grow faster and have more circular orbits \cite{2009Icar..203..644R}. 
While the overall architecture of the formed planetary systems resembles approximately the Solar System, the 'Mars-problem'
still remains.
One scenario to overcome the problem of the too small Mars is the {\it Grand Tack} scenario where during the early 
Solar System the giant planets Jupiter and Saturn migrated first far inward and then turned around to move out to their present
locations \cite{2011Natur.475..206W}. 
We will not discuss this scenario any further in contribution and refer the reader to excellent reviews 
\cite{2012AREPS..40..251M,2014prpl.conf..595R}.
 
 
\begin{svgraybox}
The formation of terrestrial planets from planetesimals proceeds in different steps. In the first phase the gravitational attraction
between the growing planetesimals leads to a fast runaway growth which is followed by a slower oligarchic growth phase,
at the end of which an ensemble of about 50 Moon to Mars sized objects has formed, spatially  well separated.
On timescales of tens of millions of years these
planetary embryos evolve under their mutual gravitational force and form through a sequence of collisions and impacts 
terrestrial type planets and the cores of giant planets. 
\end{svgraybox}

%
%
%
\section{The formation of massive planets by core accretion}
\label{lect:03}
In this section we describe how the growth of massive, gaseous planets is believed to proceed.
The massive planets of the Solar System are about 100 times (Saturn) or 300 times (Jupiter) more massive
than the Earth, which is the most massive object in the terrestrial planet region. 
Their mean densities are roughly comparable to that of water or even lower in the case of Saturn which implies that
they consist of a huge amount of gas in addition to a possible central solid core. 
The question arises, how it is possible to collect large amounts of Hydrogen and Helium to form those planets
under the conditions in the protoplanetary nebula. These lightest elements are highly volatile and difficult
to condense. Two main pathways for the formation of massive planets have been discussed \cite{2009ApJ...695L..53B}. 

In the first scenario, the {\it core accretion} model, an initial seed object forms onto which the gas can later accumulate.
It is a bottom-up process, where initially a solid core forms along a similar evolutionary path as for the embryos in the formation of
terrestrial planets. Upon reaching a certain critical mass the gravity of the core becomes high enough that a rapid,
runaway gas accretion onto the core is possible, leading eventually to the formation of a gaseous planet. 
In the second scenario, the {\it gravitational instability} (GI) model, the formation pathway is similar to that of star formation
and it is a top-down process. The scenario is believed to operate if the initial gas density of the protoplanetary disk
is so high that a dynamical gravitational instability ensues that leads directly to the collapse of a local patch of the disk. 

The physical structure of the gaseous planets in the Solar System has been studied extensively from the ground and 
through space missions.
For the Solar System, the favored scenario is the core accretion model because it explains straightforward the existence of cores
in the centers of the massive planets and the large amount of solids, with Jupiter enriched about $1.5-6$ times above solar
composition and Saturn about $6-14$ times \cite{2004ApJ...609.1170S}.
Additionally, the atmospheres of Saturn and Jupiter are enriched in metals and in noble gases,
in particular Argon \cite{2014arXiv1405.3752G}.
The formation via the GI process may have operated in the creation of the observed
distant, directly imaged extrasolar planets. 
In this section we first focus on the core accretion pathway, while the GI model will be discussed below
in the next section \ref{lect:04}. The status of knowledge about the internal structure of the Solar and the extrasolar giant planets
are reviewed in \cite{2010SSRv..152..423F} and recently in \cite{2014arXiv1405.3752G}. 
A classic review of this phase of planet formation is presented in \cite{1993ARA&A..31..129L},
while modern reviews are given in the relevant chapters of the {\it Protostars and Planets} series, here in 
PPVI \cite{2014prpl.conf..643H,2014prpl.conf..619C}.
\subsection{Background}
\label{subsec:gas01-back}
Planetary growth in the core accretion model consists of three phases. 
In the first phase a solid object forms in a manner similar
to the processes that have led to the planetary embryos in the case of the terrestrial planets. In the previous section we have analysed
the isolation mass of a growing body which is given by the amount of solid material that can be accreted directly from the 
{\it feeding zone} of an embryo.
In contrast to the terrestrial region of the protoplanetary disk the abundance of solids is much higher at the location of
Jupiter because of the low temperatures in the disk which allows for the condensation of many additional molecules, most
importantly water.
H$_2$O is the most abundant molecule in the Universe and in the disk it freezes out to ice
beyond the so-called {\it snowline}. The exact condensation temperature depends on the ambient gas pressure but for conditions in the
protosolar nebula $T$ has to be smaller than $(150-170)$K for ice to condense. For a passively heated disk where the stellar irradiation
dominates the heat input to the disk, one finds for a Solar type star
$r\sub{cond} \approx 2.7$\AU\, for the condensation radius. Hence, in the standard Hayashi-model of the protosolar nebula,
for distances to the star larger than $r\sub{cond}$, 
the amount of solids available for embryo formation is typically assumed to be 4 times higher than in the inner regions
\cite{1981PThPS..70...35H}. While newer calculations indicate possibly a lower value \cite{2014A&A...570A..36M}, we still
use the standard value here, and then surface density for solids $\Sigma\sub{s}$ is given by
\beq
   \Sigma\sub{s}(\mathrm{rock/ice}) = 30 \left(r/1 \mathrm{AU}\right)^{-3/2}  \, \mbox{g/cm$^2$ \quad for} \quad r > 2.7 \mbox{AU}  \,.
\eeq     
Using the formula given in the previous section for the {\it isolation mass} we obtain at the current distance of Jupiter ($a_{\rm Jup} = 5.2$\AU)
the following estimate
\[
    M_{\rm iso} \approx \, (5 - 9)  M_{\oplus}  \,,
\]
which is considerably higher than in the terrestrial region.
For the subsequent phases in the formation of giant planets the core mass is an important quantity, as it determines
whether the embryo can capture a sufficient amount of gas in a short timescale to become a giant.

The first phase of gas accretion proceeds in a slow hydrostatic manner.
The question arises: How much of an atmosphere can a growing planet hold? To answer this question, we assume that the minimum requirement
for an atmosphere is that the escape velocity from the planet is larger than the sound speed in the ambient disk $v_{\rm esc} > c\sub{s}$.
In the discussion below we follow P.~Armitage \cite{2010apf..book.....A} and use 
\[
    m\sub{p} = \frac{4 \pi}{3} \rho\sub{m} R\sub{p}^3, \quad \quad
    v_{\rm esc} = \sqrt{\frac{2 G m\sub{p}}{R\sub{p}}},  \quad \mbox{and} \quad  c\sub{s} = \frac{H}{r} u\sub{K} \,,
\]  
where $m\sub{p}, R\sub{p}$ denote the mass and radius of the growing planet (embryo) with density $\rho\sub{m}$,
$v_{\rm esc}$ is the escape velocity from its surface, 
$c\sub{s}$ is the sound speed and $H$ the vertical half-thickness of the disk.
For an icy body at 5 AU around a Solar mass star we will have some atmosphere (where $v_{\rm esc}$ begins to be larger than $c\sub{s}$) for
	$m\sub{p} \, \gtrsim \, 5 \cdot 10^{-4} M_{\rm Earth}$. This is a very small mass planet,
but such an atmosphere has no dynamical importance.

Let us consider the situation where the atmosphere or rather envelope takes up a sizable fraction, 
$f_{\rm env}$, of the planet such that $M_{\rm env} > f_{\rm env} \, m\sub{p}$.
For an isothermal atmosphere with $c\sub{s}(atm) = c\sub{s}(disk)$, and where the outer density matches
that of the ambient disk ($\rho_0$) one obtains at
$r=5\AU$ with $\rho_0 = 2 \cdot 10^{-11}$g/cm$^3$, $c\sub{s} = 7 \cdot 10^4$cm/s for the minimum mass of a planet to hold
an atmosphere of about 10\% of its mass ($f_{\rm env} = 0.1$) \cite{2010apf..book.....A} 
\beq
         m\sub{p} \gtrsim 0.2 M_{\oplus} \,.
\eeq
As a comparison, at the location of the Earth, one obtains for $f_{\rm env} = 0.1$ at 
$r=1$\AU\ using $\rho_0 = 6 \cdot 10^{-10}$g/cm$^3$, $c\sub{s} = 1.5 \cdot 10^5$cm/s, and $\rho\sub{m}=3$g/cm$^3$
\beq
         m\sub{p} \sim  M_{\oplus} \,,
\eeq
in very rough agreement with the actual conditions in the Solar System.
Planets are assembled over the life time of the disk and the mass of the embryo will be given by the isolation mass
as defined in eq.~(\ref{eq:lect02-iso}).
To hold an atmosphere, $M_{\rm iso}$ must be larger than the critical mass to acquire an atmosphere.
An analysis \cite{2010apf..book.....A} shows that inside $r_{\rm cond}$ the isolation mass is always smaller than the mass to hold an atmosphere,
while outside $r_{\rm cond}$ the isolation mass is larger. This relation explains qualitatively very well the fact that
the inner planets in the Solar System consist of terrestrial planets with very little atmosphere while outside we find
the gas or gas/ice giants with a very extended envelope.
Further details will depend on the actual disk model, and the final location where gas giants
are found eventually will depend on the amount of migration that occurred.
To understand the final phase of the assembly of massive planets, we will have to study in more detail the internal structure
of the growing giants.
\subsection{The growth to a giant}
\label{subsec:gas02-giant}
To calculate the structure of a growing star or planet one assumes typically that the overall
evolution of the mass growth proceeds on a timescale that is very long in comparison to
the adjustment timescale of the interior. The latter can be estimated on the basis of the sound crossing time
of an object that is of the order $\tau_{\rm cross} \sim (G \bar{\rho})^{-1/2}$, where $\bar{\rho}$ denotes the mean density
of the body. For a growing planet  $\tau_{\rm cross}$ is only a few minutes.
Indeed, this is much smaller than typical accretion time scales, and the evolution of the growing planet
can be well described by a sequence of {\it hydrostatic models}, similar to stellar evolution.
 
Hence, the basic equations that describe the inner structure of a growing planet are given by
the standard stellar structure equations \cite{1990sse..book.....K}, adapted to
the planet formation process \cite{2005AREPS..33..493G} 
\bea
   \mbox{\small mass conservation:} \hspace{1.0cm} \quad  \frac{d m}{d r} & = &
     4 \pi r^2 \rho 
    \label{eq:masscon}  \\
   \mbox{\small hydrostatics:} \hspace{1.1cm} \quad  \frac{d p}{d r} & = &
     - \frac{G m(r)}{r^2} \, \rho   \label{eq:hydrostat} \\
   \mbox{\small radiative diffusion:}  \hspace{0.7cm} \quad  \frac{L(r)}{4 \pi r^2}
    & = &
     - \frac{4 \, \sigma T^3}{3 \, \kappa\sub{R} \rho} \, \frac{d T}{d r}  \\
   \mbox{\small energy generation:}  \hspace{1.0cm} \quad  \frac{d L}{d r}
    & = &
     - 4 \pi r^2 \rho \left(  \epsilon  - T \frac{\partial S}{\partial t} \right)
     \label{eq:energy}
\eea
with: $m(r)$ mass interior to the radius $r$; $L(r)$ luminosity at radius $r$; 
$\epsilon$ internal energy generation, $\sigma$ Stefan-Boltzmann constant, and 
$\kappa\sub{R}$ Rosseland-opacity. 
Equations (\ref{eq:masscon} - \ref{eq:energy}) are essentially the same equations as for stellar interiors with the difference that the
luminosity is not due to nuclear burning of hydrogen but due to the contraction and cooling of the gas that
changes the entropy $S$, as given by the last term in the energy equation. Impacting planetesimals 
give an additional contribution.
In case of convection (for a super-adiabatic stratification) the adiabatic temperature gradient is used.

To calculate the structure of the whole planet, from the very center to the surface, equations of states for matter under
extreme conditions are required. A simple estimate for the {\it minimum} central pressure, $p\sub{c}$, can be obtained from the hydrostatic
equation using the assumption of a constant density inside the body, $\rho(r) = \bar{\rho}$. 
Integrating eq.~(\ref{eq:hydrostat}) then yields
\beq
\label{eq:pcentral}
       p\sub{c} =  \frac{1}{2} \, \bar{\rho} \, \frac{G M}{R} \,,
\eeq
where $M$ and $R$ are the total mass and radius of the planet.
For the observed mass and radius of Jupiter 
($M_{\rm Jup} = 1.9 \cdot 10^{27}$kg $= 318 M_{\rm Earth}$ and $R_{\rm Jup}= 7 \cdot 10^7$m $=11.2 R_{\rm Earth}$) the mean density is given by
$\bar{\rho} = 1.33$ g/cm$^3$, and eq.~(\ref{eq:pcentral}) gives $p\sub{c} \approx 2.5 \cdot 10^{12}$ Pa $ = 2.5 \cdot 10^7$ bar.
This result shows that to describe the interior structure of (massive) planets the conditions of matter under
extreme conditions has to be known.

The obtained result for Jupiter's $p\sub{c}$ is slightly beyond that what present day experimental setups can reach
(about $0.6 \cdot 10^{12}$ Pa in diamond anvil cells \cite{2012NatCo...3E1163D}),
and due to the assumption of constant density the estimated value is the minimum $p\sub{c}$ that Jupiter can have.
Hence, the equations of state (EOS) that can be used to describe the inner cores of giant planets are constructed through a
combination of theoretical calculations and experimental results.
A summary of frequently used EOS is presented in \cite{2010SSRv..152..423F}, see also \cite{2014prpl.conf..763B}.
 
A simple equation of state used for the outer regions (envelope and atmosphere) is given by the
ideal gas law
\beq
\label{eq:gas-pgas}
      p = \frac{R_{\rm gas} \, \rho T}{\mu} \,,
\eeq
with the gas constant, $R_{\rm gas}$, and the mean molecular weight, $\mu$, in units of $m\sub{H}$.

To solve the structure equations (\ref{eq:masscon} - \ref{eq:energy}) for a growing planet suitable {\it boundary conditions} have to be chosen. These 
are now different from the standard conditions of stellar growth because the planet is still embedded in the protoplanetary disk that acts as a 
mass and heat reservoir.

The inner boundary is given in this situation not by the central values but rather by the size and luminosity of the core.
The size, $R_{\rm core}$, is given by solving the structure equations for the core for a given composition and core mass, $M_{\rm core}$.
As a first approximation a constant density can be chosen for the core. 
The luminosity, $L_{\rm core}$, has different contributions.
The main part comes from the residual kinetic energy of the accreted planetesimals after they have fallen through
the envelope and land on the core. Additional core luminosity comes from radioactive decay in the core,
core contraction and cooling of the core.

The outer boundary conditions at the planet's radius, $R$, depend on the evolutionary status of the planet and the ambient disk.
Mainly 3 different options have been considered \cite{2015IJAsB..14..201M} \\
1) {\it Attached or nebular phase}:\\ 
For small masses ($M$ lower than about 10-20 $M_{\rm Earth}$) the protoplanet is still deeply embedded in the
disk such that its radius is much smaller than the disk scale height, $H$. Hence, its envelope is
smoothly attached to the nebula. The planet's radius in this case is given by the smaller of the accretion radius,
$R_{\rm A} = G M\sub{p}/c\sub{s}^2$,  and the Hill radius, $R\sub{H} = (M\sub{p}/3 M_*)^{1/3} a\sub{p}$ (eq.~\ref{eq:terr-Hill}),
where a smooth transition is used between these two values. The accretion radius is half of the Bondi-radius which is often used in spherical 
accretion problems. The temperature and pressure at the outer radius, $R$ are then identical to the disk temperature at that location. 
\\
2) {\it Detached or transition phase}: \\
For larger masses, no solution satisfying the attached state exists, and the proto-planet contracts
to a radius much smaller than $R\sub{H}$. The luminosity follows from radial infall, via a spherical accretion shock,
or through accretion from a circumplanetary disk \cite{2005A&A...433..247P}.
The gas accretion is not anymore determined by planet, but by nebula conditions.
For the spherical case the pressure at the surface is determined by the accretion shock (gas in free-fall)
at the photospheric pressure.
In reality the accretion onto the planet will not be completely spherical symmetric anymore and mass will be accreted through an
accretion disk around the planet, the circumplanetary disk.
Numerical simulations \cite{1999MNRAS.303..696K,1999ApJ...526.1001L} show that accretion occurs via 2 'streams' 
that are the extensions of the spiral shock waves
generated by the planet that extend all the way into the Hill sphere of the planet, see Fig.~\ref{fig:jup-flow}.
High resolution three dimensional simulations of the flow near growing planets show that a substantial amount of material is accreted
through the polar regions \cite{2003ApJ...586..540D,2014ApJ...782...65S}.  \\ 
3) {\it Isolated or evolution phase}: the planet evolves at constant mass and 
receives irradiation from central star that is spread over the whole planetary surface.
The albedo of the planet is required for an accurate determination of the absorbed radiation.
The photosphere is given, as in stellar models at an optical depth, $\tau = 2/3$.

For the case where the luminosity is provided by the accretion of planetesimals 
the planetesimal-envelope interaction during the infall is crucial for the energy and mass deposition profile in the envelope.
To determine the trajectory of infalling particles one has to consider first the gravity and gas drag 
(i.e. the envelope increases the capture radius of the growing planet),
and the secondly thermal ablation and aerodynamical disruption. A modern example of this process is given by the
infall of the comet Shoemaker-Levy 9, where temperatures over 30,000 K were reached in the bow shock around
the infalling body. This interaction with the envelope will heat up the planetesimal that re-radiates this
energy and will loose some of its mass, and may eventually disintegrate. The accumulation of energy by the envelope
or core during these processes will determine the time evolution of the planet.
 
When performing evolutionary simulations of growing planets one extremely important finding was the fact that
beyond a critical value of the core mass, $M_{\rm crit}$, the envelope cannot be in hydrostatic equilibrium anymore and 
will collapse onto the core \cite{1974Icar...22..416P,1978PThPh..60..699M,1980PThPh..64..544M}.
This finding has given rise to the term {\it core-instability} model, sometimes also called {\it core-accretion} scenario.
In the simulations, the hydrostatic equations stated as above were solved, where it was assumed that 
all planetesimals reach the core and provide the luminosity, $L\sub{core}$, hence in the envelope $L(r) = const. = L\sub{core}$. 
The critical mass, $M_{\rm crit}$, is a function of the opacity of the accreted material and it lies around $10 M_{\rm Earth}$
for typical ISM composition.
Subsequent numerical simulations confirmed exactly these early results and the runaway mass accretion is now {\it the}
cornerstone in the core-accretion scenario. Noteworthy, classic simulations for the in-situ formation of Jupiter in the
Solar System are presented \cite{1986Icar...67..391B, 1996Icar..124...62P}. 
More elaborate models on the formation of massive planets that include the effects of accretion through
a circumplanetary disk have been formulated \cite{2005A&A...433..247P,2012A&A...547A.111M},
which have been incorporated into population synthesis models of planets. The latter article \cite{2012A&A...547A.111M}
contains a very readable introduction to that field.

\begin{figure}[t]
    \begin{center}
        \includegraphics[width=0.45\textwidth]{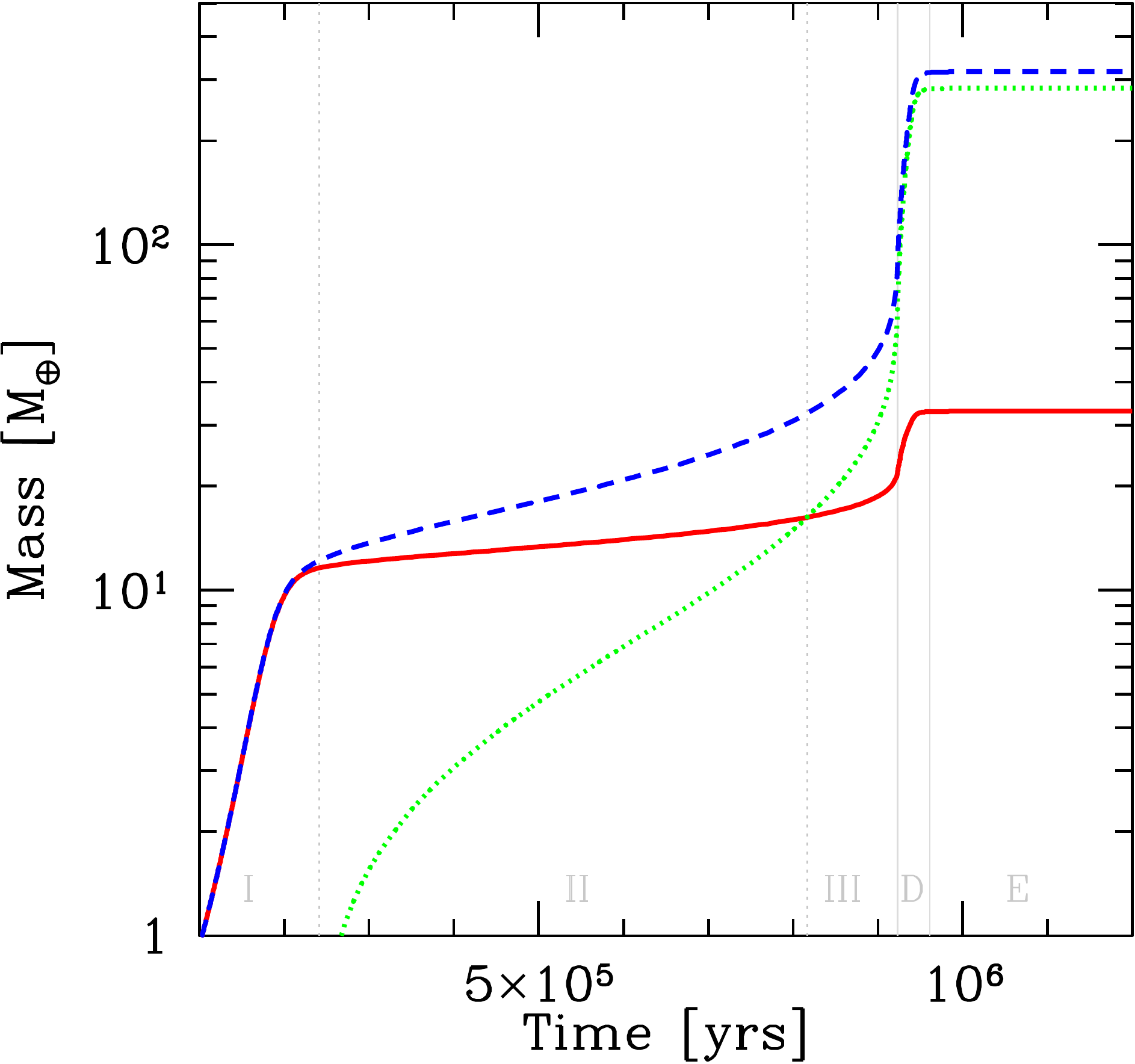} \quad
        \includegraphics[width=0.45\textwidth]{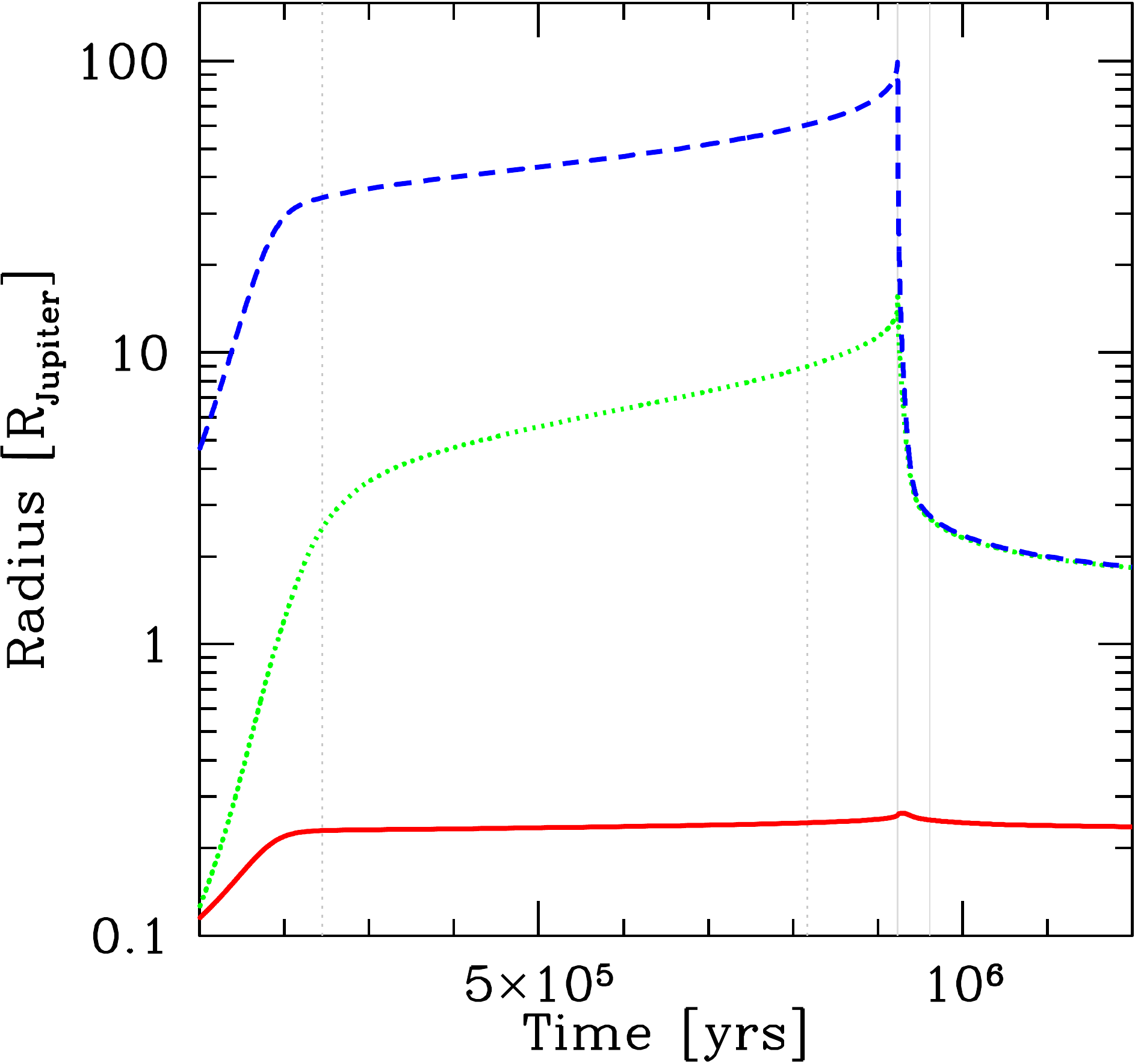} \quad
    \end{center}
    \caption{The evolution of the mass and radius of Jupiter in the Solar Nebula.
    During the whole growth the planet was fixed at its present position (in-situ formation).
    The core mass (made of solids) is given by the red line, the gas mass by the green line, and the
    total mass by the dashed blue line.
    The different phases of the accretion process have been marked by the vertical lines
    and additional roman numbers in the left panel, see text for more details.
    In the right panel, the red line refers to the core radius, the blue dashed line to the total radius,
    and the green line to the capture radius of planetesimals.
       Courtesy: Chr. Mordasini, see also \cite{2012A&A...547A.111M}
    \label{fig:gas-evolution}
     }
\end{figure}

Initially difficult to understand the physical mechanism of the core-instability process,
Stevenson \cite{1982P&SS...30..755S} presented a simple analytical model of the process, that
helped to uncover the underlying principles, and that
is summarized nicely in \cite{2010apf..book.....A}. 
For typical parameters in the protosolar disk one obtains \cite{1982P&SS...30..755S}
\begin{equation}
            M_{\rm crit} \simeq  20 \kappa\sub{R}^{3/7} \, M_{\oplus} \,,
\end{equation}
where $\kappa\sub{R}$ denotes the Rosseland mean opacity in units of [cm$^2$/g].
For $M_{\rm core} > M_{\rm crit}$ the core contracts rapidly and runaway gas accretion sets in.
Full numerical simulations are required to obtain the exact evolution of this phase that includes the 
details on the boundaries as given above in particular the contact with
the ambient disk and the solid particle accretion process.

An example of such a calculation for the in-situ formation of Jupiter in the Solar nebula \cite{2012A&A...547A.111M}
is shown in Fig.~\ref{fig:gas-evolution}. 
The vertical lines and roman numbers indicate different evolutionary phases:
I-III) refer to the attached phase with the protoplanet in contact with the disk, where
I) denotes the assembly of the core, 
II) refers to the the continued slow core accretion and hydrostatic gas accumulation, isolation of solid particles,
and in III) the core reaches the critical mass with the onset of rapid gas accretion. 
D) is the detached phase, and E) the long term isolation, or evolution phase.
The authors \cite{2012A&A...547A.111M} point out that the depicted evolution serves as a test case that
was described very similarly in early simulations \cite{1986Icar...67..391B,1996Icar..124...62P}.
The overall evolution indicates that the typical time scale to form Jupiter is about a million
years. However, there are many factors that have an impact on the overall evolutionary timescale.
Details will depend on the:
\begin{itemize}
\item Opacity \\
Low values of $\kappa\sub{R}$ will allow faster envelope growth. It it determined essentially by the
amount of dust present in the envelope because the gas opacities alone are much too low.
\item Convection \\
The onset of convection in the envelope will enhance the efficiency of energy transport and 
hence change the time scale for accretion.
\item Chemical composition \\
The chemical composition influences the opacity and the mean molecular weight, $\mu$.
The latter has through the equation of state (\ref{eq:gas-pgas}) a direct impact on the pressure and hence the 
radial stratification through the requirement of hydrostatics.
\item Accretion rate \\
The accretion onto the core, $\dot{M}_{\rm env}$, is often estimated from one-dimensional models.
However, only through two-dimensional (circumplanetary disk) and three-dimensional simulations (polar accretion)
can the actual accretion rate been obtained.
\item Migration \\
The migration of the growing planet through the disk has an influence on the mass
accretion rate. For example, new reservoirs for the solid particle as well as gas accretion 
will be opened.
\end{itemize}

One very important ingredient in determining the overall growth time scale is the formation time
of the core, which is determined by the accretion of solids from the disk. 
Obviously this is directly proportional to the solid particle density in the disk, $\Sigma_{\rm solid}$.
The simulations indicate that for $\Sigma_{\rm solid} <  10 $g/cm$^2$ the time scale for core formation is
too long such that the gas reservoir of the disk will also be depleted given the typical disk
lifetime of a few million years.
On the other hand, for $\Sigma_{\rm solid} > 10 $g/cm$^2$ there are too many heavy elements
in comparison to today's Jupiter composition.
Given that the surface density of solids for the MMSN is only 5 g/cm$^2$ (at the location of Jupiter),
there appears to be a timescale problem in that the growth time for the core formation is too long for Jupiter,
and the problem becomes even more severe in case of Uranus and Neptune formation.
 
\begin{figure}[t]
    \begin{center}
        \includegraphics[width=0.65\textwidth]{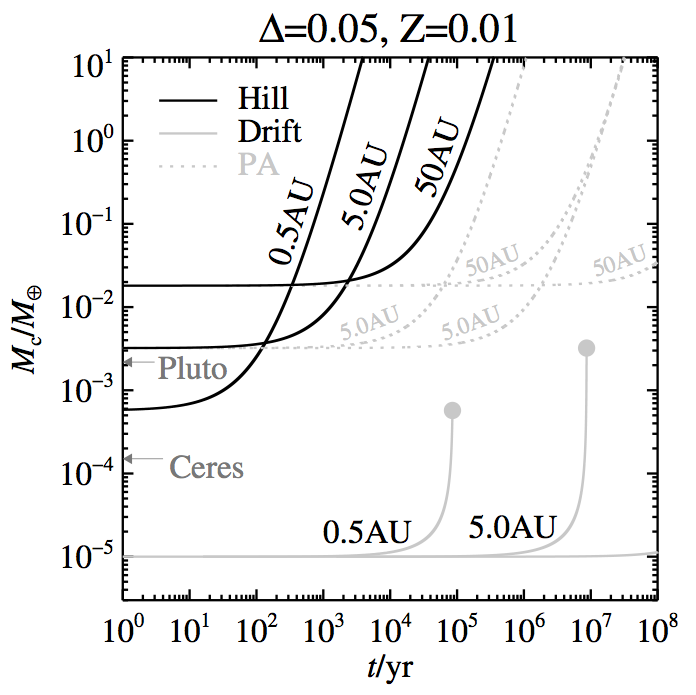}
    \end{center}
    \caption{The growth of a planetary core as a function of time at different radii in the disk.
    Curves are shown for the Hill regime ($m\sub{p} < M\sub{t}$, with the transition mass $M\sub{t}$, see eq.~\ref{eq:gas-mt}),
    the drift regime ($m\sub{p} > M\sub{t}$) and the standard pebble accretion (PA).
    The result shown are for a headwind parameter $\Delta \equiv \Delta v\sub{K}/c\sub{s} = 0.05$
     (see eq.~\ref{eq:gas-headwind})
     and a dust to gas mass ratio $Z \equiv \Sigma\sub{p}/\Sigma = 0.01$.
       Taken from \cite{2012A&A...544A..32L}
    \label{fig:pebble-accretion}
}
\end{figure}
A new option that is presently discussed in more detail is the idea to enhance the growth of planetary cores
by rapid accretion of small pebbles \cite{2012A&A...544A..32L}. In contrast to the larger planetesimals,
these typically cm to dm-sized particles feel the gas drag. When pebbles approach a growing core which has some atmosphere
already collected within its environment, the gas drag will slow them down and they will spiral deeper into
the Hill sphere towards the planet and become eventually accreted by it.
Depending on the mass of the growing core two different types of accretion regime can be distinguished \cite{2012A&A...544A..32L}.
First we define a Bondi-radius for the growing object as $R\sub{B} = G M\sub{p} / \Delta v\sub{K}$, where $\Delta v\sub{K}$ is the
velocity difference between the Keplerian rotation and particle speed due to the gas drag.
By equating the Bondi-radius and the Hill radius one can define a transition mass, $M\sub{t}$, where both are equal.
For the MMSN one finds 
\begin{equation}
\label{eq:gas-mt}
   M\sub{t} \approx  0.016 M_\oplus \, \left( \frac{r}{30 AU} \right)^{3/4} \, \left( \frac{ \Delta v\sub{K}}{0.1c\sub{s}} \right)^3 \,.
\end{equation}
For $m < M\sub{t}$ ($R\sub{B} < R\sub{H}$) accretion occurs in the so-called drift regime,
while for $m > M\sub{t}$ (i.e. $R\sub{B} > R\sub{H}$) it occurs in the Hill regime.
For the growth of massive planets the more interesting phase will be the Hill regime with $m > M\sub{t}$.

Classically, the accretion rate of planetesimals onto a planetary core (100\% sticking) is given by 
$\dot{M}\sub{p} = \Sigma \Omega R\sub{p}^2 F_{\rm grav}$ \cite{1987Icar...69..249L} (see also eq.~\ref{eq:terr-dmdt}), where
$R\sub{p}$ is the planetary radius and $F_{\rm grav}$ the gravitational focusing factor (eq.~\ref{eq:terr-enhance}).
Since only a fraction $\alpha\sub{p} = R\sub{p}/R\sub{H}$ of the whole Hill sphere is captured for a growing core 
\cite{2004ARA&A..42..549G,2012A&A...544A..32L},
the  mass accretion will be changed to $\dot{M}\sub{p} (\text{planetesimals}) =  \Sigma \Omega \alpha\sub{p} R\sub{H}^2 F_{\rm grav}$.
On the other hand, since pebbles feel the gas drag, all objects that reach the Hill sphere will be accreted and the
accretion rate becomes  $\dot{M}\sub{p} (\text{pebbles}) =  \Sigma \Omega R\sub{H}^2 F_{\rm grav}$.  
Clearly, for a growing core with $m > M\sub{t}$, the Hill radius will be much larger than its actual,
physical radius, such that $\alpha\sub{p} \ll 1$. Then the ratio  $\dot{M}\sub{p} (\text{pebbles})/ \dot{M}\sub{p} (\text{planetesimals})$
will be much larger than unity, reducing the accretion time scale for the core considerably.
Indeed, the simulations of Lambrechts \& Johansen \cite{2012A&A...544A..32L} show that under the assumption of a steady
influx of new material (from outside) and pebbles well settled to the midplane of the disk, the growth time for
a 10 $M_{\rm Earth}$ core is less than $10^5$ yrs at 5 AU, and less than $10^6$ yrs at 50 AU, see Fig.~\ref{fig:pebble-accretion}.
Clearly, these growth times are much shorter than those expected by pure growth of planetesimals and would
ease the timescale problems for massive planet formation considerably. The exact efficiency of this problem will depend on
several physical mechanisms, such as the so-called headwind parameter
\beq
\label{eq:gas-headwind}
    \Delta  =  \Delta v_\phi/c\sub{s}  \,,
\eeq
the vertical pebble concentration, $Z=\Sigma\sub{p}/\Sigma$, a possible migration of the core and of course the structure of the envelope
within the Hill sphere of the planet.
In eq.~(\ref{eq:gas-headwind}) $\Delta v_\phi$ is the velocity difference between the pebbles, which have
a Keplerian velocity, and the gas in disk that moves slower due to the radial pressure gradient.
The details need to be worked out, and also the question how the formation of lower (Neptune) mass planets fit into this
new model.
\subsection{The final mass}
\label{subsec:gas03-massfinal}
In the previous section we have seen that in the final stages the planet grows in a runaway fashion, and 
the question arises what determines the final mass of a planet.
The simplest answer to this might be fact that the amount of gas that is available in the disk is
necessarily finite which sets eventually a natural mass limit. While the limited mass reservoir in the disk certainly plays
a role, there is an important additional factor that does inhibit mass growth, and this is the opening of a gap.
We might expect that when the Hill radius of the planet exceeds the vertical scale height of the disk, $R\sub{H} \gtrsim H$,
this will have a significant impact on the disk structure, and alter the mass accretion rate. We will analyse this from
two perspectives, a particle based approach and a hydrodynamical one.

In the first {\it particle approach}, the velocity change experienced by an individual (gas) particle that passes by a growing planet in the disk
can be calculated most easily in the {\it impulse approximation} \cite{1979MNRAS.186..799L}. 
Here, it is assumed that the motion of the particle is primarily Keplerian in most
part of the orbit and only perturbed in the vicinity of the planet.
This creates an additional gravitational force (impulse) on the particle that leads to a deflection of its trajectory
and a slight change in its azimuthal velocity, and hence angular momentum.
Due to the collisions with other nearby disk particles this angular momentum change is immediately 'shared' between them.
Integrating the angular momentum exchange over all particles, i.e. over the whole disk, one can calculate the
total rate of angular momentum input to the disk mediated by the planet
\beq
\label{eq:impulse}
      \dot{J}_{\rm grav} = - \frac{8}{27} \left(\frac{r\sub{p}}{\Delta r_0}\right)^3
      \left(\frac{m\sub{p}}{M_*}\right)^2 \, \Omega\sub{p}^2 \, \Sigma(r\sub{p}) \, r\sub{p}^4 \,,
\eeq
where $r\sub{p}$ denotes the distance of the planet from the star, $\Delta r_0$ is the closest approach of a disk particle to the planet,
and the index $p$ denotes that all quantities are evaluated at $r\sub{p}$.
The minus sign indicates that the inner disk loses angular momentum while the outer gains it, i.e.
in the planet region the disk will be 'pushed' away from the planet, see below.
In spite of the simple approximation is result nearly exact.
A nice pedagogical treatment of the derivation is given in the lecture notes of the {\it Saas-Fee Advanced Course~31}
by P.~Cassen \cite{2006expl.conf..369C}.  From eq.~(\ref{eq:impulse}) it is clear that in the 
vicinity of planet ($\Delta r_0 \rightarrow 0$) there is a strong increase of $\dot{J}$, and the
total amount deposited will depend on the choice of $\Delta r_0$.

In the second {\it hydrodynamic approach}, the continuum behaviour of the gas in the disk is analyzed
and the angular momentum deposition follows from more complex wave phenomena.
A planet embedded in a disk produces disturbances in the disk's density distribution. These are sound waves that
spread out from the planet's position because its presence impacts the dynamics of the ambient gas.
The Keplerian shear flow in the disk turns these sound waves into a spiral wave pattern with an outer
trailing arm and an inner leading one, as shown in Fig.~\ref{fig:pladisk-basic} below.
The spiral wave pattern is stationary in a frame corotating with the planet and
hence, at a certain radial distance from the planet they reach a supersonic speed with respect to the
Keplerian disk flow. At this point the waves turn into shock waves, dissipate energy and deposit angular momentum to
the disk. In the outer disk the spiral arm is faster than the disk material and it deposits positive angular momentum,
i.e. the outer disk material gains angular momentum and hence speed, and moves away from the planet, because in a Keplerian disk
the angular momentum increases as $\propto r^{1/2}$. Inside of the planet the situation is
reversed, the spiral arm is slower than the disk and negative angular momentum is deposited which leads to an inward
motion of the disk material. In total, the disk matter is receding from the location of the planet lowering the density
in its vicinity. The effect increases with planet mass and eventually an annular gap in
the density is formed at the location of the planet, which happens even without considering direct mass accretion onto the planet.
This lowering of the ambient density decreases the available mass reservoir and the gas mass accretion onto the planet will
be reduced accordingly. 

From eq.~(\ref{eq:impulse}) it is clear that the deposition rate of angular momentum will increase with the mass of
the growing planet, and in the absence of other effects there will be no mass left at the planet location.
However, there are two main competitors, viscosity and pressure, working against this continuing gap deepening.
The viscous disk torques at the location of the planet are given by
\beq
\label{eq:gas-jdotvisc}
        \dot{J}_{\rm visc} = \dot{M}_{\rm disk} j\sub{p}  \, = \,  3 \pi \Sigma(r\sub{p}) \nu \,  r\sub{p}^2 \Omega\sub{p}
\eeq
with the kinematic disk viscosity $\nu$ and the specific angular momentum of the planet, $j\sub{p} = r\sub{p}^2 \Omega\sub{p}$.
In writing eq.~(\ref{eq:gas-jdotvisc}) we have assumed a stationary accretion disk with a globally constant mass accretion rate
\cite{1981ARA&A..19..137P}.
The {\it viscous criterion} for gap formation is then $\dot{J}_{\rm grav} \geq \dot{J}_{\rm visc}$.
If one assumes for the smallest distance the Hill radius (\ref{eq:terr-Hill}) of the planet, $\Delta r_0 = R\sub{H}$, then
\beq
\label{eq:crit-viscous}
     q  \geq q_{\rm visc} \simeq \frac{10 \nu}{\Omega\sub{p} \, r\sub{p}^2} \,,
\eeq
where $q$ is the planet to star mass ratio, $q = m\sub{p}/M_*$.
A second, {\it pressure criterion} of gap formation is obtained from the condition that the Hill sphere of the
growing planet is larger than the disk thickness, i.e. $R_{\rm H} \geq H$ which
gives
\beq
\label{eq:crit-pressure}
    q  \geq q_{\rm Hill} \simeq 3 \left(\frac{H}{r}\right)^3\sub{p} \,.
\eeq
For typical parameter of the protoplanetary disk both criteria yield similarly
a limiting mass for gap formation between Saturn and Jupiter.
This explained the mass of Jupiter in the Solar System quite well.
Using two-dimensional hydrodynamical simulations a more general criterion for gap formation has been derived
\cite{2006Icar..181..587C} 
\beq
\label{eq:gap-crida}
    \frac{3}{4} \frac{H}{R_{\rm H}} + \frac{50}{q \, Re}  \leq 1  \,,
\eeq
with the Reynolds number $Re = r\sub{p}^2 \Omega\sub{p}/\nu$. This last criterion assumes that the density of the disk at the location of the planet
has been reduced to 10\% in comparison to the unperturbed disk density. Additional, new estimates on the depth and width of the gap
have been developed more recently, they will be discussed in more detail in section~\ref{lect:05} below.

\begin{figure}[t]
    \begin{center}
        \includegraphics[width=0.45\textwidth]{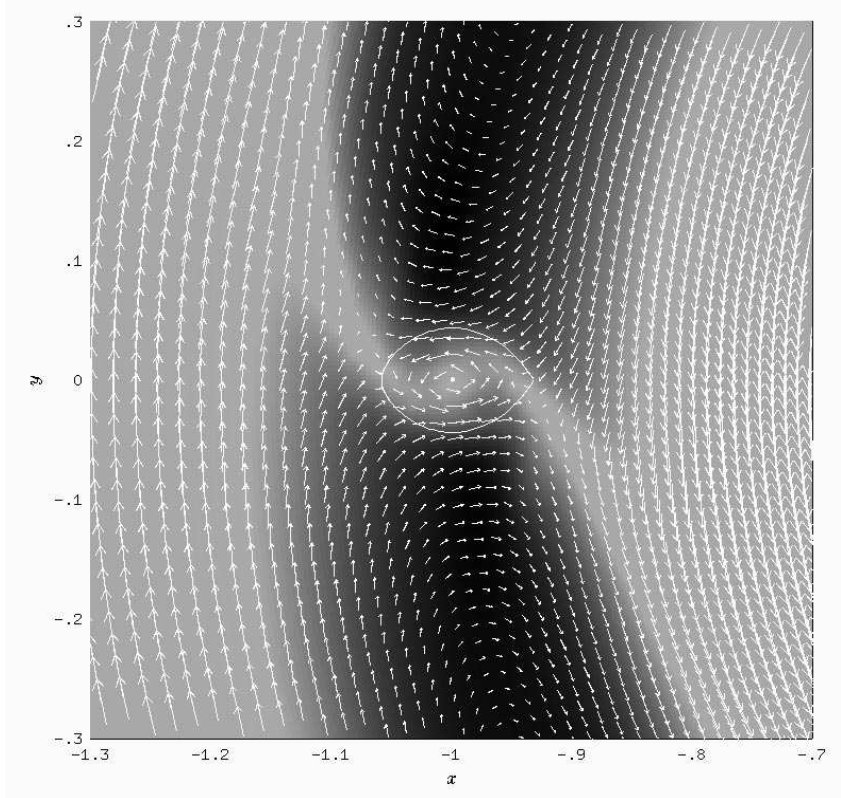}  \quad
        \includegraphics[width=0.45\textwidth]{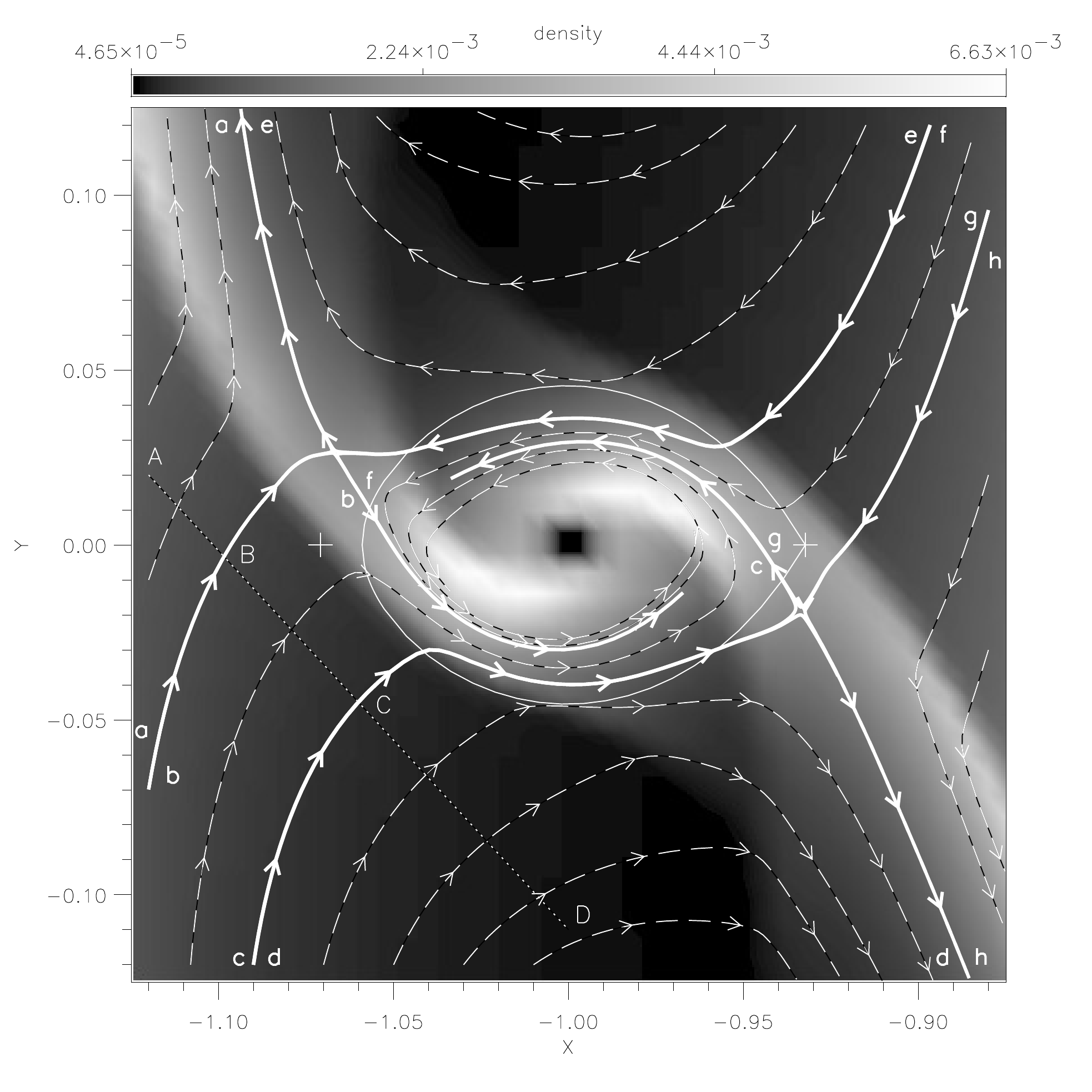}
    \end{center}
    \caption{The gas flow around a Jupiter type planet embedded in a protoplanetary disk.
    The planet is at location $x = -1, y=0$ (in units of $5.2$\,AU), and the star is located at the origin. 
    Both panels show a density image (in Cartesian coordinates) of the Roche lobe region near the planet.
    The motion of the planet around the star would be counter clockwise in an inertial frame of reference.
    {\bf Left}) The flow field around the planet, displayed in a reference frame corotating with the planet.
    The solid white line indicates the Roche lobe of the planet. Taken from \cite{1999MNRAS.303..696K}.
    {\bf Right}) Density contours with sample streamlines given by the dashed lines.
     The left (right) plus sign marks the L2 (L1) point. Critical streamlines that
     separate distinct regions are the solid white lines. Material approaching the planet within these critical lines
     (i.e. between 'b' and 'c' on the outside, and between 'f' and 'g' on the inside)
      can become accreted onto the planet, while material at the outside (or inside) either
     circulates or enters into the horseshoe region and crosses it. Taken from
       \cite{1999ApJ...526.1001L}.
    \label{fig:jup-flow}
    }
\end{figure}
The finding that the mass of Jupiter coincides with the mass required to open a significant gap in the disk has often been taken
as an indication that it is the gap creation that will eventually limit the final mass of a planet.
This is after all in agreement with the fact that Jupiter is the most massive planet in the Solar System.
However, the discovery of several extrasolar planets with masses much above Jupiter's indicated that
there must be a way to increase the mass even further in spite of the gap formation.
Indeed, hydrodynamical simulations showed that the gap is not as impermeable as thought, because mass can enter the horseshoe region and either be
accreted onto the planet or move from outer disk (beyond the planet) to inner disk or vice versa. The detailed flow field of the
gas in the close vicinity of the planet is depicted in Fig.~\ref{fig:jup-flow} for a Jupiter mass planet.
Clearly, even though a clear gap has formed material can still be accreted onto the
planet - the mass entering within the critical while lines on the right panel. For a Jupiter mass planet the mass accretion rate
onto the planet is of the order of the equilibrium disk accretion rate, $\dot{M}_{\rm disk} = 3 \pi \Sigma \nu$ which leads
to a doubling time of a few $10^5$ yrs for typical disk masses \cite{1999MNRAS.303..696K}.
Increasing the mass from 1 to 6 $M_{\rm Jup}$ the accretion rate onto the planet drops by nearly one order of magnitude \cite{1999ApJ...526.1001L}.
The results indicate that one Jupiter mass is not the limiting mass for planets in agreement with the observations,
but beyond 5-6 $M_{\rm Jup}$ growth times become very long. 
Despite this apparent limitation, it turns out that further mass accretion is nevertheless possible
because massive planets will induce a significant eccentricity in the outer disk
such that periodically the planet will enter into the disk allowing for more mass accumulation onto the planet
\cite{2006A&A...447..369K}. 
%
\begin{figure}[t]
    \begin{center}
        \includegraphics[width=0.65\textwidth]{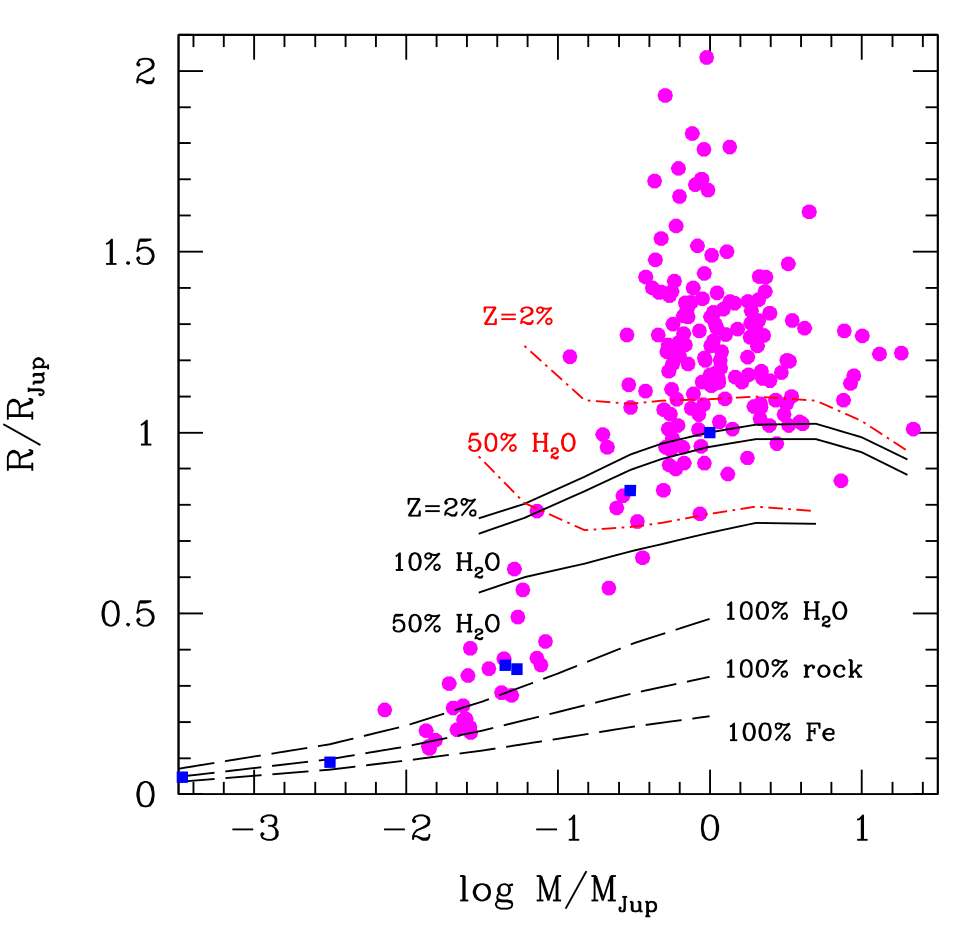}
    \end{center}
    \caption{Mass versus radius of known exoplanets, including Solar System planets (blue squares, Mars to Neptune) and
    transiting exoplanets (magenta dots). The curves correspond to interior structure and evolution
    models at 4.5 Gyr with various internal compositions, and for a mass range in 0.1 $M_{\rm Earth}$ to 20 $M_{\rm Jup}$. 
    The solid curves refer to a mixture of H, He and heavy elements, as indicated by the labels.
    The long dashed lines correspond to models composed of pure water, rock or iron from 
    The 'rock' composition here is olivine (forsterite Mg$_2$ SiO$_4$) or dunite. 
    Solid and long-dashed lines (in black) refer for non-irradiated models while dash-dotted (red) curves correspond to
    irradiated models at 0.045 AU from a Sun. Taken from \cite{2014prpl.conf..763B}.
    \label{fig:mr-diagram}
    }
\end{figure}
\subsection{Interior structure of planets}
\label{subsec:gas03-interior}
After having studied the formation of planets let us very briefly comment on the information on the internal composition
that can be drawn from the sample of observed exoplanets. The transit method
allows for a determination of the planet's radius, or at least the ratio of planetary to stellar radius, $R\sub{p}/R_*$, because 
the reduction in flux during the transit is directly proportional to the square of this quantity. The Kepler mission
allowed the determination of planetary radii for over thousand exoplanets and even from the ground there have been  
over 200 transit detections. To know in addition the mass of the planet, the radial velocity signal is required.
Unfortunately, this has only been possible for a small fraction of the Kepler planets but for many of the ground based transit detections.
Knowing the mass and radius of the planet, the mean density of the planet is determined and rough estimates as to its composition 
can be made. The important mass-radius (M-R) diagram can be constructed and compared to the theoretical models for the planet interiors.
To calculate the interior models the structure equations as written above (\ref{eq:masscon} - \ref{eq:energy})
need to be solved for long evolutionary times,
after the disk has dispersed. What determines the final radius of a planet with a given mass is primarily its {\it composition}.
Different constituents directly alter the mean density and will determine the important equation of state (EOS), that relates pressure
to density (and temperature). Hence, the EOS defines the compressibility of the material and has direct influence on the 
planetary radius.
The second factor is the {\it age} of the planet, as this determines the remnant heat that is still
incorporated within and has not been lost by some cooling process. A third factor is the {\it distance from the star} as this
determines the amount of external heating that is received by the planet for example by direct irradiation from the central star
or the strength of the tidal interaction and dissipation.

In order to obtain accurate models, details of additional physical processes have to be considered that are often not known very well.
The equation of state has to be known within a regime up to about 20,000 K and 70 Mbar which is only partly accessible by
experiments. Shock-wave and compression experiments give only data up to a few Mbar, and theoretical 
calculations based on Quantum Molecular Dynamics simulations have to be performed in addition.
Knowledge about possible phase transitions has to be acquired. For the cooling rate of planets the amount of radioactive
elements need to be known, the efficiency of convection or plate tectonics (viscosity of the material) to be determined.
For a recent summary of the present status of the field see the PPVI review \cite{2014prpl.conf..763B}, where 
Fig.~\ref{fig:mr-diagram} is taken from. In the figure, the $M-R$ diagram is shown for a sample of detected transiting exoplanets
and some Solar System planets together with calculated theoretical curves. Because the transit probability increases strongly
with shorter periods, i.e. shorter distances from the star, the exoplanets sample refers basically to 'hot' planets.

As seen in the diagram, these can be divided in two major groups: hot Neptunes (Super-Earths) and hot Jupiters. The first 
group which is the actually most abundant in absolute numbers, as discovered by the Kepler-mission,
is under-represented here due to the mentioned lack of radial velocity data of the Kepler planets. 
Clearly, as can be inferred from the reference objects in the Solar System, the locations of the points give indeed a good
first indication of the composition of the planets, indicating that many of the Super-Earths may consist of rock and iron
material while the Jupiter mass objects are gaseous planets that will consist primarily of Hydrogen and Helium.   
Sometimes very exotic options make it into the public media. One example is the so called 'diamond' planet 55 Cnc e,
for which some radius estimates indicated a value that matched exactly that of a planet with a composition of 100\% diamond material.
However, a mixture of carbon, silicates and iron
material will match the observations equally well \cite{2012ApJ...759L..40M}.
An additional feature that can be inferred from the Fig.~\ref{fig:mr-diagram} is that the majority of the hot Jupiter planets are much 
larger than predicted even if purely solar composition is assumed, i.e. they are inflated. Assuming irradiation from the 
star does indeed increase the radius somewhat but is by no means sufficient to explain the observed large radii.
Observationally, it has been found that the amount of inflation decreases clearly with distance from the central star \cite{2011ApJS..197...12D},
implying that the central star is responsible for the effect. Possible suggested mechanisms are irradiation, tidal friction between the orbiting
planet and the star during the circularization process of eccentric planets, or electrical currents generated through the interaction of ionized
particles with the planetary magnetic field \cite{2011ApJS..197...12D}. However, the main cause is not known as of today.

\begin{svgraybox}
Information drawn on the inner structure of the massive planets in the Solar System has given rise to the core accretion
scenario of giant planets. Within this model, first solid cores of a few Earth masses are forming in a  manner similar
to the assembly of terrestrial planets. Once grown big enough, the ambient gas will be accreted onto the core,
hence the terminology core accretion scenario. This
evolutionary phase can be described by the classical stellar evolution equations augmented by suitable boundary conditions
accounting for the fact that the planet is still embedded into the disk and the luminosity is created by the infall
of solid material. As it turns out, once the core has grown to a critical mass,
the gas accumulation proceeds in a runaway fashion such that on a timescale
of a few million years a gas giant can be created at the location of Jupiter's orbit. Going to larger distances from the star
the evolutionary timescales (for core formation and gas accretion) become longer than the typical lifetimes of the disk.
A solution to this timescale problem may be given by the pebble accretion scenario where solid,
cm-sized particles are continuously accreted such that the critical core mass can be reached very fast, reducing the
formation time significantly.
Information on the interior composition can be obtained for transiting extrasolar planets if additional radial velocity data
allow for a mass determination. The observations show that, due to star-planet interactions, most of the Jupiter type planets
are significantly inflated in their radii. 
\end{svgraybox}

%
%
%
\section{Planets formed by gravitational instability}
\label{lect:04}
Having discussed the formation of planets via the core accretion (CA) scenario in the preceding section we will now
turn to the alternative scenario of planet formation, the gravitational instability (GI) of the disk.
While CA is the preferred scenario for the planets in the Solar System, GI is a possible
a pathway considered for planets at large distances from their host stars.
First, we will present observational examples of directly imaged planets
that are indeed located at large distances from their host stars.
Then we will consider the question under what conditions a disk can fragment directly to form planets.
For such an analyses primarily two methods have been used. First a {\it linear stability analyses} and secondly
full nonlinear numerical simulations of self-gravitating disks. We will describe the main findings below.

\subsection{Background}
\label{subsec:background}
\begin{figure}[t]
    \begin{center}
        \includegraphics[width=0.95\textwidth]{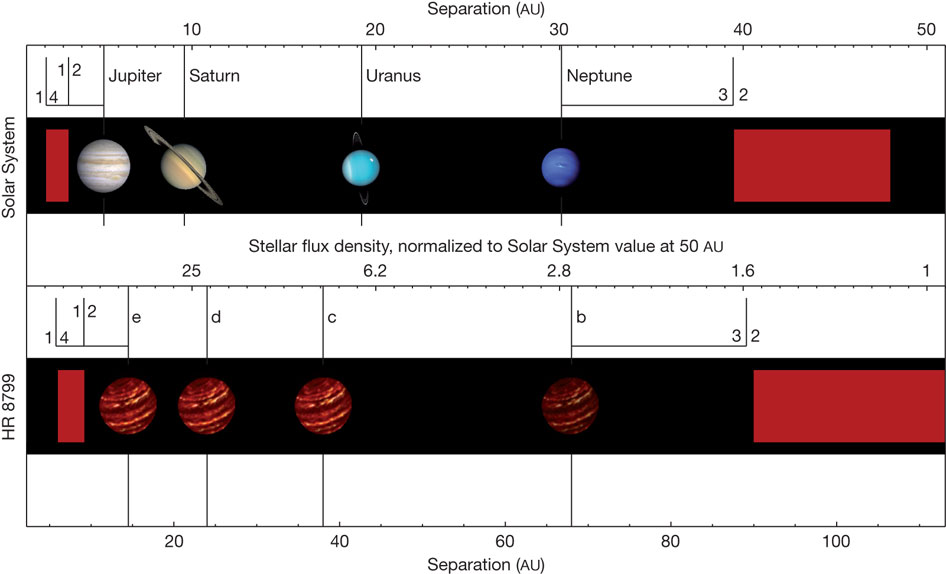} 
    \end{center}
    \caption{The structure of the planetary system HR~8799, that contains 4 massive planets
    all discovered by direct imaging, in comparison to the outer Solar System. The $x$-axis has been compressed
    according to the luminosities of the host stars by the factor $\sqrt{L_{\rm HR8799}/L_\odot}$, with $L_{\rm HR8799}=4.9 L_\odot$.
    This means that the planets in HR~8799 are about two times farther away from the star but have the same
    equilibrium temperature as the Solar System planets because of the higher luminosity of the host star in HR~8799.
    The red rectangles indicate the regions where debris material is orbiting the stars.
    The lines marked with 1:4, 1:2 and 3:2 indicate the locations mean-motion resonances with respect to the planets. 
       Taken from \cite{2010Natur.468.1080M}.
    \label{fig:HR8799}
    }
\end{figure}

The most prominent example of directly imaged planets is the system HR~8799 in the constellation Pegasus. For this system, in 2008 the
detection of 3 planets was announced, discovered using adaptive optics at the Keck telescopes in Hawaii \cite{2008Sci...322.1348M}.
The system was observed at different epochs such that the motion of the individual planets across the plane of the sky was detected.
Hence, this discovery marks clearly a breakthrough in exoplanetary science because for the very first time the actual motion of
planets around another star was directly detected, following the classical terminology of a planet being a 'wandering star'.
Coincidentally, the position of this first directly imaged 'real' planetary system in the sky lies very close to the first
planet discovered by the RV method, 51~Peg. The system is observed nearly face on, and
the observed motion of the planets agrees very well with the Keplerian motion about the host star that has a mass of $1.5 M_\odot$.
The 3 planets are 24, 38, and 68 AU away from the star and have estimated masses of $10, 10$ and $7 M_{\rm Jup}$. These masses
are upper limits set by dynamical stability arguments where it must be assumed that they are engaged in a resonant
configuration \cite{2010ApJ...710.1408F}. Later, in 2010 a fourth planet located at only 14 AU distance from the star was discovered
\cite{2010Natur.468.1080M}. In Fig.~\ref{fig:HR8799} the layout of the whole planetary system HR~8799 is displayed.
It is shown in a special scaling such that the similarities to the outer Solar System planets become apparent.
Even the spatial distribution of the debris material is similar. So, one may speculate about the existence of terrestrial planets in
that system. Otherwise there have only been a few systems where directly imaged planets in the few Jupiter mass range have been
found, for example Fomalhaut b, $\beta$ Pictoris b or Gliese 504 b, even though the status of Fomalhaut b has been debated.

Despite these interesting similarities the large absolute distances of the planets in HR~8799 pose a challenge for their formation,
due to severe timescale and stability problems.
Comparing the effectiveness of various formation scenarios to form planets at large distances such as core accretion (with or without migration),
outward scattering from the inner disk, or gravitational instability 
it was suggested that the last scenario is the most likely \cite{2009ApJ...707...79D}.
For example, in the classical core accretion scenario the growth time of Neptune at its present orbit will
be of the order \cite{2010apf..book.....A} 
\beq
          \tau_{\rm grow} = \frac{m\sub{p}}{d m\sub{p}/dt} \approx 5 \cdot 10^{10} \, F_{\rm grav}^{-1} \quad \mbox{yr}  \,,
\eeq
where we assumed $\Sigma_{\rm part} = 1$ g/cm$^3$ in eq.~(\ref{eq:terr-dmdt}). 
This is a very long timescale unless the gravitational focusing factor, $F$, is very large ($\sim 10^4$).
Considering the larger distances of the planets in HR~8799 and the youth (60 million yrs) of the host star
it appears unlikely that those planets have formed at these large distances by core accretion, although pebble
accretion may help in this case (see above). And, since the formation via scattering often led to unstable
systems it was concluded that formation via gravitational instability was the most viable mechanism to form
the HR~8799 system \cite{2009ApJ...707...79D}, possibly related to later periods of mass fall-in from the
envelope \cite{2010ApJ...714L.133V}. In models explaining HR~8799, the resonant structure of the planets places
further constraints as it seems to require some migration process. 

\subsection{Linear stability analyses}
\label{subsec:self04-GIlinear}
To study the linear stability of a thin disk rotating around a central object we first assume that the disk is
infinitesimally thin and follow the evolution in a two-dimensional setup.
In the context of galactic dynamics, the classical studies are given by Toomre \cite{1964ApJ...139.1217T}
and Lin \& Shu \cite{1964ApJ...140..646L},
and a local shearing sheet analysis is presented in \cite{1987gady.book.....B}.
To give an idea on how such a stability analyses is performed we sketch briefly the procedure. 
The set of hydrodynamical equations in cylindrical coordinates ($r, \varphi$) for a disk confined in the $z=0$ plane are given for example 
in \cite{1998ApJ...504..945L} and read
\beq
\label{eq:hyd-sigma}
 \pdoverdt{\Sigma} + \nabla \cdot (\Sigma {\bf u} ) =  0   
\eeq
\beq
\label{eq:hyd-urad}
 \pdoverdt{(\Sigma v)} + \nabla \cdot (\Sigma v {\bf u} )
  =
\Sigma \, r  \Omega^2
        - \pdoverd{P}{r} - \Sigma \pdoverd{\psi}{r}
\eeq
\beq
\label{eq:hyd-uphi}
 \pdoverd{(\Sigma r^2 \Omega )}{t}
   + \nabla \cdot (\Sigma r^2 \Omega {\bf u} )
         =
        - \pdoverd{P}{\varphi} - \Sigma \pdoverd{\psi}{\varphi}
\eeq
where $P$ is the 2D vertically integrated pressure,
$\vec{u} = ( u_r, u_\varphi) = ( v, r \Omega)$ the 2D velocity,
and $\psi$ the gravitational potential 
\beq
  \psi = \psi_{*} + \psi\sub{d} + \psi_{\rm p} \,, 
\eeq
that is given here as the sum of the stellar potential, $\psi_* = - G M_*/r$, the disk contribution
\beq
\label{eq:psi-disk}
            \psi\sub{d}(r,\varphi) = - G \int_{\rm disk} \
              \frac{\Sigma(r') r' dr' d \varphi'}{\sqrt{r^2 + r^{'2} - 2 r r' \cos \varphi'}} \,,
\eeq
and possibly a planetary contribution, $\psi_{\rm p}$, that is neglected in this discussion.
In eq.~(\ref{eq:psi-disk}) the integration has to be performed over the whole disk.
For the pressure we assume that it is given as a function of the surface density, $P = P(\Sigma)$, as is the 
case for the isothermal or adiabatic equations of state.
To study the stability we start from an axisymmetric equilibrium state
\beq
     r \Omega_0^2 - \frac{1}{\Sigma_0} \pdoverdr{P_0} - \pdoverdr{\psi_0}  = 0  \,, 
       \label{eq:GGW}
\eeq
where the subscript $0$ refers to the unperturbed basic state which is a function of the radius alone.
Now the system is perturbed by adding a small perturbation
\beq
\label{eq:ansatz}
     f(r,\varphi,t)  = f_0(r) + f_1(r,\varphi,t) \,,
\eeq
with $f \in \{\Sigma, v, \Omega, \psi\}$.
Here, we only consider perturbations within the plane of the disk.
The ansatz (\ref{eq:ansatz}) is substituted into the full time dependent hydrodynamical equations
(\ref{eq:hyd-sigma}) to (\ref{eq:hyd-uphi}) which are then {\it linearized}, i.e. two main assumptions are applied:
{\it a}) The functions $f_1$ are assumed to be small compared to their equilibrium counterparts, i.e.  $f_1 \ll f_0$.  
This implies that terms that are quadratic in the perturbations $f_1$ can be neglected with respect their linear counterparts.
{\it b}) The stratification of the background varies only slowly, which implies that the radial derivatives of the basic
functions are assumed to be small in comparison to those of the
perturbed functions, i.e. $\partial f_0/\partial r \ll \partial f_1/\partial r$.
After these simplifications the non-linear hydrodynamic equations have been transformed into a set of linear
equations for the perturbed quantities, that can in principle be integrated numerically. However, better insight is
obtained by further analysis.

Because the basic state has neither a time nor an azimuthal dependence, one can quite generally
expand the perturbations in a Fourier series 
\beq
\label{eq:fourier}
      f_1 = \tilde{f}_1 (r)  e^{i \left(m \varphi - \sigma t \right)} \,,
\eeq 
where $m$ denotes the azimuthal wave number of the disturbances and $\sigma$ the frequency of the temporal variations.
As written in eq.~(\ref{eq:fourier}), the perturbation functions $\tilde{f}_1$ depend now only on the radius,
and the time and azimuthal derivatives become
$\pdoverdt{} \, \Rightarrow  - i \sigma$ and $\pdoverdphi{} \Rightarrow  i m$, respectively.
With the expansion (\ref{eq:fourier}) the linearized equations become as set of ordinary differential equations in radius
\bea
\label{eq:lin5-1}
  \tilde{\Sigma}_1 ( \sigma - m \Omega_0) & = &
    - i \Sigma_0 \tilde{u}_1'  + \Sigma_0 m \tilde{\Omega}_1 \\
\label{eq:lin5-2}
  \tilde{u}_1 ( \sigma - m \Omega_0) & = &
     i 2 r \Omega_0 \tilde{\Omega}_1  - i \frac{c^2_{\rm s_0}}{\Sigma_0} \tilde{\Sigma}_1'
     - i \tilde{\psi}_1'  \\
\label{eq:lin5-3}
  \tilde{\Omega}_1 ( \sigma - m \Omega_0) & = &
     - i \frac{\kappa_0^2}{2 r \Omega_0} \tilde{u}_1
  -  \frac{c^2_{\rm s_0}}{\Sigma_0}
      i m \tilde{\Sigma}_1 + \frac{1}{r^2} i m \tilde{\psi}_1  \,,
\eea
with the radial derivative $f' = \partial f/\partial r$.
Here, $\kappa_0$ denotes the {\it epicyclic frequency}
\beq
\label{eq:kappa}
              \kappa_0^2 \equiv  \frac{1}{r^3}
         \pdoverdr{} \left[ (r^2 \Omega_0)^2 \right]
           \,  =  \, 4 \Omega_0^2  + 2 \Omega_0 r \pdoverdr{\Omega_0} \,.
\eeq 
For a disk in pure Keplerian rotation, $\kappa_0 = \Omega\sub{K}$, a relation which is also approximately
fulfilled in self-gravitating disks. The sound speed in the disk is denoted with $c_{\rm s}$. 
Now, to simplify matters, we expand also the radial direction in a Fourier-series,
i.e. the radial dependence is given by $\propto e^{i k r}$, which implies that 
$\pdoverdr{} \, \Rightarrow  i k $.

\begin{figure}[t]
    \begin{center}
        \includegraphics[width=0.65\textwidth]{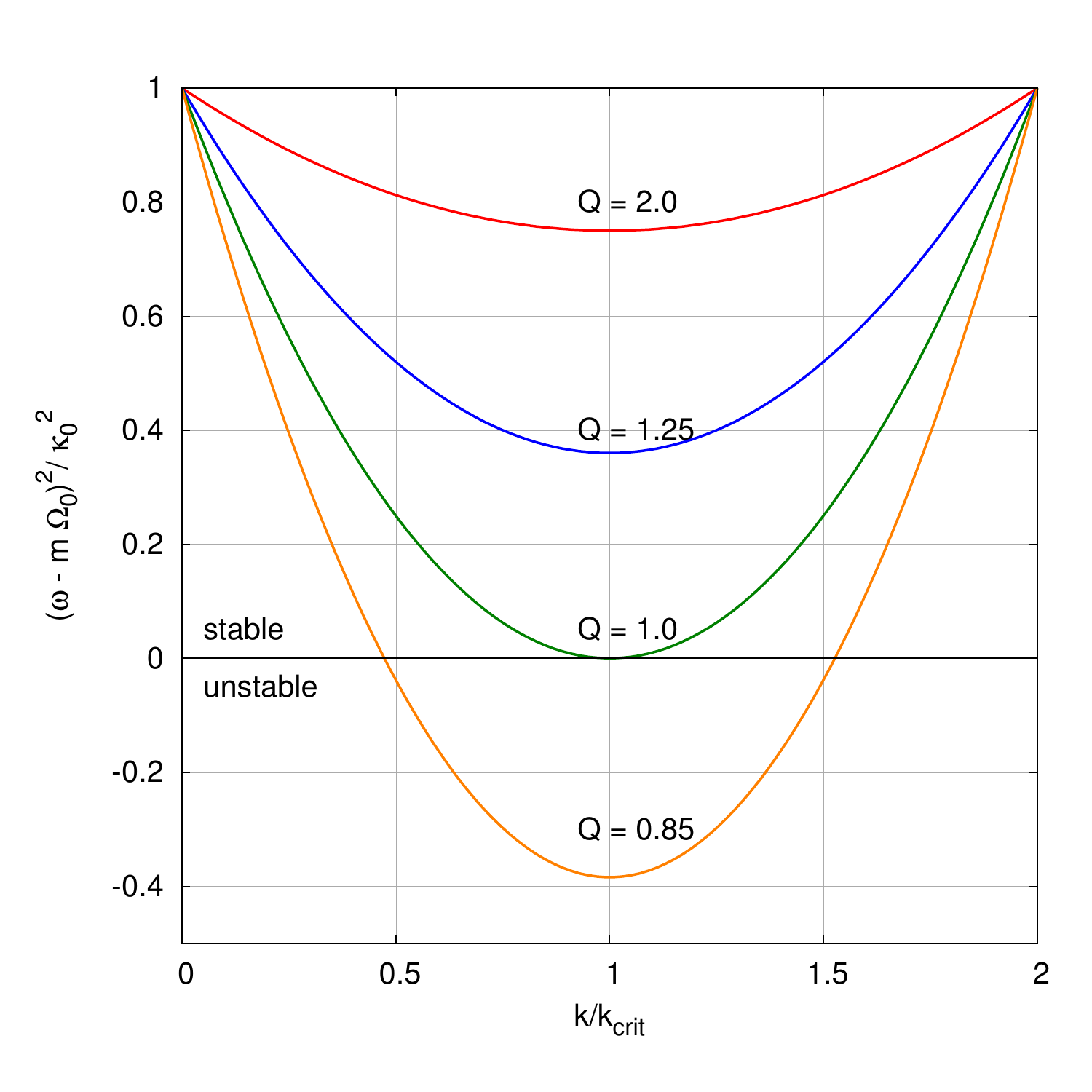}
    \end{center}
    \caption{The normalized  dispersion relation (\ref{eq:dispnorm}) for perturbations in an infinitesimally thin disk.
     The critical wavenumber, $k\sub{crit}$ is given by eq.~(\ref{eq:kcrit}) and the Toomre number, $Q$ by  eq.~(\ref{eq:Q}).
     For $Q=0$ marginal stability is reached.
     For large $k$, i.e. small wavelengths, the disk is stabilized by pressure (sound waves) and for small $k$ by
     a reduced density $\Sigma_0$. Increasing rotation $\kappa_0$ stabilizes as $Q$ rises.
    \label{fig:dispersion}
    }
\end{figure}

Finally, we make the so-called {\it tight winding approximation}, which means $k r \gg m$,
i.e. the radial wavelength ($1/k$) is small against the azimuthal ($r/m$). This implies that all terms containing
the azimuthal wavenumber $m$ on the right hand side in eqs.~(\ref{eq:lin5-1}) to (\ref{eq:lin5-3}) can be neglected.
The last problem relates to the evaluation of the perturbation of the disk potential, $\tilde \psi_1$.
This is obtained not from the Poisson-integral (\ref{eq:psi-disk}) but rather from the linearized Poisson-equation
\beq
\label{eq:poisson}
   \nabla^2 \psi_1 =  4 \pi G \Sigma_1 \delta(z) \,, 
\eeq
where we assume a matter distribution that is only non-zero within the plane of disk ($z=0$), i.e. $\delta(z)$ denotes
the $\delta$-function.
Integrating eq.~(\ref{eq:poisson}) over a small volume around the disk one obtains \cite{2010apf..book.....A}
\beq
  \tilde{\psi}_1 = - \frac{2 \pi G \tilde{\Sigma}_1}{|k|} \,,
\eeq
and eqs. (\ref{eq:lin5-1}) to (\ref{eq:lin5-3}) turn into the {\it dispersion relation} \cite{1964ApJ...140..646L}
\beq
\label{eq:dispersion}
   (\sigma - m \Omega_0)^2 = \kappa_0^2  + c^2_{\rm s_0} k^2 - 2 \pi G |k| \Sigma_0 \,.
\eeq 
Remembering the time dependence of the perturbations, $\propto e^{i \sigma t}$, it is clear that
in general perturbations are:
{\it a}) Stable, for $\sigma^2 > 0$ because in this case $\sigma$ is real and the disturbances oscillate in time
with a frequency $\sigma$, or they are
{\it b}) Unstable, for $\sigma^2 < 0$ because in this case $\sigma$ is imaginary and the disturbances
can grow exponentially, leading to an unlimited growth, i.e. instability. 
The point of marginal stability is given by $\sigma = 0$.
From our relation (\ref{eq:dispersion}) one can see that the epicyclic oscillations ($\kappa_0$ - term)
are stabilizing at all spatial scales, this is the classical {\it Rayleigh stability criterion}. 
Indeed, for individual particles orbiting a central object that are slightly perturbed, $\kappa_0$ is the oscillation frequency
of the particle around its equilibrium position. In the case of a spread out gas the propagation of sound waves and self-gravity
come into play. 
For the sound waves ($c_{\rm s}$ - term) the stabilizing effect is larger for larger $k$, i.e.
for smaller spatial scales. 
On the other hand, the last term in eq.~(\ref{eq:dispersion}) refers to the effect of the self-gravity
of the disk which is always destabilizing due to the minus sign and proportional to the local surface density, $\Sigma_0$.

Let us now consider axisymmetric disturbances with $m=0$. 
Since for stability the frequencies must be real, $\sigma^2 \geq 0$, the most unstable oscillations are those where
$\sigma^2$ is minimal. Because $\sigma$ is a function of $k$ the most unstable {\it critical} wavelength,
$k_{\rm crit}$, can be calculated from $d \sigma^2 / d k =0$, and we obtain
\beq
\label{eq:kcrit}
          k_{\rm crit} = \frac{\pi G \Sigma_0}{c_{\rm s}^2} \,,
\eeq
where we drop for simplicity the index $0$ at the sound speed.
Substituting this into the dispersion relation (\ref{eq:dispersion}) and investigating the point of 
marginal stability by setting $\sigma^2(k_{\rm crit}) =0$ we obtain after rearranging
\beq
\label{eq:Q}
        Q \equiv  \frac{c\sub{s} \kappa_0}{\pi G \Sigma_0} = 1 \,,
\eeq
where we defined the {\it Toomre parameter} $Q$. This relation implies that $Q=1$ defines the borderline between
stable and unstable configurations, the marginal state.
Indeed, from the dispersion equation (\ref{eq:dispersion}) that is quadratic in $k$ one can
show that in the case of axisymmetric disturbances the following inequality must be satisfied for stability
\cite{1992pavi.book.....S}
\beq
\label{eq:toomre}
  \text{For stability:} \quad \quad         Q  >  1    \quad
   \quad      \mbox{(Toomre-Criterion)}  \,.
\eeq
As just discussed above and directly seen from (\ref{eq:toomre}), 
for a given $\kappa_0 \approx \Omega\sub{K}$ the disk will be stabilized by higher
temperatures (increase in $c\sub{s}$) implying thicker disks, while a larger surface density will lead to
destabilization. Hence, whenever $Q$ is of the order unity the disk is prone to instability. 
Using these definitions for $k\sub{crit}$ and $Q$ the dispersion relation (\ref{eq:dispersion}) can be rewritten as
\beq
\label{eq:dispnorm}
        \left( \frac{\omega - m \Omega_0}{\kappa_0} \right)^2 =
           1 + \frac{1}{Q^2} \, \left( \frac{k^2}{k\sub{crit}^2} - 2 \frac{|k|}{k\sub{crit}} \right) \,.
\eeq
This function is displayed graphically in Fig.~\ref{fig:dispersion} for different $Q$ values.
Obviously, the minima of the parabolas occur always at $k = k\sub{min}$ and for $Q=1$ it coincides with marginal stability. 

After having obtained now a useful criterion for disk instability, it remains to be seen what happens actually to
an unstable disk. Before doing so, let us consider the situation in the Solar System.
For the protosolar nebula at 10~AU with $H/r \approx 0.05$ (i.e. $c\sub{s} \approx  0.33$km/s), a value $Q=1$
requires $\Sigma_0 \simeq 10^3$ g/cm$^2$, which is much larger than the MMSN value ($\approx 50$ g/cm$^2$).
This implies that for the Solar System the gravitational instability could only have worked in
an early evolutionary phase, when the mass of the protosolar nebula was still high. 
Using the critical wavelength $\lambda_{\rm crit} = 2 \pi /k_{\rm crit}$ the mass of such a fragment can be estimated to be around
\beq
   M\sub{p} \sim \pi \Sigma_0 \lambda_{\rm crit}^2  =  \frac{4 \pi c\sub{s}^4}{G^2 \Sigma_0}
       \sim  2 M_{\rm Jup} \,
\eeq
which lies in the range of the most massive gas giant in the Solar System. The idea 
of Solar System planet formation via gravitational instability goes back to Kuiper \cite{1951PNAS...37....1K}
or Cameron \cite{1978M&P....18....5C}. 
%
\begin{figure}[t]
    \begin{center}
        \includegraphics[width=0.66\textwidth]{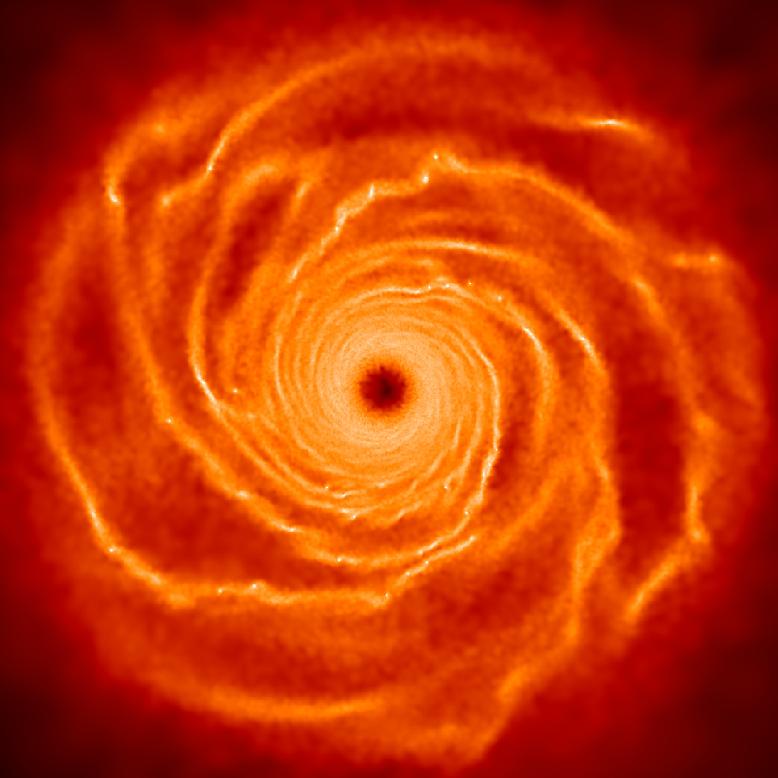}
    \end{center}
    \caption{An example of an SPH simulation of a self-gravitating disk.
    During the evolution spiral arms are forming that are later fragmenting
    into a number of individual planet like objects. For details of the simulation see
    section \ref{subsec:self04-nonlin} below.
       Courtesy: Farzana Meru.
    \label{fig:SPH-meru}
    }
\end{figure}
\subsection{Fragmentation conditions}
\label{subsec:self04-fragment}
Many non-linear hydrodynamic simulations have been performed to study the fate of unstable disks.
In such simulations, typically a young protostar with a mass in the range of $(0.5-1.0) M_\odot$ is surrounded
by a disk having a mass of a few tenth of the stellar mass. 
When approaching the stability limit, which is usually done by increasing the disk mass with respect
to the stellar mass, the main outcome 
is the formation of spiral arms of low order azimuthal wavenumber, very similar to galactic disks
\cite{1998ApJ...504..945L,2005ApJ...619.1098M}. 

To determine those areas which are most susceptible to fragmentation let us consider an accretion disk
with a constant mass flow through the disk, $\dot{M} = 3 \pi \nu \Sigma$. 
In an accretion disk the effective viscosity is given by
\beq
\label{eq:diskvisc}
     \nu = \alpha c\sub{s} H = \alpha c\sub{s}^2/\Omega  \,,
\eeq
and we find for the Toomre-$Q$
\beq
           Q  \propto  \frac{c\sub{s}^3}{\dot{M}} \,.
\eeq
Assuming that $\dot{M}$ does not vary too strongly with radius this implies that $Q$ falls off with radius
because the disks become cooler at larger distances from the star.
Hence, the most unstable region lies in the outer parts of the disk.
This trend is clearly seen in early 3D grid-based simulations by A.\,Boss \cite{1997Sci...276.1836B} where he studied the evolution of isothermal
and adiabatic disks with a mass of about $140 M_{\rm Jup}$ within 10~AU around a Solar mass star.
In both cases clumps formed near the outer boundary.
Similar results were obtained with the Smooth Particle Hydrodynamics (SPH) method. Using over one million particles in isothermal simulations
is was shown that fragments of planetary mass can form easily \cite{2002Sci...298.1756M}
and on very short timescales of only a few hundred years. 
A typical example on how such a simulation looks like is shown in Fig.~\ref{fig:SPH-meru}.
This strong reduction in assembly-time
clearly shows the great interest in the GI-mechanism as a possible path to giant planet formation. 
On the other hand for adiabatic simulations it was found that the forming clumps were
sheared out and dispersed \cite{2004ApJ...609.1045M}. 
 
These results clearly indicate that the outcome, whether or not fragments are forming, will depend not only on the present
value of the temperature but on the disk thermodynamics, that determines how the matter reacts upon compression.
The internal temperature of the disk is determined by the balance of heating and cooling processes. If the cooling is
higher than the heating, the disk will be unstable, if it is lower, the disk will be stable.
For the disk we have the following {\it heating mechanisms} operating:
\begin{itemize}
\item internal dissipation in shock waves, for example produced by spiral arms 
\item effective viscosity, produced by the turbulent motion within the disk. This is typically modeled via 
  standard viscous dissipation for example in an $\alpha$-disk model as in eq.~(\ref{eq:diskvisc})
\item heating by external sources such as the central star, cosmic rays or nearby stars. This will be more important in the
   outer parts of the disk
\end{itemize}
On the other hand the following cooling mechanisms can be considered
\begin{itemize}
\item  equation of state (EOS) \\
   The EOS determines the behavior of the gas upon compression.
   A medium that can be strongly compressed without heating up will be more susceptible to instability
   than a medium that heats up strongly. The following cases are often considered in running disk models: \\
   EOS1 - (locally) isothermal.
    Here the disk temperature is a given function of radius that cannot change. This is equivalent to strong cooling, i.e.
    the gas cannot heat up upon compression \\
   EOS2 - locally isothermal for low gas densities, then adiabatic above some suitable $\rho_{\rm crit}$. This
   type of EOS models the turnover from an optically thin gas of low density and a denser medium that heats
   up upon compression. This type frequently used in star formation simulations \cite{2004RvMP...76..125M}
\item simple cooling laws \\
   Quite generally the cooling time is defined by
  \beq
     t_{\rm cool} =  \frac{e_{\rm th}}{d e_{\rm th}/dt} \,,
   \eeq
  where $e_{\rm th}$ is the thermal energy of the disk per surface area. 
  Simple approximations are often used in the context of planet formation \\
  $t_{\rm cool} \Omega = const.$ \, ($t_{\rm cool}$ is a  fixed fraction of the local rotational period.) \\
   $t_{\rm cool} = const.$ \, ($t_{\rm cool}$ is fixed throughout)
\item radiative cooling \, (from disk surfaces) \\
  The Rosseland mean opacity is proportional to $\kappa\sub{R} \propto Z \, T^\epsilon$, where the
     magnitude of $\kappa\sub{R}$ is given by the amount of dust particles embedded in the disk, with
     the metal abundance $Z$. If one assumes that
     the energy is locally radiated away from the two disk surfaces with the flux, $F_{\rm eff} = \sigma\sub{B} T_{\rm eff}$, then 
\beq
\label{eq:cool-opac}
              t_{\rm cool} \simeq \frac{e_{\rm th}}{2 \sigma\sub{B} T^4_{\rm eff}}
     \propto T/T_{\rm eff}^4 \propto T^{-3+ \epsilon} \, Z
\eeq
   where we have assumed an optically thick case where the midplane temperature is related to the surface
   temperature via 
\beq 
\label{eq:T-eff}
       T^4_{\rm eff} = T^4_{\rm mid} / \tau
\eeq
    using the mean vertical
   optical depth, $\tau \sim \Sigma \kappa\sub{R}$. 
   Typically: $-3 < \epsilon <3$, such that $t_{\rm cool}$ grows with lower temperature
\end{itemize}
The last option, radiative cooling in combination with a realistic EOS is certainly the most realistic approximation one can
make for flat two-dimensional disks but often the simple cooling laws are used.

As pointed out, the likelihood for fragmentation depends on the cooling timescale of the gas. For rapid cooling the system will
be more likely to fragment than systems where the cooling rate is longer. Often, a simple {\it $\beta$-cooling} law
is applied in numerical simulations with
\beq
\label{eq:beta}
           t_{\rm cool} = \beta \, \Omega^{-1} \,,
\eeq
and a constant value of $\beta$.
In a linear analysis for the local shearing sheet it was then shown \cite{2001ApJ...553..174G} 
that the instability is then determined directly by the value of $\beta$, in particular
\bea
\label{eq:stab-crit}
     t_{\rm cool} \leq 3 \Omega^{-1} \quad  & \Rightarrow & \quad  \mbox{fragmentation} \\
     t_{\rm cool} \geq 3 \Omega^{-1} \quad  & \Rightarrow & \quad  \mbox{no fragmentation} \,.
\eea
This implies that $\beta_{\rm crit} =3$ is the critical value \cite{2001ApJ...553..174G}.
A simple estimate for $\beta$ in accretion disks can be obtained from
thermodynamic equilibrium where it is assumed that the internally produced heat is radiated away locally
leading to the following cooling behaviour \cite{1981ARA&A..19..137P} 
\beq
   \frac{e_{\rm th}}{t_{\rm cool}} = \Sigma \nu \left(\frac{d \Omega}{d r}\right)^2 \,.
\eeq
Here, it was assumed that the heat generation, produced for example by gravo-turbulence for marginally stable
self-gravitating disks, can be written as an effective viscous dissipation with
kinematic viscosity $\nu$. For an ideal gas with the thermal energy $e_{\rm th} = c\sub{v} \Sigma T$,
where $c\sub{v}$ denotes the specific heat at constant volume,
an $\alpha$-viscosity as in eq.~(\ref{eq:diskvisc}), and
Keplerian rotation, one finds for the equilibrium state
\beq
\label{eq:tau-cool}
    \Rightarrow \quad \quad
    t_{\rm cool} \simeq  \frac{4}{9} \frac{1}{\gamma ( \gamma -1) \alpha} \, \Omega^{-1} \,.
\eeq
As to be expected, the cooling ability of the gas directly determines the level of viscosity in the disk.
locally. For $\alpha \sim 10^{-2}$ and $\gamma = 1.4$ we find with $e = \Sigma c\sub{v} T$ a cooling time
of $t_{\rm cool} \sim 12$ periods. This is roughly the timescale 
for changes of the thermal structure of an accretion disk.

One should keep in mind however that eq.~(\ref{eq:tau-cool}) describes the equilibrium situation for the disk
and local variations are to be expected in realistic cases. Additionally, the simple $\beta$-cooling that went into
the derivation of (\ref{eq:tau-cool}) is not a realistic cooling, as it does not depend on the density of the gas.
When a gas clump is compressed it is to be expected that the cooling time rises and the clump will heat up preventing
further collapse, i.e. for a realistic modeling more sophisticated cooling prescriptions, such as (\ref{eq:cool-opac})
will have to be applied. Applying the $\beta$-cooling prescription (\ref{eq:beta}) 
simulations performed in the 2D shearing-sheet approximation 
using grid-based numerical models give results in rough agreement with
the above fragmentation condition \cite{2001ApJ...553..174G}, as do corresponding
global 3D disk models using the SPH method \cite{2003MNRAS.339.1025R}.
However, using very high resolution simulation in the 2D isothermal setup, 
it was shown that even for $\beta \approx 20$ fragments could form due to the
stochastic nature of the turbulent flow \cite{2012MNRAS.421.3286P}. 

As just mentioned, for fragmentation to occur in disks a cooling time shorter than $\approx 3 \Omega^{-1}$ is required.
From the definition of $Q$ (\ref{eq:Q}), using representative disk conditions with $H/r = 0.05$, one can 
obtain estimates on the cooling time. Assuming a disk close to possible instability by setting $Q=1.5$, and using 
eq. (\ref{eq:T-eff}) for the optical depth one finds the following relation for the cooling time \cite{2010apf..book.....A}
\beq
\label{tcool-disk}
         t_{\rm cool} \Omega  \sim  10 \, \tau \, \left( \frac{r}{5 AU} \right)^{-1/2} \,.
\eeq 
Noticing that in optically thick disks the optical depth can easily reach $10^2$ and more, this last relation seems to imply that
there is no possibility for fragmentation at typical planet formation locations in the Solar System. 
This finding represents the fact that the fragmentation requirement, high mass (implying large $\Sigma$ and high $\tau$) 
and low cooling times (low $\tau$) contradict each other. The presented arguments are based on the assumption of
vertical energy transport by radiation (relation \ref{eq:T-eff}). Including vertical convective energy transport the disk can cool faster
and the situation improves somewhat. For lower temperatures the opacities are lower and the cooling time is reduced again.
Including additionally external perturbation, for example by a passing star that compresses the outer regions in the disk,
a gravitational instability may be triggered.
In summary one may conclude that, if at all, fragmentation can only be expected at very large radii from the star 
beyond 50 AU or so \cite{2005ApJ...621L..69R}.
\subsection{Non-linear simulations}
\label{subsec:self04-nonlin}
At the end of this section we would like to summarize briefly the results of a few recent numerical simulations concerning the
fate of gravitational unstable disks. This is not an exhaustive review as there have been many simulations concerning this
topic and the issue is not conclusively decided at present. 
A  summary of the field was given
in the Protostars and Planets V proceedings \cite{2007prpl.conf..607D} where different numerical methods were directly compared
using similar initial initial conditions. The codes used in the comparison included particle methods such 
as SPH (the public codes {\tt GASOLINE} and {\tt GADGET}) as well as the grid codes that used finite difference or finite volume
methods either based on upwind schemes ({\tt Indiana Code}) or Riemann-solvers ({\tt FLASH}). It was noted that the outcome of the
simulations depends crucially on numerical aspects such as spatial resolution (number of grid-points, or particles),
regularization methods (smoothing length, artificial viscosity, flux-limiters), solver for self-gravity (smoothing criteria)
and so on. Hence, it was concluded that in order to obtain reliable results it seems unavoidable to compare different methods
on the same physical problem. Using appropriate criteria for the individual codes similar results could be obtained.
One should keep in mind however, that the results obtained with different codes on this problem will never be identical
because of the chaotic nature of the problem. Hence, the results can only be compared in a statistical sense.
%
\begin{figure}[t]
    \begin{center}
        \includegraphics[width=0.66\textwidth]{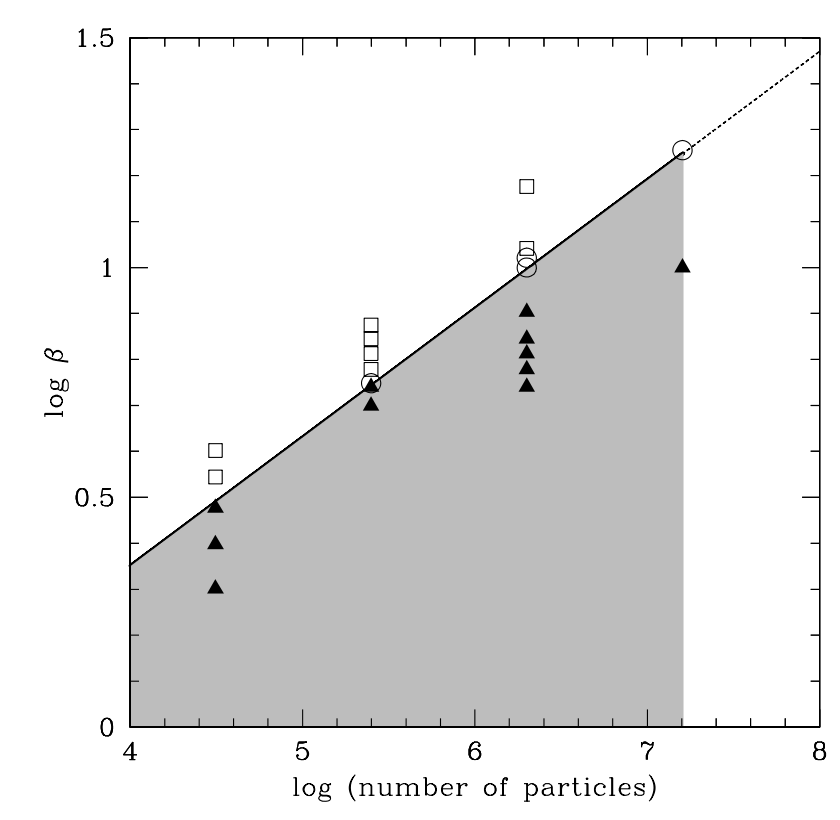}
    \end{center}
    \caption{The cooling rate $\beta$ against numerical resolution (here the used particle number)
    for SPH-simulations of self-gravitating disks, here a $0.1 M_\odot$ disk around a one solar mass star.
     The symbols denote the numerical outcome: non-fragmenting (open squares), fragmenting (solid triangles) 
    and borderline (open circles) simulations. 
    The borderline simulations are those that initially fragment but whose fragments are sheared apart.
    The solid black line separates fragmenting and non-fragmenting cases and the gray region marks the 
    fragmentation regime. 
       Taken from \cite{2011MNRAS.411L...1M}
    \label{fig:beta-conv}
    }
\end{figure}
\subsubsection{The convergence issue}
\label{subsubsec:self04-converge}
The sample study of Meru \& Bate \cite{2011MNRAS.411L...1M} illustrates one of main numerical problems very well.
They considered the situation of a $0.1 M_\odot$ disk around a $1 M_\odot$ star spanning a factor of ten in 
radii, ranging from $0.25$ to $25$ (in dimensionless units). For the cooling they used the $\beta$-prescription
(\ref{eq:beta}) with different values for $\beta$ within $2.0 \leq \beta \leq 18$. They performed SPH-simulations
using a huge range of different particle numbers from $N = 31,250$ to $N= 16$ million particles. 
A typical result of such a simulation is displayed in Fig.~\ref{fig:SPH-meru}.
The results indicate that the value of $\beta$ below which fragmentation occurs depends strongly on the
particle number used in the simulations. As shown in Fig.~\ref{fig:beta-conv} the critical value of $\beta$ increases
monotonically with $N$ as given by the solid line. The hatched region indicated the fragmentation regime while above the
line the disks are stable. Clearly, there is no real indication of convergence in these simulations. 
Because the aforementioned SPH simulations are fully three-dimensional and hence require a very high particle number for numerical
convergence, an alternative option that has been used frequently is the two-dimensional grid code {\tt FARGO} \cite{2000A&AS..141..165M}
that has been empowered with a self-gravity module in 2D \cite{2008ApJ...678..483B}.  
Comparing both codes, SPH and {\tt FARGO}, Meru \& Bate \cite{2012MNRAS.427.2022M} find that the question of convergence hinges
at the treatment of numerical viscosity, for both codes.
They argue that convergence can be reached yielding critical $\beta$-values of about 20 to 30, making fragmentation easier.
A detailed analyses and additional SPH simulations using an improved cooling prescription as well improved artificial viscosity
(and consequently artificial dissipation) found numerical convergence and
fragmentation occurring for cooling times between $\beta = 6$ and $\beta = 8$ for an adiabatic index $\gamma = 5/3$ \cite{2014MNRAS.438.1593R}.
They also find that fragmentation can occur only at very large distances beyond at least 50 AU, in agreement with the above expectations
\cite{2005ApJ...621L..69R}, and as seen in other studies \cite{2010ApJ...710.1375K,2013A&A...552A.129V}. 
These cases demonstrate again the numerical problems that are still present in such self-gravitating disk simulations, and 
indicate the necessity to compare different codes and methods. 

\begin{figure}[t]
    \begin{center}
        \includegraphics[width=0.66\textwidth]{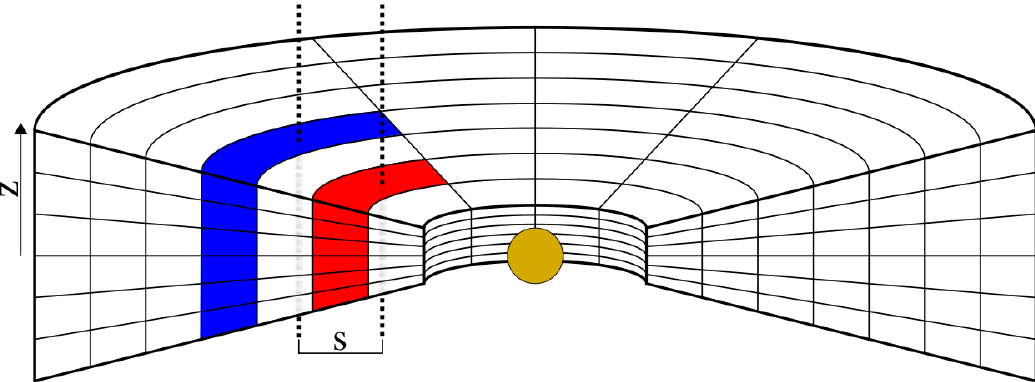}
    \end{center}
    \caption{
                \label{fig:disk_model_disk}
                Geometry of a protoplanetary disk around a central star to illustrate the requirements
                for the calculation of the disk's self-gravity in 2D simulations.
                The goal is to calculate the gravitational force exerted by a vertical slice of the disk (blue)
                on another vertical slice of the disk (red), that are separated by the projected distance $s$.
                As seen in the drawing, two vertical integrations have to be performed along the dashed lines 
                that go through the cell centers assuming cylindrical coordinates.
                To obtain the total force between the two segments, this value has to be multiplied by the corresponding areas,
                see also Eq.~(\ref{eq:force_sg}).
       Taken from \cite{2012A&A...541A.123M}
    }
\end{figure}
\subsubsection{The gravitational potential}
\label{subsubsec:self04-potential}
Care has to be taken when treating the disk in a flat 2D geometry only.
Even though the disk may be vertically thin, the gravitational potential has to be modified to account correctly for the vertical
extent of the disk \cite{2012A&A...541A.123M} as illustrated in Fig.~\ref{fig:disk_model_disk}.
The potential at a point $\vec{r}$ generated by the remaining part of a self-gravitating disk is given by
\begin{equation}
        \label{eq:psi-sg}
        \Psi_\mathrm{sg}(\vec{r})
        = - \int_{\rm Disk}  \frac{G \rho(\vec{r}')}{\left| \vec{r} - \vec{r}' \right|} \,d\vec{r}' \,.
\end{equation}
Here, in contrast to the earlier expression (\ref{eq:psi-disk}) the full vertical stratification of the disk
has been taken into account. As the standard 2D hydrodynamic equations (\ref{eq:hyd-sigma}) to (\ref{eq:hyd-uphi}) are obtained by an
averaging process over the vertical direction, this has to be done for the gravitational potential as well. 
This is typically approximated by a suitably chosen smoothing parameter. 
For this purpose it is convenient to analyze the force between to individual elements (segments) of the disk.
The potential at the location $\vec{r}$ generated by a disk element located at $\vec{r}'$ which is a
projected distance $s$ away is given by
\begin{equation}
   \label{eq:self-psi-sg}
        \Psi_\mathrm{sg}(\vec{r}) = - \iiint \frac{G \rho(r', \varphi', z')}{\left( s^2 + (z-z')^2 \right)^\frac{1}{2}} \,dz'\,dA' \,.
\end{equation}
Here $dA'$ is the surface element of the disk in the disk's midplane (the $z=0$\,plane) that is located at the point $\vec{r}'$,
separated by the projected distance $s$.
The force at the position $\vec{r}$ due to the rest of disk is calculated from the gradient of the potential (\ref{eq:self-psi-sg}),
and vertical averaging leads to the following force density \cite{2012A&A...541A.123M}
\begin{align}
\label{eq:force_sg}
        F_\mathrm{sg}(s)
        &= -\int \rho(r,\varphi,z) \frac{\partial \Psi_\mathrm{sg}}{\partial s} \,dz  \nonumber  \\
        &= -Gs \iiiint \frac{\rho(r',\varphi',z')\,\rho(r,\varphi,z)}{\left( s^2 + (z-z')^2 \right)^\frac{3}{2}} \,dz'\,dA'\,dz \,,
\end{align}
where the integral is over the vertical extent (at position $\vec{r}$) and all other locations of the disk.
Even under the assumption that the vertical structure of the disk is given analytically (e.g. a simple Gaussian), the computation of the
force (\ref{eq:force_sg}) is extremely costly. Hence, the potential (\ref{eq:self-psi-sg}) is approximated often by the so-called 
$\epsilon$-potential 
\begin{equation}
        \label{eq:self-pot2d}
        \Psi^\mathrm{2D}_\mathrm{sg}(s)  = - \iint \frac{G \Sigma(\vec{r}')}{\left(s^2 + \epsilon_\mathrm{sg}^2\right)^\frac{1}{2}} \,dA' \,,
\end{equation}
where $\epsilon_\mathrm{sg}$ is the smoothing length for self-gravitating disks that takes into account the unresolved vertical extend
of the disk. 
The force acting on each disk element is then calculated from the gradient of $\Psi^\mathrm{2D}_\mathrm{sg}$.
This formulation extends eq.~(\ref{eq:psi-disk}) for numerical simulations and the smoothing length does not only
ensure that the numerical evaluation remains finite but is physically necessary due to the finite disk thickness.
By comparing the results obtained with (\ref{eq:self-pot2d}) to the exact formulation (\ref{eq:self-psi-sg}) (using a Gaussian
vertical density profile) one obtains a good match with $\epsilon\sub{sg} \approx H$ \cite{2012A&A...541A.123M}. 

\subsubsection{Fragmentation outcome and longterm evolution}
\label{subsubsec:self04-evolution}
As a massive protoplanetary disk may still be embedded somewhat in their envelope it is important to model the disk evolution
in combination with infall onto the disk. Such studies have been performed through 2D numerical simulations of viscous disks
that include radiative cooling, stellar irradiation, and mass infall \cite{2012ApJ...746..110Z}.
The results show that fragmentation is possible when mass infall is included. However,
two major problems have been identified: \, {\bf a)} Mass challenge: To be able to fragment the disk must have a high mass ($M_{\rm disk} > 0.3 M_\odot$)
and requires a high infall rate. The fragments then grow very fast and typically end up as Brown Dwarfs (BD) with
masses well above the planetary regime. \, 
{\bf b)} Migration challenge: Once formed the fragments migrate inward rapidly on a timescale $< 1000$yrs.
Only large masses survive, and small ones get tidally disrupted. The authors conclude that that GI will lead to massive BDs or binaries.
The migration problem will be discussed also in the following section. Similar findings have also been found in full 3D simulations 
of fragmenting disks \cite{2013MNRAS.436.1667T}.

Important further aspects concern the subsequent evolution of the clumps after they have formed. As just mentioned they migrate inward
very rapidly and can be tidally disrupted upon reaching the inner region of the accretion disk close to the central star.
This can lead to accretion events such as FU Ori outbursts that occur in the early phase of star formation when
the disk is still very massive \cite{2006ApJ...650..956V}. As an alternative, the {\it tidal downsizing scenario}
considers another fate of the clumps.
Upon inward migration the Roche radius of the massive clumps is shrinking continuously such that they might lose
their envelopes by tidal effects through mass overflow across the $L_1$ Lagrange point, see Fig.~\ref{fig:terr-lagrange}.
The final outcome could be objects with much smaller mass,
sometimes even rocky planets \cite{2010MNRAS.408L..36N}. In this scenario the above mentioned mass challenge would be resolved
as the clumps loose the majority of their mass, and the gas giants could be assembled on a short timescale.

After having discussed two different pathways (by CA or GI) in forming giant planets,
it is interesting to ask if one can possibly distinguish them. The answer lies in the thermal evolution of the young planets.
Planets formed by gravitational instability are thought to have higher entropy, larger radii, 
and possibly higher effective temperatures than core-accretion objects, because of the stored heat in the accumulated gas
that did not have time to cool during the rapid accretion process. Following the longterm thermal evolution it has been
shown that for planets with mass less than about 5 $M_{\rm Jup}$ differences in the effective temperature
persist for a few million years only, while for planets of higher mass it can take up to about 100 million years before
equilibration \cite{2012ApJ...745..174S}. However, the feasibility of this distinction hinges on details of the 
models. For example, the accretion of additional solid material during the subsequent evolution will enhance the cooling of
the planets formed by GI and reduce the differences to the CA planets, implying that it might be difficult to
separate the two options observationally.

As a final remark, concerning possible difficulty of producing gaseous planets directly through the GI process
we would like to point out that even the formation of planetesimals may be enhanced in the early phases of very massive disks.
As shown in the previous section, small bodies experience gas drag and will drift towards pressure maxima occurring in the disk.
As it turns out, in marginally stable disks the spiral arms may be sufficiently long lived to support the collection of
particles in them such that the concentrated dust particles can collapse to their own self-gravity to form directly
planetesimals in the early phase of planet formation \cite{2006MNRAS.372L...9R}.

\begin{svgraybox}
Forming planets by direct gravitational instability of the disk has long been considered as an alternative option 
to the core instability pathway. An advantage would be the associated short time scale of the formation process.
As stated by the well-known Toomre criterion, for the mechanism to work the disk must be sufficiently massive and 
and at low temperature.
In order for the collapse to proceed the gas must be able to cool sufficiently fast to get rid
of the thermal energy. The required cooling times are only a few (3-5) dynamical timescales.
These conditions, high disk mass and short cooling times can only be fulfilled in the very outer parts of
a protoplanetary disk, probably only beyond a distance of about 50 AU. Hence, it represents a possible formation
scenario for the distant planets detected by the direct imaging method.
\end{svgraybox}

%
%
%
\section{Planet-disk interaction}
\label{lect:05}
In the preceding section we have implicitly assumed that the planets remain on their orbits during the growth phase.
This in-situ formation process is based on the situation in the Solar System in which the large planets reside further out
where they could accrete large amounts of gas, while the terrestrial planets are located further in where, due to the higher
disk temperatures, only small amounts of gas were available to be accreted onto the rocky cores.
Comparing this situation with the observed semi-major axis distribution of the extrasolar planets 
(as shown in Fig.~\ref{fig:intro-mass-distance}) it was clear from the very
beginning, in fact already after the discovery of the very first object, 51~Peg, by Mayor \& Queloz \cite{1995Natur.378..355M}, that 
the presence of close-in planets is difficult to explain by an in-situ formation process. This is because
close to the star  the amount of solids available for core formation is not sufficient and the disk temperature is
too high to condense the gas. For an alternative point of view see for example \cite{2000Icar..143....2B}. 
This difficulty led to the immediate suggestion that 51~Peg had formed in the standard way at larger distances,
similar to Jupiter in the Solar System, and subsequently drifted inward driven by planet-disk
interaction \cite{1996Natur.380..606L}.
Such a migration process is a natural consequence of the planet formation process because
the gravitational interactions between the planet and the ambient disk will necessarily lead to forces acting on the planet
causing it to change its orbital elements.
In this section we will describe the physical processes of disk-planet interaction in more detail. Recent reviews
of the topic are presented in the Protostars and Planets (V and VI) proceedings \cite{2007prpl.conf..655P,2014prpl.conf..667B}
or in \cite{2013LNP...861..201B} and \cite{2012ARA&A..50..211K}. 

\subsection{Basic concepts}
\label{subsec:pladisk01-basic}
\begin{figure}[t]
    \begin{center}
        \includegraphics[width=0.95\textwidth]{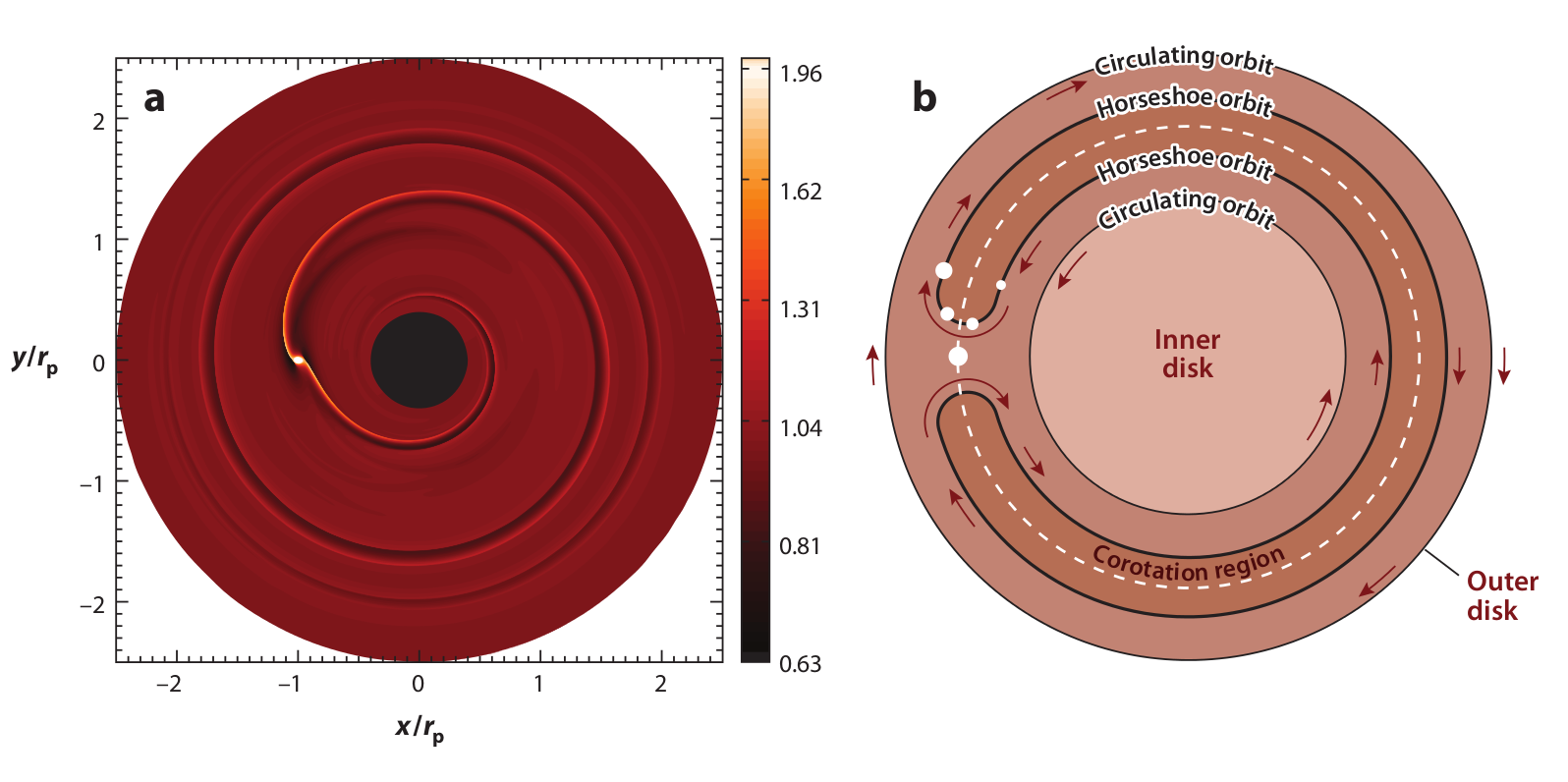} 
    \end{center}
    \caption{The influence of a planet that is embedded in a protoplanetary disk. The planet moves counterclockwise
    around the central star which is located in the center of the two plots.
    The left panel ({\bf a}) shows the surface density distribution (in dimensionless units) 
    of a 10 $M_{\rm Earth}$ planet on a circular orbit that is embedded in
    a disk of constant density. Shown is the configuration at about 20 orbits after insertion of the planet. 
    The planet is a unit distance away from the star at the location $x=-1, y=0$.
    The right panel  ({\bf b}) shows a schematic plot of the motion of gas particles (smaller white dots) on orbits close
    to the planet. The motion is depicted using red arrows in a reference frame corotating with the planet.  
    The horseshoe region that is corotating with the planet is shown in the dark brown color, while the
    inner and outer circulating orbits are shown in the lighter color.
       Taken from Kley \& Nelson \cite{2012ARA&A..50..211K}
     }
    \label{fig:pladisk-basic}
\end{figure}

Embedding a planet into the disk leads to characteristic disturbances in the density of the disk. 
Due to the shearing motion in the disk the sound waves emanating from the planet's location are sheared out
and spiral waves are forming. As shown in Fig~\ref{fig:pladisk-basic} on the left panel, 
the inner spiral is leading and the outer one trailing, where the planet moves counterclockwise around the star.
%
These spiral waves represent over densities in the gas distribution and generate a net force acting on the planet
that changes its orbit.
The second contribution to the forces comes from the corotation region of the planet which is depicted in the right panel
of Fig.~\ref{fig:pladisk-basic}. For single particle trajectories the dynamics on both sides of the planet would be
identical. However, for fluid motion there exists a small asymmetry between the U-turns of the gas in front 
and behind of the planet which generates again a net force on the planet.
For a planet of mass $m\sub{p}$ on a {\it circular orbit} its angular momentum is given by
\beq
\label{eq:j_planet}
    J_{\rm p} = m\sub{p} r\sub{p}^2 \Omega\sub{p}\,,
\eeq
where $r\sub{p}$ is the distance of the planet from the star and $\Omega\sub{p}$ its orbital speed.
The density asymmetries generate a torque, $\Gamma\sub{p}$, acting on the planet that changes the angular momentum according to
\beq
\label{eq:jdot}
    \dot{J}\sub{p} = \Gamma\sub{p} \,.
\eeq
Here, $\Gamma\sub{p}$ denotes the $z$-component of the total torque that needs to be calculated adding the individual
contributions of the whole disk
\beq
\label{eq:gamma_tot}
      \Gamma\sub{p} = - \int\sub{disk} \Sigma ( \vec{r}_{\rm p} \times \vec{F}) \, df  \, = \,
      \, \int_{\rm disk} \Sigma ( \vec{r}_{\rm p} \times \nabla \psi_{\rm p}) \, df  \, = \,
      \, \int_{\rm disk}  \Sigma \frac{\partial \psi_{\rm p}}{\partial \varphi} df \,.
\end{equation}
In eq.~(\ref{eq:gamma_tot}) we assume a flat two-dimensional disk,
where $\vec{F}$ denotes the gravitational force per unit mass between the planet (located at $\vec{r}_{\rm p}$) 
and a small surface element, $d f$, of the disk which has the surface density, $\Sigma(r,\varphi)$.
$\psi_{\rm p}$ denotes the gravitational potential of the planet. 
From eq.~(\ref{eq:gamma_tot}) we can infer that it is exactly the asymmetry in the azimuthal direction
that changes the angular momentum of the planet.
From eqs.~(\ref{eq:j_planet}, \ref{eq:jdot}) and using Keplerian rotation 
\beq
 \label{eq:kepler}
           \Omega\sub{K}(r)  =  \sqrt{\frac{G M_*}{r^3}} \,,
\eeq
we can calculate the change in the planet's distance as
\beq
\label{eq:migrat}
   \frac{1}{r_{\rm P}} \,  \frac{d r_{\rm P}}{dt}
   \,  \equiv  \,  \frac{1}{\tau_{\rm M}}
   \  =  \, 2 \, \frac{\Gamma_{\rm p}}{J_{\rm P}} \,,
\eeq
where we introduced the {\it migration timescale}  $\tau_{\rm M}$.
To infer the importance of the planetary migration, this timescale needs to be compared to the overall evolution time of disk.

Obviously, the analysis of the migration process comes down to the calculation of the torques acting on 
a planet. As we shall see below in Sect.~\ref{subsubsec:eccentricity} this statement refers to planets on circular orbits only,
otherwise the power acting on the planet has to be considered as well.
In principle, according to eq.~(\ref{eq:gamma_tot}), this appears to be an easy evaluation. The problem lies in the
fact that the total torque is always the difference between a positive and negative contribution that have nearly the 
same magnitude. As a consequence the total torque depends delicately on physical aspects of the disk such as viscosity, or heat diffusion.
As we shall see, this does not only influence the speed of the migration but can change even the overall direction.

The calculation of the total torque can be performed by different approaches. For lower mass planets that perturb the density
of the disk only mildly one can perform a {\it linearization}  of the hydrodynamic equations around the
unperturbed disk without the planet, similar to the methods shown in the previous section. Here, the basic state is given by an  
axisymmetric disk that is in pure rotation around the star with $\Omega(r)$ and has a given $\Sigma(r)$ profile
and possibly temperature gradients.
The planet is then added as a small disturbance via an additional potential that rotates with the speed $\Omega\sub{p}$.
Linearizing the equations yields a new perturbed disk structure, whose back reaction on the planets gives the torque,
and hence the migration rate.
In the azimuthal direction the potential can be expanded in a Fourier-series due to the periodicity. Using this procedure,
the strongest perturbations in the disk are excited at particular resonances where the pattern speed of individual Fourier-components
of the perturbed potential, $\omega = m  (\Omega(r) - \Omega\sub{p})$, matches a natural frequency of oscillation in the disk.
The first resonance is the corotation resonance that occurs for material at the location of the planet which has
$\Omega = \Omega\sub{p}$. These are related closely to the horse-shoe drag that we will discuss below.
Other resonances are excited at the those locations where the pattern speed $\omega$ matches the epicyclic frequency $\kappa$,
see eq.~\ref{eq:kappa}. These are the {\it Lindblad resonances} that may occur inside of the planetary orbit or outside. 

In the second approach the torques are obtained from full non-linear hydrodynamic simulations of embedded planets
in disks. Here, the disk is initialized with an equilibrium profile of density and temperature and then the planet is
added as a perturbing object to the flow. In one type of models the initial planetary orbit is fixed in time and only
the disk structure changes due to the action of the embedded planet.
The simulations are then evolved in time until the system has reached a quasistationary equilibrium.
The torque on the fixed planet is continuously monitored and the theoretical migration speed can be calculated
by eq.~(\ref{eq:migrat}). In other models the planet is allowed to change its orbital elements according to the
forces acting on it, and the change in orbital elements is directly calculated. 
Will not go into details of these methods but just quote the main results

\begin{figure}[t]
    \begin{center}
        \includegraphics[width=0.46\textwidth]{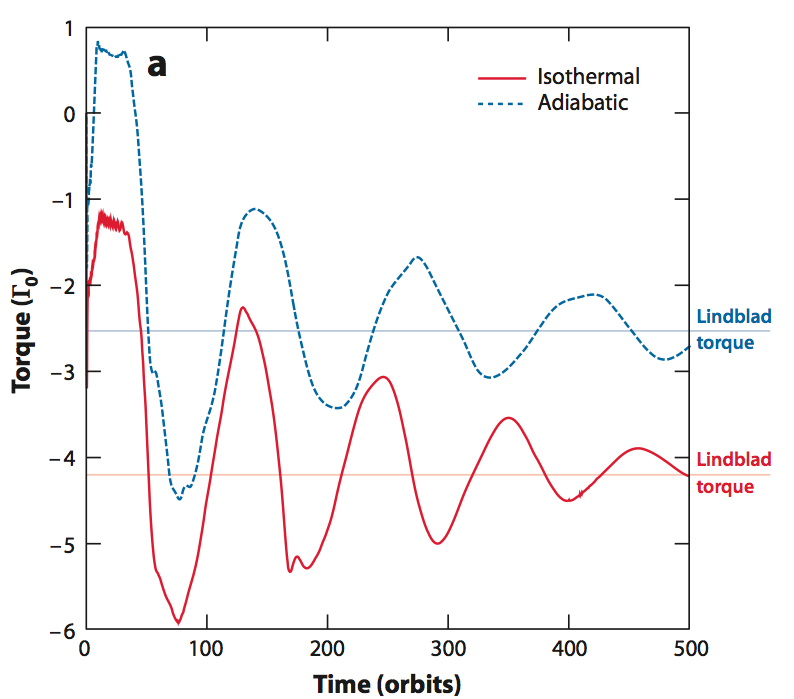} \quad
        \includegraphics[width=0.44\textwidth]{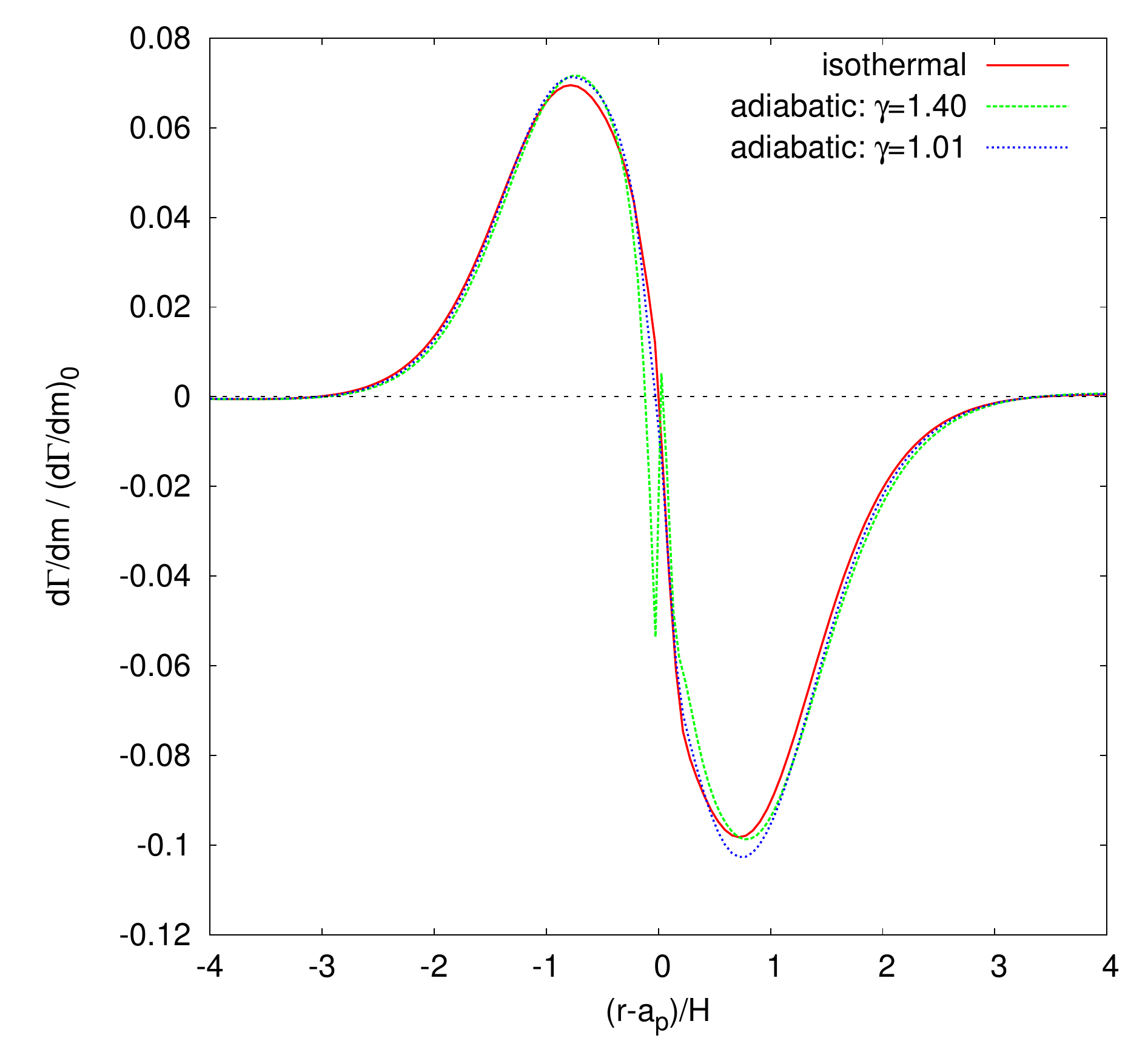} 
    \end{center}
    \caption{
    The torque acting on a low-mass planet on a fixed circular orbit that is embedded in a disk
    that has a constant surface density and very low viscosity.
    Shown are results from 2D hydrodynamic simulations for an adiabatic and a locally isothermal disk. 
    The {\bf left panel} shows the time evolution of the total torque acting on planet.
   At the start of the simulations, when the gas performs its first U-turn near the planet, 
   the torques reach a maximum (fully unsaturated values). Then they drop to reach the final saturated values through 
  a series of oscillations. The horizontal lines refer to the 2D isothermal and adiabatic Lindblad torques \cite{2010MNRAS.401.1950P}.
   The {\bf right panel} shows the radial torque density in the final state, again for an isothermal and two
   adiabatic simulations for $\gamma=1.01$ and $\gamma=1.4$ (adapted from \cite{2012A&A...546A..99K}).
   The units $\Gamma_0$ and $(d \Gamma/d m)_0$ are stated in the main text, see eqs.~(\ref{eq:dgamm0}) and (\ref{eq:gamma0}).
     }
    \label{fig:pladisk-Lindblad}
\end{figure}
\subsection{Type I migration}
\label{subsec:pladisk-type-I}
In this section we deal with {\it lower mass planets}, up to about a few Earth masses that do not perturb the
disk structure significantly. In particular, these planets do not open gaps in the disk.
The torques can be then analysed through linear, as well as non-linear approaches. This migration regime where
the disk is only weakly disturbed is termed {\it type I migration}.
In this case both contributions are important, spiral arms (Lindblad torques) as well as the corotation region.  

As seen from the left panel in Fig.~\ref{fig:pladisk-basic} the spiral arms are the most prominent features 
visible in the perturbed density distribution in disk as generated by the planet.
Clearly, from a purely mechanical point of view, and loosely speaking, the inner spiral arm pulls the planet forward
which leads to an acceleration of the planet and a gain of angular momentum. On the other hand, the outer spiral pulls the planet
back which leads to a loss in angular momentum. This feature is reflected in the mass weighted radial torque density
distribution that is displayed in the right panel of Fig.~\ref{fig:pladisk-Lindblad}. The part inside of the planet
is positive while the part outside is negative. 
The total torque is given by the integral over this distribution 
\beq
    \Gamma\sub{tot} = 2 \pi \int \frac{d \Gamma}{d m} (r) \, \Sigma(r) \, r dr
\eeq
and, as already hinted above, it is a balance between positive
inner and negative outer torques.
The torque density, $d \Gamma(r)/dm$, scales with the planet to star mass ratio squared and as $(H/r)^{-4}$,
and we rescale our results in units of
\begin{equation}
\label{eq:dgamm0}
     \left( \frac{d \Gamma}{d m}\right)_0 =  \Omega\sub{p}^2 \, r_{\rm p}^2 \, \left(\frac{m_{\rm p}}{M_*}\right)^{2}
     \, \left(\frac{H}{r_{\rm p}}\right)^{-4},
\end{equation}
where quantities with index $p$ are evaluated at the planet's position, e.g. $\Omega\sub{p} = \Omega(r\sub{p})$.
We note that in evaluating the pressure scale height one has
to choose either the isothermal or adiabatic sound speed such that $H_{\rm adi} = \sqrt{\gamma} H_{\rm iso}$. Using this scaling the torque
densities for all simulations, isothermal and adiabatic, are identical (right panel in Fig.~\ref{fig:pladisk-Lindblad}).

From a  wave point of view the spiral arms can be treated as the superposition of the sounds waves emanating from the planet.
The torques generated can then be analyzed by the mentioned linearization procedure
which leads to the Lindblad resonances where $m (\Omega(r) - \Omega\sub{p}) = \pm \kappa(r)$,
with the epicyclic frequency $\kappa$. 
The plus sign, i.e., $\Omega(r) = \Omega\sub{p} + \kappa(r)/m$, refers to inner Lindblad resonances (interior to $r\sub{p}$)
where the disk rotates faster than the planet. The minus sign refers to the outer Lindblad resonances, i.e.,
$\Omega(r) = \Omega\sub{p} - \kappa(r)/m$ (exterior to $r\sub{p}$ and the corotation region in Fig.~\ref{fig:pladisk-basic}). 
For a Keplerian disk the epicyclic frequency matches exactly the Keplerian speed, i.e. 
$\kappa = \Omega\sub{K}$ (\ref{eq:kepler}), and the locations of the Lindblad resonances are given by
\beq
\label{eq:lindblad}
                 r\sub{L} = \left( \frac{m}{m \pm 1} \right)^{2/3} \, r\sub{p} \,.
\eeq
These resonances are named after Bertil Lindblad, who studied the dynamics of galaxies.
In case of a radial pressure gradient in the disk the rotation rate is not exactly Keplerian and the resonances are
shifted slightly such as to avoid a divergence at the planetary radius for large $m$ \cite{1980ApJ...241..425G,1993ApJ...419..155A}. 
This causes the torque density to peak at a certain distance, typically a pressure scale length, $H$, away from the
planet, as shown in the right panel of Fig.~\ref{fig:pladisk-basic}.

Detailed results for the Lindblad and corotation torques in locally isothermal disks with a fixed $T(r)$ stratification have
been obtained for lower mass planets, of a few $M_\oplus$, through full 3D analyses.
These were obtained either analytically by linear analysis \cite{2002ApJ...565.1257T} or through fully non-linear
simulations \cite{2010ApJ...724..730D}.
The results for the total torque can be summarized as \cite{2010ApJ...724..730D}
\begin{equation}
\label{eq:gennaro}
        \Gamma_{\rm tot}  = - (1.36 + 0.62 \beta_\Sigma + 0.43 \beta\sub{T}) \, \Gamma_0 \,,
\end{equation}
where $\beta_\Sigma$ and $\beta\sub{T}$ denote the gradients of the density and temperature in the disk that are defined through
the following power laws 
\begin{equation}
        \Sigma(r) = \Sigma_0 \, r^{-\beta_\Sigma}  \quad \quad \mbox{and} \quad \quad   T(r) = T_0 \, r^{- \beta\sub{T}} \,.
\end{equation}
The fit formula (\ref{eq:gennaro}) has been obtained from 3D local isothermal simulations with varying gradients in $\Sigma$ and $T$,
and for a viscosity of $\alpha = 0.004$ \cite{2010ApJ...724..730D}. 
For the definition of the $\alpha$-viscosity see eq.~(\ref{eq:diskvisc}). 
The torque normalization is given by 
\begin{equation}
\label{eq:gamma0}
       \Gamma_0 =  \left(\frac{m_{\rm p}}{M_*}\right)^{2}
       \, \left(\frac{H}{r_{\rm p}}\right)^{-2}   \, \left( \Sigma_{\rm p} \rp^2 \right) \, \rp^2 \Op^2 \,,
\end{equation}
which reflects the scaling of the torque with the physical properties of the disk. Here, $\Sigma_{\rm p}$ denotes
the {\it unperturbed} surface density at the location of the planet.
As seen, the torque scales with square of the planet mass, $m\sub{p}$. This is due to the fact that the perturbations in the disk 
induced by the planet scale directly with $m\sub{p}$, and the back reaction on the planets adds the second factor.
The opening angle of the spiral arms is given by the temperature in the disk, here through the scale height, $T \propto H^2$.
Lower temperatures lead to tighter spirals and consequently larger torques because the torque maxima are closer
to the planet (see Fig.~\ref{fig:pladisk-Lindblad}). The direct scaling with the disk's density 
follows immediately from the definition of the torque in eq.~(\ref{eq:gamma_tot}).
The migration rate follows then from eq.~(\ref{eq:migrat}).
The torque formula, eq.~(\ref{eq:gennaro}) shows that for the typical gradients in the disk ($\beta_\Sigma$ and $\beta\sub{T}$ 
larger than zero) the migration will be directed inwards.
While this is encouraging in explaining the close-in observed exoplanets, the
short timescale of the migration causes problems, however.
Using the slope values for the minimum mass solar nebula (MMSN), $\beta_\Sigma = 1.5$ and $\beta\sub{T} = 0.5$,
we obtain $\Gamma_{\rm tot} = 2.5 \Gamma_0$ and then from Eqs.~(\ref{eq:j_planet}), (\ref{eq:migrat}) and (\ref{eq:gennaro}) 
we find for the migration time
\beq
       \tau_{\rm mig}  \approx  \frac{1}{5} \,  
        \left(\frac{m_{\rm p}}{M_*}\right)^{-1} 
        \left(\frac{m_{\rm d}}{M_*}\right)^{-1} 
       \, \left(\frac{H}{r_{\rm p}}\right)^{2}   \, \Op^{-1} \,,
\eeq
where we used $m\sub{d} = \Sigma\sub{p} r\sub{p}^2$ as a proxy for the disk mass.
Looking at the speed of migration we find for a $1 M_{\rm Earth}$ planet at 5 AU a migration timescale of 
a few $10^5$ yrs. This is shorter than typical lifetimes of protoplanetary disks, and this is a general result for
type I migration in locally isothermal disks. Consequently, to prevent this too rapid inward radial drift,
during the past years much effort has been put into finding mechanisms to slow it down.

The key to the solution lies in the corotation, or horseshoe region.
Looking at the left panel of Fig.~\ref{fig:pladisk-Lindblad} one notices that directly after insertion of the planet
the torques acting on the planet are much smaller than in the final equilibrium case. 
This is true for the isothermal, as well as the adiabatic model. In the latter, the torques are even reversed 
(positive) initially which would imply an outward migration. This torque reduction or reversal is caused by the flow dynamics
in the corotation region. Directly after the insertion of the planet the flow begins to perform U-turns in the vicinity of the
planet. As indicated in Fig.~\ref{fig:pladisk-basic} by the white dots, material from the inside moves outwards
and by doing so has to gain a certain amount of angular momentum from the planet. On the other side the process is reversed
and the material gives angular momentum to the planet. Clearly, if there is an asymmetry in the U-turn
motion in front and behind the planet there will be a net transfer of angular momentum to the planet.
Because in this U-turn motion only material within the horseshoe region is involved, the
asymmetry is driven by radial gradients of particular disk properties across the horseshoe region,
and the resulting torque is called the {\it horseshoe drag} \cite{1991LPI....22.1463W}. The oscillatory settling of the torque is due to the
motion of the material along their horseshoe orbits. The oscillation period is given approximately by the
libration time of matter in the horseshoe because after each libration period the matter reaches again the planet and receives
a new kick. It is the mixing of the matter in the horseshoe region that tends to wipe out the initial gradients of the flow,
and eventually the corotation torques saturate \cite{1991LPI....22.1463W}.
From linear analyses \cite{1979ApJ...233..857G}, it is known that the corotation torque is
proportional to the radial gradient in the specific vorticity (sometimes also called {\it vortensity})
\beq
\label{eq:vortensity}
     \tilde{\omega} \, =  \, \frac{\omega_z}{\Sigma} \, = \,  \frac {(\nabla \times \vec{u})|_z}{\Sigma} \,,
\eeq
and proportional to the radial gradient in the entropy, $S$, see \cite{2008ApJ...672.1054B,2008A&A...485..877P}.
The quantity $\omega_z$ in eq.~(\ref{eq:vortensity}) is the $z$-component of the vorticity of the disk flow. 
For inviscid and barotropic ($p=p(\rho)$) flows one can show that $\tilde{\omega}$ and $S$ are constant
along streamlines, i.e.
\beq
     \frac{d}{dt} \left(\frac{\omega_z}{\Sigma}\right) = 0
\quad \mbox{and} \quad
     \frac{d S}{dt} = 0 \,.
\eeq
As a consequence any initial gradients (in $S$ and $\tilde{\omega}$) are wiped out in the corotation region
in the course of the simulations and the horseshoe drag vanishes \cite{2001MNRAS.326..833B}
and only the (negative) Lindblad torques remain, exactly as seen in the nearly inviscid simulation
presented in Fig.~\ref{fig:pladisk-Lindblad}, see also \cite{2002A&A...387..605M}.
Hence, the necessary physical mechanisms to maintain those gradients are viscosity and radiative diffusion, or radiative cooling 
\cite{2008ApJ...672.1054B,2008A&A...485..877P}.

\begin{figure}[t]
    \begin{center}
        \includegraphics[width=0.75\textwidth]{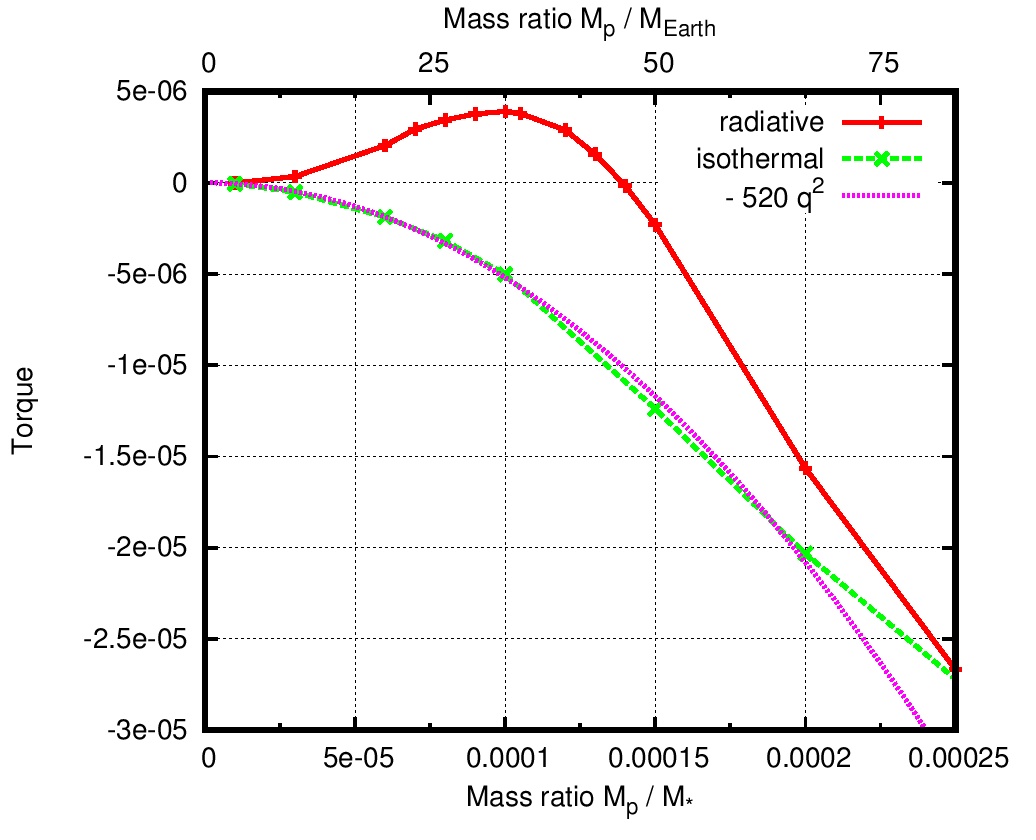}
    \end{center}
    \caption{
    The torque acting on embedded planets in 2D, viscous disks using $\alpha = 0.004$. Results are shown for a locally isothermal disk (green)
    and radiative models (red) that include viscous heating, radiative transport in the midplane, as well as radiative cooling
    from the disk surfaces. The purple line shows a parabola that scales with the square of the planet mass.
    Adapted from \cite{2008A&A...487L...9K}.
     }
    \label{fig:iso-rad-migrat}
\end{figure}

Indeed, through a series of 2D and 3D hydrodynamical simulations it has been shown in more realistic disks models 
which include radiative transport that the migration of planets can be slowed down and even directed outwards,
see Paardekooper \& Mellema \cite{2006A&A...459L..17P} for the very first simulation of this kind.
Subsequently, model sequences using varying planet masses in 2D \cite{2008A&A...487L...9K} and 3D \cite{2009A&A...506..971K} for 
active, viscous disks with radiative transport have shown that there exists is a typical mass range in which the total torque is positive
and the migration can be directed outwards. For maximum efficiency of the torque desaturation the viscous and radiative 
diffusion timescales should lie within the libration and U-turn timescales of the gas in the horseshoe region. Hence,
it depends on the mass of the planet, and disk properties such as viscosity and opacity. 
An example of such a study for the 2D case with embedded planets located at 5~AU is shown in Fig.~\ref{fig:iso-rad-migrat}.
For this disk regime the mass range with the fastest outward migration lies between 25 and 40 $M_{\rm Earth}$.
To study the overall evolution of planets in disks it is useful to apply formulae that provide 
reliable estimates of the torque and can be computed fast and easily.
These were provided in \cite{2010ApJ...723.1393M, 2011MNRAS.410..293P} and confirmed by full 3D radiative simulations
\cite{2011A&A...536A..77B}. The torque formulae indicate that for small planets of a few Earth masses the migration should
be directed inward, a feature hardly visible in Fig.~\ref{fig:iso-rad-migrat}. Indeed, the inward migration of planets below
a mass of about 3 $M_{\rm Earth}$ has been verified in 3D high resolution simulations \cite{2014MNRAS.440..683L}.
Very recently, the migration of low mass planets in this mass range has been addressed by including the effect of accretional
heating from the (solid) material added to the planet in the simulations. 
Taking this additional heat source into account, it has been found that the migration can be directed
outwards for very low mass planets \cite{2015Natur.520...63B}.
This {\it heating torque} effect might play an important role in preventing the too rapid inward migration of
planetary cores towards their host stars.
These studies indicate that the migration direction and rate of an embedded planet depends on the details of the disk flow 
in the close vicinity of the planet. This requires more high resolution simulations similar to the nested-grid simulations
presented in \cite{2003ApJ...586..540D}, or the more recent example by \cite{2015ApJ...811..101F}.

As mentioned above, the torques acting on a planet depend on the physical disk properties (viscosity and opacity).
During the evolution of the disk these will change in space and time. To account for this, sequences of 2D axisymmetric disk models 
that include radiative cooling and viscous and stellar heating have been computed for different evolutionary states
of the host star \cite{2015A&A...575A..28B} as a function of the disk's metallicity, accretion rate, and lifetime.
The parameterized disk structures in conjunction with the torque formulae \cite{2011MNRAS.410..293P} form an ideal setup
for studying the longterm evolution of planets in disks.
%

\begin{figure}[t]
    \begin{center}
        \includegraphics[width=0.66\textwidth]{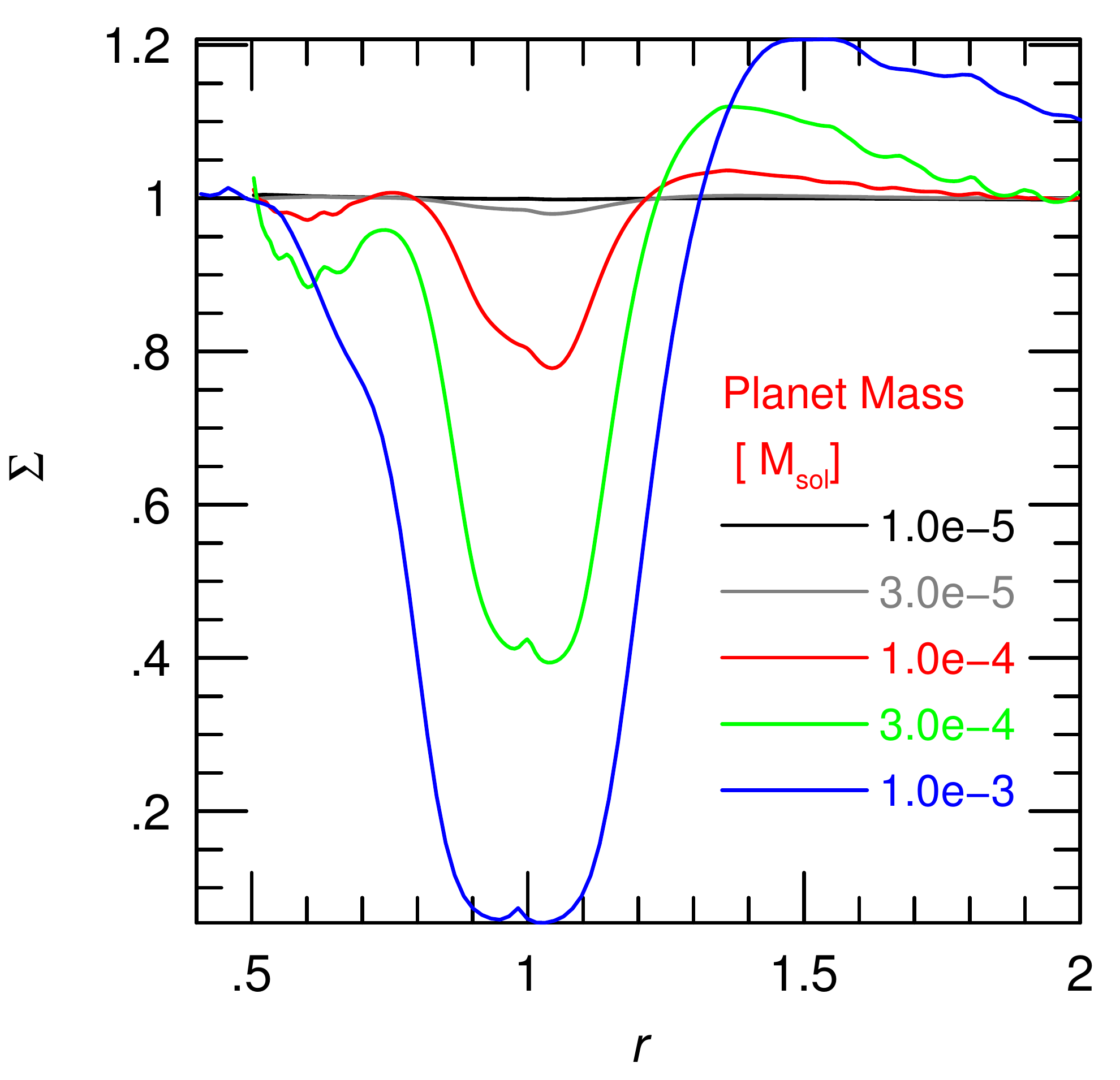}
    \end{center}
    \caption{
    The radial surface density profile of embedded planets with different masses in
    2D, viscous disks using $\alpha = 0.004$. Results are shown for a locally isothermal disk with an aspect ratio
    $H/r = 0.05$. The radius is normalized to the position of the planet and the density to the unperturbed value.
     }
    \label{fig:sigma-mass}
\end{figure}

\subsection{Type II migration}
\label{subsec:pladisk-type-II}
The analysis just presented above refers to the situation of embedded low-mass planets
that do not change the disk's original structure significantly. However, for larger planet masses,
the interaction becomes increasingly nonlinear, and the density profile in the disk will be modified.
In the following, we describe the consequences of this process in more detail.

Before evaluating the impact on migration we will first look at the process of gap formation.
An example on how the presence of a planet impacts the surface density profile of the disk is presented
in Fig.~\ref{fig:sigma-mass}, where results of 2D hydrodynamical simulations of isothermal disks are shown
for various planet masses.
As explained in section \ref{subsec:gas03-massfinal} of this chapter, planets that reach a certain mass will tend to
open a gap in the disk. As explained there, the equilibrium gap structure is determined by a balance between gap opening 
(tidal forces) and gap closing (pressure, viscosity) forces/torques. The thermal condition implies that the radius of the
Hill sphere is larger than the disk scale height and the viscous condition follows from the balance of viscous to
tidal torques. For the necessary planet masses to open a gap we find
\beq
\label{eq:gap-opening}
     q > q_{\rm therm}  \equiv  3 \left( \frac{H}{r} \right)^3\sub{p} = 3 h\sub{p}^3  \quad \text{and} \quad
     q > q_{\rm visc}  \equiv  30 \pi \alpha h^2 \,, 
\eeq
where $q = m\sub{p}/M_*$ refers to the planet to star mass ratio.
For a typical sample disk with aspect ratio $h= 0.05$ and viscosity $\alpha = 10^{-2}$ 
one obtains
$q_{\rm therm} \approx 1.25 \cdot 10^{-4}$, or $m\sub{p} \geq 0.13 m_{\rm Jup}$ and 
$q_{\rm visc} \approx 2.4 \cdot 10^{-3}$, or $m\sub{p} \geq 2.5 m_{\rm Jup}$.
Hence, for these parameters the criteria imply that the gap opening will be driven initially by the thermal criterion rather than the viscosity.
However, from eq.~(\ref{eq:gap-opening}) we also infer that low mass planets of a few Earth masses will also open up gaps for very low 
disk viscosity \cite{2002ApJ...572..566R}. 
This has been seen in numerical simulations \cite{2009ApJ...690L..52L} where a strongly reduced
migration rate was noticed as well.
As also mentioned above, the two criteria (\ref{eq:gap-opening}) have been combined to a single one
\cite{2006Icar..181..587C} in eq.~(\ref{eq:gap-crida}). 
While eq.~(\ref{eq:gap-opening}) gives some estimate under what conditions gaps are to be expected
it does not say anything about the depth of a gap that is carved out by the planet.
Recently, some progress has been made in this direction and fits to numerical simulations 
have shown \cite{2013ApJ...769...41D} that the determining parameter is given by
\beq
             K \equiv  q^2 \, h^{-5} \, \alpha^{-1} \,.
\eeq
The depth of a gap is typically defined as the ratio of the surface density at the location of the gap, $\Sigma_{\rm gap}$, 
to the unperturbed density, $\Sigma_0$, of the disk without the planet.
The simulations show that $\Sigma_{\rm gap}/\Sigma_0$ scales as $K^{-1}$ \cite{2013ApJ...769...41D}. This can be understood
from a simple analytical argument \cite{2014ApJ...782...88F}, that we summarize here.
Assuming that the Lindblad torques are generated within the gap, they are proportional to $\Gamma_0$ (see eq.~\ref{eq:gamma0})
with $\Sigma\sub{p}$ replaced by $\Sigma_{\rm gap}$. This replacement can be justified because in linear theory the prime contribution to the 
torque comes from a radial region separated by one scaleheight $H$ from the planet as shown in Fig.~\ref{fig:pladisk-Lindblad}
. Hence,
\beq
       \Gamma\sub{L}  \sim  q^2  h^{-3} \Sigma_{\rm gap} \, \Omega\sub{p}^2  r\sub{p}^4  \,.
\eeq
This torque has to be balanced essentially by the viscous torque
which scales as the derivative of $\Sigma \nu (d\Omega/dr)$. Substituting now $d \Sigma/dr$ with $\Sigma_0/r$ and $\Omega\sub{K}$ 
for $\Omega$ the viscous torque scales as
\beq
       \Gamma_{\rm visc}   \sim  \Sigma_0  \nu  \Omega'sub{K} r^2 \,.
\eeq
Equating these two equations one obtains with $\nu = \alpha H^2 \Omega\sub{K}$
exactly $\Sigma_{\rm gap}/\Sigma_0 \propto K^{-1}$ \cite{2014ApJ...782...88F}.
As this estimate is based on linear theory the scaling agrees well for up to Neptune sized planets.
For larger planets slightly different exponents have been suggested \cite{2014ApJ...782...88F}.
Through a more careful analysis of the angular momentum conservation the 
scaling
\beq
\label{eq:gap-depth}
            \frac{\Sigma_{\rm gap}}{\Sigma_0} =  \frac{1}{1 + 0.04 K}
\eeq
has been suggested \cite{2015ApJ...806L..15K,2015ApJ...807L..11D}, which agrees well with the simpler
scaling for large values of $K$. While eq.~(\ref{eq:gap-depth}) gives good agreement with numerical
simulations for values up to $K \approx 10$, the deviations are larger in the range $K \sim 10 - 1000$ \cite{2015ApJ...807L..11D},
due to non-linear effects. 

\begin{figure}[t]
    \begin{center}
        \includegraphics[width=0.75\textwidth]{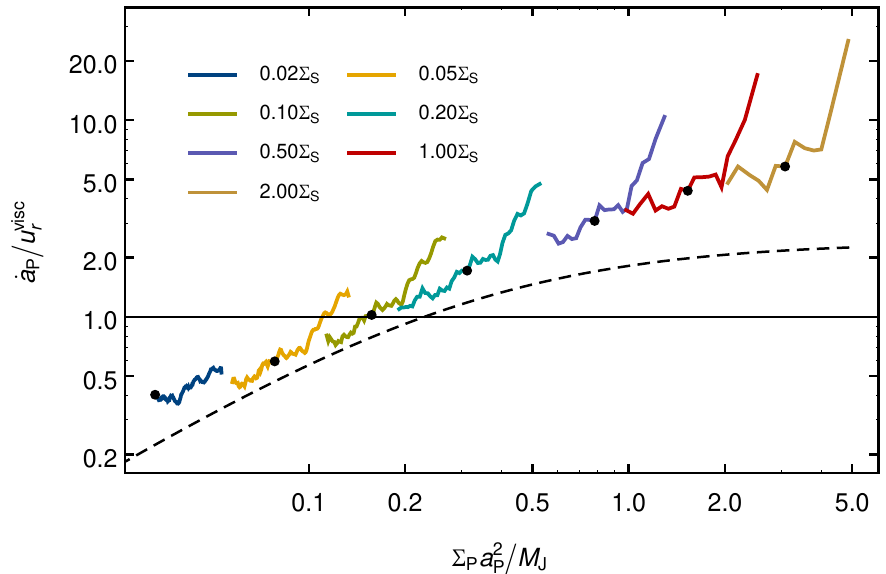}
    \end{center}
    \caption{
    The migration speed, $\dot{a}\sub{p}$, of a massive Jupiter mass planet through locally isothermal disks with different surface densities.
    The speed is normalized to the viscous inflow velocity of the disk, $u_r^{visc} = 3/2 \, \nu/r$, where $\nu = \alpha c\sub{s} H$.
    Each of the models, as denoted by the different colors has a specific, constant mass accretion rate where the unit surface density,
     $\Sigma\sub{S}$, belongs to $\dot{M} = 10^{-7} M_\odot/$yr. All planets start at unit distance ($r=1$) at the top right end
    of the individual curves. The black dots refer to the location $r=0.7$ The dashed line refers to the results by \cite{2014ApJ...792L..10D}
    where a globally constant viscosity $\nu$ was used.
    Figure taken from \cite{2015A&A...574A..52D}.
     }
    \label{fig:mig-typeII}
\end{figure}

After having looked at the conditions for gap formation we will now analyze the impact on the migration of the planet. 
For gap opening planets the corotation region will be reduced in mass leading to a lowering of the corotation torques.
For larger planet masses the gap widens such that the Lindblad torques will be reduced as well,
and the migration rate will be considerably slowed down.
Given by the strong lowering in the density (Fig.~\ref{fig:sigma-mass}) the perturbations are now nonlinear and the corresponding
regime of migration was coined {\it type II migration} \cite{1997Icar..126..261W},
in order to distinguish it from the previous linear type I regime.
The main assumption of type II migration assumes that after having opened an annular gap in the disk,
the planet will remain always in the middle of this gap during the migration, i.e.
the planet has to move with the same speed as the gap.
Because the disk (with its gap) evolves on a viscous timescale it was expected that the planet is coupled entirely to the viscous
evolution of the disk \cite{1986ApJ...309..846L} and has to move with the same timescale
\beq
\label{eq:migII}
        \tau_{\rm mig,II} = \tau_{\rm visc} = \frac{r\sub{p}^2}{\nu} = \frac{r\sub{p}^2}{\alpha c\sub{s} H} 
     = \frac{1}{\alpha h^2 \Omega\sub{p}} \,.
\eeq
In this type II migration regime the rate is independent of the planet mass, and depends only on disk and stellar parameter.
This can only be valid if the disk mass, $m\sub{d} \approx  \pi \Sigma_0 r\sub{p}^2$, is larger than the planet mass $m\sub{p}$.
Otherwise the planetary inertia will require a reduction factor such that $\tau_{\rm mig,II} = m\sub{d}/m\sub{p} \tau_{\rm visc}$
\cite{1999MNRAS.307...79I,1995MNRAS.277..758S}. 
Numerical simulations of a migrating Jupiter type planet over long timescales found a good overall qualitative agreement with
these expectations \cite{2000MNRAS.318...18N}.

Recently however, the assumption (\ref{eq:migII}) has been brought into question and it has been argued that
the migration of gap-opening planets is not locked to the viscous evolution of the disk \cite{2014ApJ...792L..10D}.
By performing sequences of numerical simulations of massive planets in the disk it was found that the migration rate
is not given by eq.~(\ref{eq:migII}) but can be faster as well as slower than this rate, a feature that was noted already earlier
\cite{2007ApJ...663.1325E}.
The reason for this behaviour lies in the fact that a gap does not separate the inner and outer disk completely as is typically assumed.
During the migration of the planet, mass is transferred from one side of the planet to the other side across the
planetary orbit and hence the gap will always be dynamically recreated. If the planet migrates faster than the viscous speed
then mass is transferred from the inside to the outside and the other way around for migration speeds that are slower than
the viscous speed.
More recently, models of migrating planets in accreting disks have been constructed \cite{2015A&A...574A..52D} 
and very similar results have been found. As shown in Fig.~\ref{fig:mig-typeII} the typical migration rate of a massive planet can be
faster or slower as the viscous rate. From a physical perspective this result was expected because the planet can only be moved through
the disk by torques acting on it, such that the actual migration speed is always given by eq.~(\ref{eq:migrat}).

The total torque acting on a migrating planet can be written as the product of a stationary torque and
a dynamical correction factor \cite{2015A&A...574A..52D}.
The first contribution, the stationary torque $\Gamma_{\rm still} = f_{\rm still} \Gamma_0$, refers to the situation of a planet that does not move through the disk
but remains at its actual distance from the planet. Here, $\Gamma_0$ is the normalization from above, see eq.~(\ref{eq:gamma0}), and
$f_{\rm still}$ is a correction factor that depends for a given disk parameter (viscosity, thickness) on the mass of the planet. For
smaller mass planets that can be treated in the linear regime, $f_{\rm still}$ is a constant, given for example by eq.~(\ref{eq:gennaro}).
However, larger planets clear out a gap such that $f_{\rm still}$ becomes lower \cite{1997Icar..126..261W}.
Simulations of embedded planets in accreting disks with non-zero $\dot{M}$ have been performed that first 
keep the planet fixed on its orbit in order to calculate $f_{\rm still}$ \cite{2015A&A...574A..52D}. Then the planets were released and the subsequent
evolution followed.
Interestingly, for locally isothermal disks it was found that the normalized torque $\Gamma/\Gamma_0$ was constant {\it during}
each migration process, which implies that one can write, $\Gamma = f_{\rm mig} \, f_{\rm still} \, \Gamma_0$, with a constant $f_{\rm mig}$.
This means that the factor $f_{\rm mig}$ does not depend on the local disk mass but only on the planet mass and disk viscosity
\cite{2015A&A...574A..52D}. It is typically smaller than one, only for rapid type III migration it is larger.

\subsection{Other regimes of migration}
\label{subsec:mig-others}
Having studied the standard form of migration of planets in viscous, laminar disks we will now briefly describe other forms of
planet migration that consider more aspects of disk physics that have hitherto been neglected.
In addition to the mentioned type I and II migration regimes there exists a regime dominated by rapid, or runaway migration
\cite{2003ApJ...588..494M}, for which subsequently the new  
term {\it type III migration} \cite{2004pfte.confE...2A,2004AGUFM.P32A..06W} has been established.
Because accretion disks are driven by internal turbulence either hydrodynamically or via a combination with magnetic fields,
the gas motion is not laminar but rather shows random fluctuations. The interaction of these variable flow field with embedded
planets leads to an additional type of migration, sometimes termed {\it stochastic migration}. Finally, in more massive disks
the effects of disk self-gravity cannot be neglected anymore and will influence the density structure and subsequently the 
planet migration process.  
\subsubsection{Type III migration}
\label{subsubsec:typeIII}
In contrast to most of the previous calculations (with the exception of the
type II discussion) now the torque acting on a moving planet is considered. Let us focus on a  
planet undergoing inward radial migration on a circular orbit. Material located in the vicinity of
the separatrix, between inner disk material that is on circulating streamlines 
and material on librating horseshoe orbits, will undergo a single U-turn upon a close encounter with the planet 
and will cross the entire horseshoe region. In doing so, it will directly move from the inner to the outer disk
and not be trapped in the horseshoe region. This transfer of disk material from one side of the planet to the
other implies an exchange of angular momentum with the planet. The inner material that crosses horseshoe gains angular momentum
that the planet has to loose. The exerted torque can be written as \cite{2012ARA&A..50..211K}
\beq
\label{eq:gamma_flow}
                  \Gamma_{\rm flow} =  2 \pi \, \Sigma\sub{s} \, \dot{r}\sub{p} \Omega\sub{p} r\sub{p}^3 x\sub{s} \,,
\eeq
where $x\sub{s}$ is the width of the horseshoe region and $\Sigma\sub{s}$ the surface density at the inner separatrix.
This ''flow-through corotation torque'' depends on the migration speed, $\dot{r}\sub{p}$, of the planet,
and there is a positive feedback between the migration rate and the torque, i.e. the torque increases with the migration 
speed. Hence, in principle this can lead to a runaway migration \cite{2003ApJ...588..494M,2004ASPC..324...39A}.
More detailed analysis \cite{2003ApJ...588..494M} shows that the efficiency of the migration will depend on the mass
contained within the horseshoe region. This is because the planet has to carry this material with it upon migrating inward
which acts as additional inertia that will tend to counteract the runaway process.
On the other if the horseshoe is too empty, i.e. the planet too massive, there will
be too little material left over ($\Sigma\sub{s}$ too small) to drive fast migration.
The results show that the highest efficiency requires a partially emptied horseshoe region, where the 
difference between the mass that would be contained in the horseshoe region if the disk was unperturbed by the planet
and the mass that is actually contained in this region (sometimes called coorbital mass-deficit, $\delta m$)
equals the planet mass.
For the MMSN the most promising planet mass for type III migration lies in the Saturn mass range beyond a distance
of about 10~AU. An increase in disk mass will enhance the runaway migration such that disks that are prone to type III migration
are always close to being at the limit of gravitational instability \cite{2003ApJ...588..494M}.
Due to the symmetry implied by eq.~(\ref{eq:gamma_flow}) the direction of type III migration can be directed inward or outward,
depending on the initial perturbation to the planet's motion. This feature has been demonstrated explicitly by numerical simulations
\cite{2008MNRAS.387.1063P,2008MNRAS.386..179P}, where is was also shown that even massive planets of a few Jupiter masses can enter the rapid
type III migration regime. The fast migration terminates when the planet moves into disk conditions that do no longer favor
type III migration, i.e. the coorbital mass deficit, $\delta m$, no longer matches approximately the planet mass. During type III
migration not only the planet itself has to be moved very fast through the disk but all of the material that resides bound to the
planet within its Hill sphere. For massive planets this material is stored in a circumplanetary disk orbiting the planet
and hence the details of type III efficiency will depend on the total mass within this disk \cite{2008MNRAS.387.1063P,2008MNRAS.386..179P}.
Type III migration may have played an important role during the early evolution of the Solar System,
as a potential driving mechanism for motions of Saturn and Jupiter with the Grand Tack scenario \cite{2011Natur.475..206W}. 
The fact that type III migration is more efficient for massive disks takes us directly to a discussion of migration in disks
where {\it self-gravity} can play a role.
\subsubsection{Migration in disks with self-gravity}
\label{subsubsec:selfgrav}
In situations where the total mass of the disk, $m_{\rm disk}$, is no longer negligible with respect to the star, $M_*$, 
it is not anymore sufficient to consider only the gravitational potential of the star but one has to add a contribution
from the disk, i.e. consider an additional term from the disk's self-gravity
$\Psi  = \Psi_* + \Psi_{\rm d}$, where the latter is given by an integral over the whole disk (\ref{eq:psi-disk}).
This change in the potential leads to a (slightly) different angular velocity of the disk that is now no longer Keplerian.
In turn this leads to a radial shift in the location of the Lindblad resonances between the Fourier components of the planet and
the disk material, see eq.~(\ref{eq:lindblad}).
As the torques on an embedded planet can be calculated in linear theory by summing over different resonances one may expect
a change in the migration rate for embedded planets \cite{2005A&A...433L..37P,2008ApJ...678..483B}.
Typical planet migration simulations with a 'life planet' that is allowed to move freely according to the gravitational forces of the disk
assume that only the planet feels the disk's gravity but not the disk itself, i.e. neglect the disk's self-gravity.
As pointed out \cite{2008ApJ...678..483B}, this is obviously an inconsistent situation, because planet and disk move in
different gravitational fields. Through direct 2D-hydrodynamical simulations that include the disk's self-gravity
they showed that for a disk only 3 times the MMSN this inconsistency may lead to a difference in the torque and hence
migration rate by a factor of 2. 
They point out that a self-consistent simulation can be constructed easily by omitting the axisymmetric contribution of the density
in calculating the force on the planet. By including the effects of disk self-gravity in the linear calculations
it was found that for disks a few times the MMSN the change in the torque is only about 10\% of that in a non-self-gravitating disk
\cite{2005A&A...433L..37P,2008ApJ...678..483B}. This difference can be accounted for by the radial shift in the Lindblad resonances,
leading to a migration rate is slightly enhanced over the standard value.

In the past years hydrodynamical simulations following the evolution of massive planets embedded in more massive disks have been presented
in 2D  \cite{2011MNRAS.416.1971B} and 3D \cite{2011ApJ...737L..42M}. These simulations assume that a planet might have formed by gravitational
instability at larger distances in the disk and then migrated to its present location. 
In the simulations presented in \cite{2011MNRAS.416.1971B} the disk setup is chosen such that they are marginally stable,
i.e. have a Toomre number of about $Q \sim 2$ in the main part. 
Under these conditions the disk does not fragment into individual blobs but generates stochastic spiral disturbances
that can produce an effective viscosity and a heating of the disk. This heat is assumed to be radiated away from the disk
surfaces, a process modeled by a radially varying cooling function, where the cooling time is given by
$\tau_{\rm cool} = \beta \Omega^{-1}$, so called $\beta$-cooling, see eq.~(\ref{eq:beta}).
Hence, a statistically stationary disk setup can be produced where the effective viscosity, as generated by the gravo-turbulent fluctuations
of the disk, is given by 
\beq
\label{eq:alpha-cool}
                 \alpha\sub{sg} = \frac{4}{9} \, \frac{1}{\gamma (\gamma -1) \beta} \,.
\eeq
This relation can quite generally be derived \cite{2001ApJ...553..174G} just by assuming that the turbulent heat generated acts like a viscous
dissipation $\propto \Sigma \nu (d \Omega / dr)^2$ which is balanced by the local cooling $d e_{\rm th}/dt \propto e_{\rm th}/\tau_{\rm cool}$,
with the thermal energy density $e_{\rm th}$, see section~\ref{subsec:self04-fragment}.
And this is independent of the source of the turbulence. 
For a value of $\beta = 20$ \cite{2011MNRAS.416.1971B} found in their simulations an $\alpha\sub{sg} \approx  (2-3) \cdot 10^{-2}$
in agreement with the expectation (\ref{eq:alpha-cool}).
Then massive planets are embedded in these disks at large distances (about 100 AU) and their subsequent evolution is followed
through 2D hydrodynamical calculations.
The simulation show that the planets experience a very rapid inward migration. For example, a Jupiter mass planet embedded into a disk
with aspect ratio $h=0.1$ ($\beta = 20$) migrated within less than $10^4$ yrs from 100 AU all the way to only $20$AU.
This fast migration is somewhat reminiscent to the type III migration that was discussed earlier. However, the planets do not
open up a clear gap such that the coorbital mass deficit equals the planet mass, because the gap opening times are longer than
the migration timescale. Hence, the migration is basically of type I with stochastic episodes superimposed, that allow even for brief
phases of outward migration.
These results have been supported by full 3D simulations of migrating Jupiter type planets \cite{2011ApJ...737L..42M} where
a Jupiter planet embedded at 25AU into a $0.14 M_{\odot}$ disk plunges in only $10^3$ yrs to 6AU.
These short migration timescales of clumps formed in more massive disks
represent a serious challenge to planet formation via gravitational instability scenario \cite{2012ApJ...746..110Z},
as shown in the previous section.
\subsubsection{Stochastic migration}
\label{subsubsec:stochastic}
As pointed out already just above, planets embedded in a disk that exhibits turbulent fluctuations will display not a
monotonous smooth migration but will experience stochastic fluctuations in their actual migration rate.
To study this process under realistic conditions, numerical simulations of turbulent disks with embedded planets
have been performed using full magneto-hydrodynamical (MHD) computations in 3D. Magnetized disks are susceptible to the
magneto-rotational instability (MRI) which is believed to be responsible for responsible for driving accretion at least in
sufficiently ionized disks \cite{1991ApJ...376..214B,1991ApJ...376..223H}.
Several direct simulations of planets embedded in turbulent discs have been performed for the case of
MRI unstable discs \cite{2004MNRAS.350..849N,2005A&A...443.1067N,2011ApJ...736...85U}.
In MHD simulations with embedded low mass planets (up to 30 $M\sub{Earth}$)
is was found that the planets experience a random walk process where on average the migration resembles standard type I inward
migration, however, with possible long term ($\sim$ 100 orbits) phases of outward migration \cite{2004MNRAS.350..849N}.
The fluctuation can also lead to a non-zero eccentricity of the embedded planets of the order of a few percent \cite{2005A&A...443.1067N}. 
For a Jupiter mass planet it has been found that the gap is somewhat wider than in the corresponding laminar case using the
same effective $\alpha$-value for the disk \cite{2011ApJ...736...85U,2012ARA&A..50..211K}. Here, the
(relative) turbulent fluctuations become weaker and the migration resembles that of standard laminar disks.
For intermediate planet masses, phases of sustained outward migration have been reported \cite{2011ApJ...736...85U}.
In simulations with very low planet masses (a few $M\sub{Earth}$) is was found \cite{2011A&A...533A..84B} that the structure of the (averaged)
flow in the corotation region in full MHD simulations is comparable to the laminar results, indicated the presence of corotation
torques also in turbulent magnetic disks.
To circumvent the expensive full 3D magnetohydrodynamical simulations, approximate treatments have been developed where the
random features of the MHD turbulence is modeled by stochastic forces that can be added in standard hydrodynamical 
simulations \cite{2004ApJ...608..489L,2010ApJ...709..759B}.
We will no go into more details at this point but rather refer to other review articles \cite{2012ARA&A..50..211K,2014prpl.conf..667B}.

\subsection{Eccentricity and inclination}
\label{subsec:elements}
In addition to a change in the distance from the central star, embedded planets will suffer a change in the other orbital
elements as well. Most important here are the eccentricity and inclination evolution. 
The circular and coplanar orbits in the Solar System were taken already early on, for example by
Kant and Laplace \cite{1755anth.book.....K,1776Laplace}, as evidence that the planets form in
a flat disk-like configuration. The growing sample of extrasolar planets with their huge variety in orbital elements
has shown that a large fraction of planets display significant orbital eccentricity and some systems 
show a large tilt with respect to the
stellar rotation axis, that is expected to be aligned approximately with the rotation axis of the protoplanetary
nebula.
The question arises whether these orbital characteristics of the extrasolar planets can be a result of disk-planet
interaction or have another dynamical origin.

\begin{figure}[t]
    \begin{center}
        \includegraphics[width=0.47\textwidth]{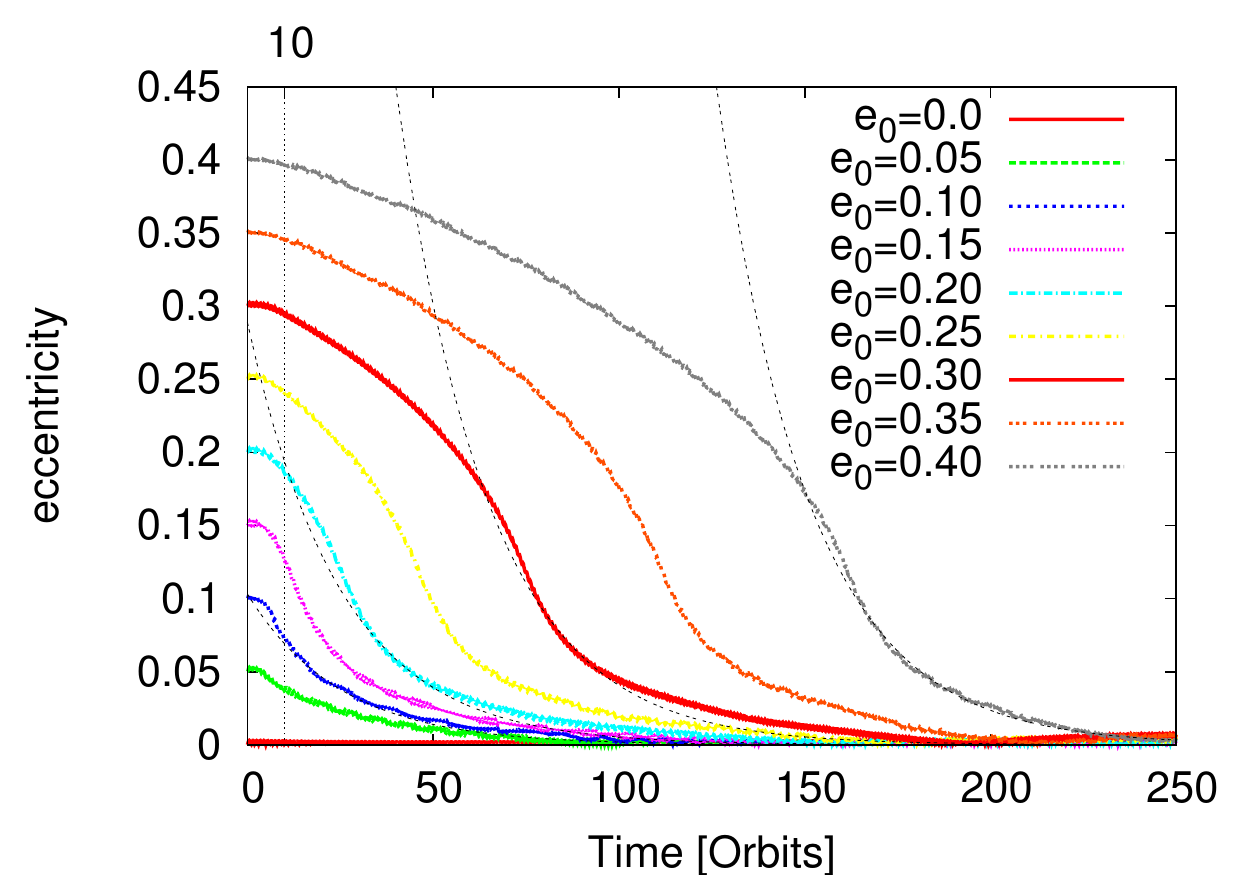}  \quad
        \includegraphics[width=0.44\textwidth]{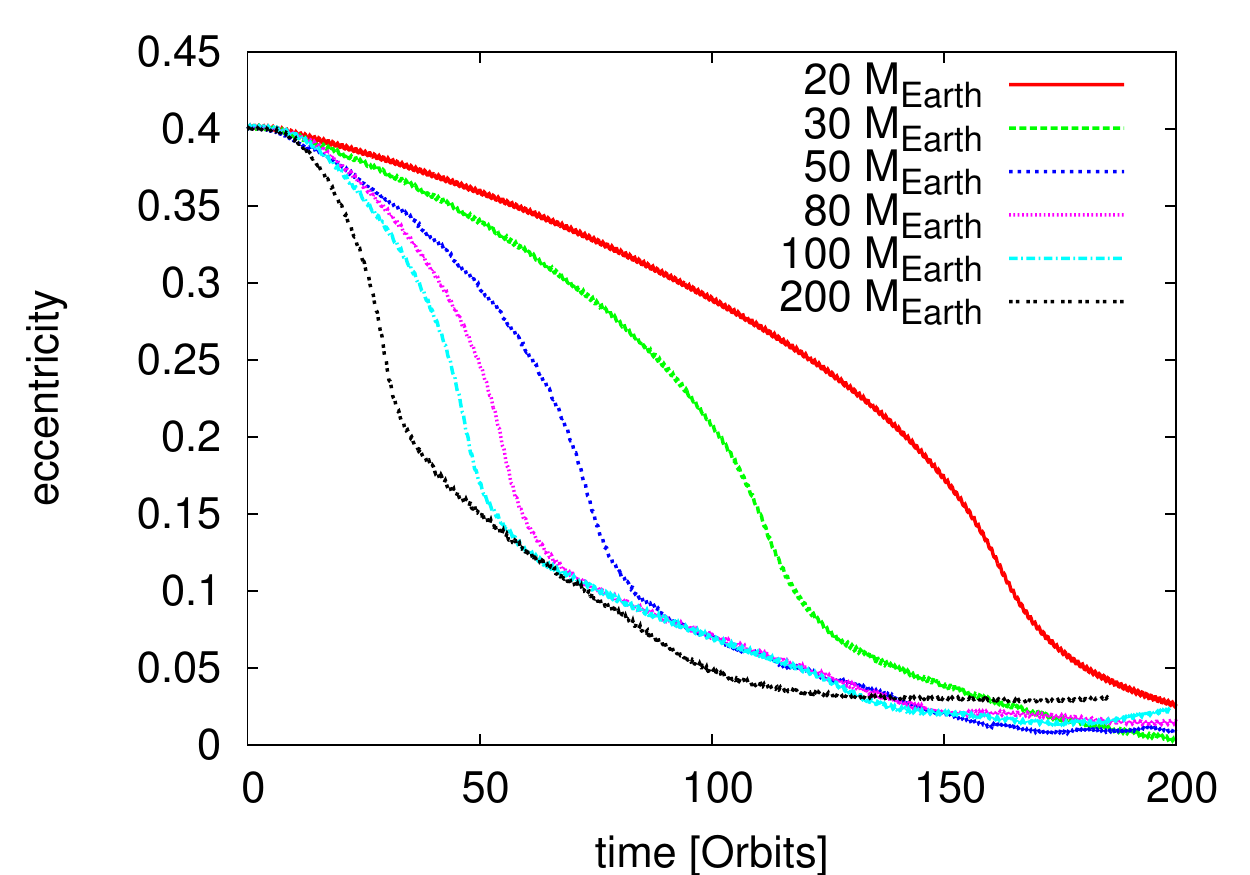}  
    \end{center}
    \caption{
    The change in eccentricity of an embedded planet due to planet-disk interaction.
    The disk models have been calculated using full 3D radiation hydrodynamical simulations. 
    {\bf Left panel} The change in eccentricity of a $m\sub{p} = 20 M_{\rm Earth}$ planet for various initial eccentricities.
     The planet mass is switched on gradually during the first 10 orbits (vertical dotted line). The dashed lines
     indicate exponential decay laws.
    {\bf Right panel} The change in eccentricity of planets with different masses all starting from the same initial
      eccentricity, $e_0=0.4$.
    Figure taken from \cite{2010A&A...523A..30B}.
     }
    \label{fig:ecc}
\end{figure}
\subsubsection{Eccentricity}
\label{subsubsec:eccentricity}
In calculating the change of the eccentricity of the planet, we have to start from the angular momentum and energy of a planet
orbiting the star. These are given by
\beq
         J\subscr{p}  =  m\subscr{p} \, \sqrt{ G M_* a} \, \sqrt{1 - e^2}   
      \quad \text{and} \quad
         E\subscr{p} \, = \, - \, \frac{1}{2} \, \frac{G M_* m\subscr{p}}{a}
\eeq
where we dropped the subscript and wrote for simplicity just $a$ and $e$ for the planetary semi-major axis and eccentricity, respectively.
Classical mechanics tells us that the angular momentum is changed by the torque and the energy by the power acting on the
planet by the disk. These are given by
\beq
      \Gamma\subscr{disk}
     =  -  \int\subscr{disk}  \Sigma \, \left. ( \vec{r}\subscr{p} \times \vec{F} ) \, \right|_z
   \, d f
      \quad \text{and} \quad
        P\subscr{disk}  \, = \, \int\subscr{disk} \,
            \Sigma \,  \dot{\vec{r}}\subscr{p} \cdot \vec{F} \, d f \,,
\eeq
where $\Gamma\subscr{disk}$ was defined already before in eq.~(\ref{eq:gamma_tot}). 
The changes in $E\subscr{p}$ and $J\subscr{p}$ are now given as
\beq
\label{eq:Jp-evol}
    \frac{\dot{J}\subscr{p}}{J\subscr{p}}
 =
  \, \,
   \frac{1}{2}   \frac{\dot{a}}{a}
   - \frac{e^2}{1-e^2}  \frac{\dot{e}}{e}
   \  =   \frac{\Gamma\subscr{disk}}{J\subscr{p}}
\eeq
and
\beq
\label{eq:Ep-evol}
  \frac{\dot{E}\subscr{p}}{E\subscr{p}}
  =
  \, \,
    - \frac{\dot{a}}{a}
   \  =   \frac{P\subscr{disk}}{E\subscr{p}} \,.
\eeq
Having calculated $\Gamma\subscr{disk}$ and $P\subscr{disk}$ from the simulations, one can directly evaluate the
changes in the orbital elements of $a$ and $e$.
Obviously, the change in the semi-major axis is given solely by $P\subscr{disk}$ and not by the torque as assumed above
in eq.~(\ref{eq:migrat}). The reason for this apparent discrepancy lies in the fact that we assumed circular orbits 
in the discussion in section~\ref{subsec:pladisk01-basic}.
Setting $e=0$  we find from eqs. (\ref{eq:Jp-evol}) and (\ref{eq:Ep-evol}) that
$P\subscr{disk} = \Omega\sub{K} \, \Gamma\sub{disk}$ such that for circular orbits one can use indeed the
torque to estimate the migration rate.

For low mass planets that do not open a significant gap within the disk, linear perturbation theory can be applied to
analyze the eccentricity evolution. In contrast to the circular case an eccentric planet generates additional resonances in
the disk \cite{1980ApJ...241..425G}. 
Similar to a planet on a circular orbit, Lindblad resonances occur at disk locations where the perturbation
frequency experienced by a fluid element equals the local epicyclic frequency,
and corotation resonances occur wherever the pattern speed equals the local angular velocity of the disk. 
External Lindblad resonances act to increase $e$, while corotation and 
co-orbital Lindblad resonances act to damp it \cite{2003ApJ...585.1024G}. The outcome depends on the relative
contribution of damping and exciting contributions and, similar to the migration, and the net result is the (small) difference
of larger individual contributions.  
Linear analyses shows that (for not too high initial values) the eccentricity is damped exponentially
on a timescale much shorter than the migration timescale, where typically 
\beq
\label{eq:tau-ecc}
          \tau_{\rm ecc}  \sim  \left(\frac{H}{r} \right)^2  \, \tau_{\rm mig} \,.
\eeq
Equation~(\ref{eq:tau-ecc}) applies for initial eccentricities
smaller than the relative scale height, $e \lesssim H/r$ \cite{2004ApJ...602..388T}.
This rapid decline in $e$ is caused primarily by the damping action of the corotation resonances.
For larger initial eccentricities with $e \gtrsim H/r$ the increased speed of the planet with respect
to the disk material makes the interaction nonlinear and the damping rate changes to \cite{2000MNRAS.315..823P} 
\beq
\label{eq:ep-damp}
        \frac{d e\sub{p}}{d t}  \propto  - \, e\sub{p}^{-2} \,.
\eeq
These results on the evolution of $e\sub{p}$ (here we added the subscript 'p' for clarity), including the turnover from the exponential to the quadratic damping in eq.~(\ref{eq:ep-damp}),
have been confirmed through full non-linear multi-dimensional hydrodynamical
simulations of embedded low-mass planets up to 30$M_{\rm Earth}$ in 2D disks \cite{2007A&A...473..329C}
and up to 200 $M_{\rm Earth}$ in 3D radiative disks \cite{2010A&A...523A..30B}.
In Fig.~\ref{fig:ecc} we display the time evolution of the planetary eccentricity with time for different initial eccentricities (left panel)
and different planet masses (right panel). From the time evolution it is clear that decay rate is indeed different for large and small
$e\sub{p}$, it changes from that of eq.~(\ref{eq:ep-damp}) to the exponential rate at the turnover eccentricity of $e\sub{p} \approx 0.1$ which
is equivalent to is $e\sub{p} \approx 2 H/r$. The same behaviour was found in the 2D simulations as well \cite{2007A&A...473..329C}.
The right panel of Fig.~\ref{fig:ecc} shows that the damping rate increases with increasing mass of the planet, at least for
large initial $e\sub{p}$. This behaviour is in conflict with theoretical expectations based on linear theory that there should be a
regime where eccentricity excitation due to planet disk interaction is possible \cite{2003ApJ...585.1024G}. The authors argue that
the damping effect of the corotation region diminishes when the planet becomes massive enough to open a gap. Hence, they expect
a possible eccentricity excitation for planet masses around Saturn or Jupiter. However, the results displayed in
Fig.~\ref{fig:ecc} from \cite{2010A&A...523A..30B} and more recent 3D simulations \cite{2013MNRAS.428.3072D}
indicate otherwise.
Only for very large planets with masses beyond 10 $M_{\rm Jup}$ planet disk interactions will lead to a significant
rise in the $e\sub{p}$ \cite{2001A&A...366..263P,2006ApJ...652.1698D}. This increase in the planet's eccentricity is caused
in that case by the back-reaction of an eccentric disk on the embedded planet where the disk's eccentricity has been generated by the
presence of the planet \cite{2006A&A...447..369K}. 
 However, very recent simulations have hinted at possible eccentricity growth for massive planets
in disks with low viscosity and small pressure \cite{2015ApJ...812...94D}, demonstrating that more studies are required
to settle this issue.
Nevertheless, the numerical results indicate that the observed high eccentricities in the extrasolar planets cannot
in general be the outcome of planet-disk interaction but are rather a result of multi-body gravitational interactions.
\subsubsection{Inclination}
\label{subsubsec:inclination}
Similar to the eccentricity changes, the disk's impact on the inclination has been studied through linear analyses
\cite{2004ApJ...602..388T} which indicate that for low inclinations with $i\sub{p} \lesssim H/r$ the inclination is
exponentially damped, i.e. $d i\sub{p} /dt \propto - i\sub{p}$. The decay occurs again on a timescale that is 
of the same order as the eccentricity damping time, i.e. by a factor $(H/r)^2$ shorter 
than the migration timescale. 
The exponential decay for the inclination has been supported in 3D hydrodynamical simulations of planets embedded in
isothermal disks \cite{2007A&A...473..329C}.  
For larger inclinations $i\sub{p} \gtrsim H/r$, these researchers find inclination damping on a longer timescale with a
behavior $d i\sub{p} /dt \propto \, - \, i\sub{p}^{-2}$ which is identical to the scaling obtained for eccentricity damping 
when $e\sub{p}$ is large. This has been verified in full 3D radiative simulations \cite{2011A&A...530A..41B}.
The different damping rate for larger inclinations is caused by the fact that for inclined orbits with
$i\sub{p} \gtrsim H/r$ the planets leave the disk region vertically twice for each orbit. 
The damping action of the disk is massively reduced and can only operate during the short time intervals while
the planet is crossing the disk. This process can then be treated by a dynamical friction approach \cite{2012MNRAS.422.3611R}.
Additionally, it has been shown that the inclination is damped by disk-planet interaction for all planet masses up to about 
$1 M_{\rm Jup}$ \cite{2011A&A...530A..41B,2009ApJ...705.1575M}.
In recent 3D simulations of planet-disk interaction for massive, inclined planets it has been shown that the 
inclination seems typically damped, however with a damping timescale
that increases for larger initial inclinations \cite{2012MNRAS.422.3611R,2013MNRAS.431.1320X,2013A&A...555A.124B}.

In summary, due to the much shorter timescales of eccentricity and inclination damping in contrast to the migration time,
it is to be expected that for isolated planets embedded in disks these quantities should be very small,
at least in laminar disks; i.e., even if planets formed in disks with non-zero $e\sub{p}$ and $i\sub{p}$,
they would be driven rapidly to a state with $e\sub{p} =0$ and $i\sub{p} =0$.

\begin{svgraybox}
Growing planets that are still embedded in the disk experience gravitational forces due to the interaction with the disk
that change the orbital parameter of the planet.
Most important is the radial migration of the planet as this implies that the locations where planets are observed today do not
necessarily coincide with their formation locations. Initial results (for isothermal disks) indicated very rapid inward migration
such that too many planets would have been lost into the star.
Recent results have shown that migration can in fact be directed inward or outward, depending on the physical
parameters of the disk. On the other hand, eccentricity and inclination are typically damped on much shorter timescales.
Hence, the large eccentricities and inclinations of the observed exoplanets are most likely not caused by planet-disk interaction
but have their origin in multi-body dynamical interactions.
\end{svgraybox}

%
%
%
\section{Multi-body systems}
\label{lect:06}
So far we have studied the growth and evolution of single planets in the disk. However, the observations indicate that
planets tend to be found in systems with several objects rather than being lonesome travelers.
In fact, through the Kepler space-mission hundreds of transiting multi-planet systems were discovered
\cite{2014ApJS..210...19B}. Out of the 1200 planetary systems observed by now (mid 2015),
about 500 are multiple planet systems (source: \href{http://exoplanet.eu/}{\tt exoplanet.eu}).
The occurrence and architecture of planetary systems has nicely been reviewed in \cite{2015ARA&A..53..409W}. 

The presence of additional objects in the system induces gravitational disturbances that can modify 
the evolution of the individual planets as well the overall evolution of the whole system.
This can lead to planets with extreme orbital elements such as high eccentricity or inclination and to the 
formation of resonant planetary systems. Furthermore, planets are found in binary star systems where their formation
pathway is made more challenging through the additional strong gravitational disturbance by the companion star. 

In this part we will discuss the dynamics of multi-planet systems, with some focus on
the formation of resonant, or near resonant systems, as well as planets in binary star systems.
 
\subsection{Resonances}
\label{subsec:multi01-resonances}
When talking about resonances, in the context of these lecture notes we will restrict ourselves to mean motion resonances (MMRs) which
are defined by an integer ratio of the mean orbital motions of two planets,
\beq
    n_1/n_2 = p/q  \quad \quad  \mbox{with} \quad \quad  q,p \in \mathbb{N}_{>0} \,,
\label{eq:reson}
\eeq
where $n_i$ denotes the mean orbital velocity of the $i$-th planet. We will denote the inner planet by $i=1$ and the outer one
by $i=2$, hence $p > q$. The {\it order} $m$ of the resonance is given by
\beq
     p =  q + m \,.
\label{eq:order}
\eeq

Among the 8 planets in the Solar System there is no MMR. Taking into account the dwarf planets, one finds that Neptune
and Pluto are in fact engaged in a 3:2 MMR, where the orbital period of Neptune is 165~yrs and that of Pluto 248~yrs, and their
semi-major axes are 29.7\,AU and 35.5\,AU, respectively. This means that for 3 orbits of Neptune, Pluto performs exactly 2 orbits
around the Sun. The high eccentricity of Pluto's orbit ($e\sub{P} = 0.25$) implies that
Pluto's perihelion lies inside the orbit of Neptune such that the two orbits are actually crossing each other.
In general, planetary orbits are only stable if the two planets never approach each other to a distance closer than their mutual
Hill radius. For crossing orbits, this condition can only be satisfied if the orbits are resonant. Hence, Pluto's orbit is protected
from being unstable by the special 3:2 resonance that avoids close encounters and maximizes Pluto's separation from Neptune.
For the Pluto-Neptune system additional stabilization arises from the fact that their orbits are mutually inclined by about
17$^\circ$. 
The fact that the orbit of Pluto is shared by many other plutinos
has led eventually to the degradation of Pluto from full planet to dwarf planet status, as decided by the International Astronomical Union
in 2006.
Similarly, in an exoplanetary system resonances are a way to protect the system from dynamical instability.

To specify a particular resonant configuration of a system, it is useful to define the resonant angles $\Phi_{k}$ 
by
\begin{equation}
\label{eq:phik}
   \Phi_{k} = p \lambda_2 - q \lambda_1 - p \varpi_1 + q \varpi_2
     - k (\varpi_2 - \varpi_1) \,,
\end{equation}
with $p>q$. In eq.~(\ref{eq:phik}) the $\lambda_i$ denote the mean longitudes and $\varpi_i$ the
longitudes of periapse of two planets.
The integers $k$ satisfy $q \leq k \leq p$ and can take $p-q+1$ possible values.
Of the $p-q+1$ resonant angles, at most two are linearly independent, and a resonance condition is given
if at least one angle will librate, i.e. its values do not extend over the whole range of 360 degrees but is limited
to a smaller range. 
Let us take as an example a system with a 2:1 resonance. Here, $p=2$ and $q=1$, such that we have
\beq
\Phi_1  =  2 \lambda_2  - \lambda_1 - \varpi_1, \quad
\Phi_2  =  2 \lambda_2  - \lambda_1 - \varpi_2
\quad \quad  (\Delta \varpi = \Phi_2 - \Phi_1) \,,
\eeq
where we defined as an additional variable the difference $\Delta \varpi$ (sometimes also $\Phi_1 - \Phi_2$) 
between the longitudes of periapse which is often used to replace one of the resonant angles $\Phi_1$ or $\Phi_2$.
The resonant angles $\Phi_{k}$ follow from an expansion of the perturbation potential between the two planets that
would move otherwise on unperturbed Kepler-orbits. Upon expanding the perturbation in a 
Fourier-series, the resonant angles are the stationary phases in the expansion.
If an angle $\Phi_{k}$ is in libration (i.e. covers a range smaller than $[0,2\pi]$), then the system is said to be in a 
MMR. For more detailed information of resonances see the excellent book 
of {\it Solar System Dynamics} by Murray \& Dermott \cite{1999ssd..book.....M}.
For a 2:1 resonance, if both angles $\Phi_{k}$ librate (i.e. $\Delta \varpi$ as well), then the system is said to be in a
state of apsidal corotation resonance (ACR), where the two lines of periapse rotate with the same speed (on average)
\cite{2003ApJ...593.1124B}. For Keplerian orbits the resonant condition (\ref{eq:reson}) for two planets 
implies a ratio of their semimajor axes of $a_2/a_1 = (n_1/n_2)^{2/3}$, which follow from Kepler's third law.

One of the most famous examples of a planetary system in a first order 2:1 resonant configuration is GJ~876. 
The initial discovery \cite{2001ApJ...556..296M} noted two massive planets
(GJ~876~b and c) with minimum masses of $0.56$ and $1.89 M\sub{Jup}$
that orbit a M-type star of only 0.3 $M_\odot$.
These two planets have periods of 30 and 60 days and their orbits are fully aligned, i.e. $\Delta \varpi$ is in libration with
$\Delta \varpi\sub{max} \approx 30^\circ$ \cite{2001ApJ...556..296M}, hence GJ~876 is in an ACR state.
Another example is the system HD 128311 where two massive planets are again in a 2:1 resonance but here only one resonant angle
is librating, while the second one is circulating, hence the system is not in an ACR state \cite{2005ApJ...632..638V}.
A system that is believed to be in a second order 3:1 resonance is HD~60532 where two massive planets orbit an F-type star \cite{2008A&A...491..883D}.

Later two additional planets have been discovered in GJ~876, one (d) very close-in on a 2d orbit around
the host star, and another fourth Uranus mass planet (e) in a 124 day orbit beyond planets b and c. This
places the three outer planets of GJ~876 in a Laplace-type 4:2:1 configuration \cite{2010ApJ...719..890R}.
This first known {\it resonant chain} amongst extrasolar planets bears some similarity to the Laplace resonance
of the Galilean satellites Io-Europa-Ganymede orbiting Jupiter but it is dynamically very different
as it exhibits a longterm chaotic dynamics \cite{2010ApJ...719..890R, 2013MNRAS.433..928M}.
Another famous example of a planetary system that is believed to be in a 3-body Laplace resonance is the directly imaged
system HR~8799 where three massive planets are orbiting a young A-star \cite{2014MNRAS.440.3140G}.
During the last years more resonant chains have been discovered using {\it Kepler}-data.
In Kepler-60 three planets with masses around $\sim 4$M$_\oplus$ are engaged in a 5:4:3
Laplace MMR \cite{2016MNRAS.455L.104G}.
The four Neptune mass planets in Kepler-223 have periods in a ratio close to 3:4:6:8 \cite{2016Natur.533..509M}. This four
planet resonant chain has been taken as another clear indication that migration of planets through the disk
must have taken place.

\subsubsection{Formation of resonant planetary systems}
\begin{figure}[t]
    \begin{center}
        \includegraphics[width=0.70\textwidth]{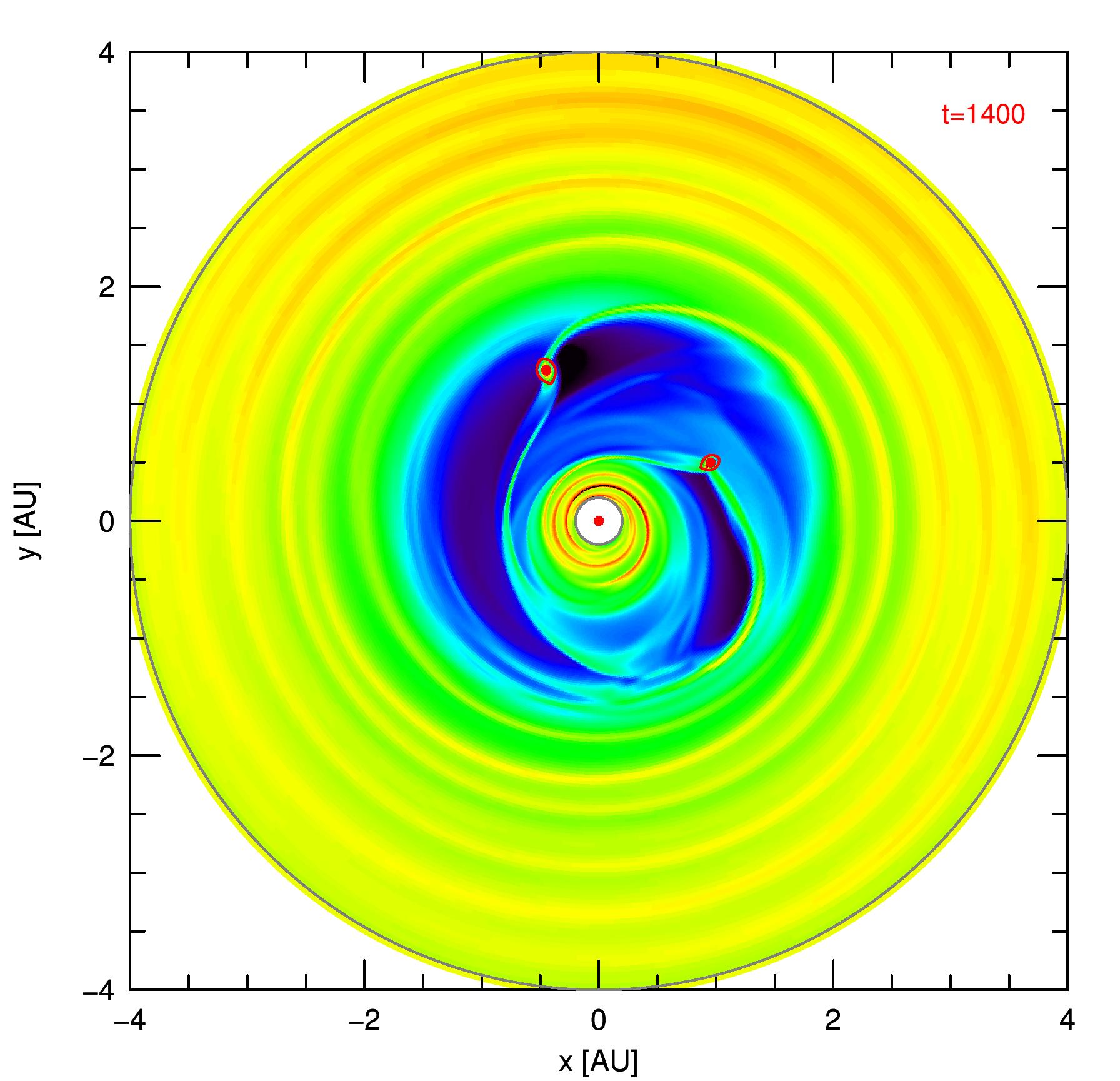} 
    \end{center}
    \caption{Outcome of a two-dimensional hydrodynamical simulation of two embedded planets in a protoplanetary disk, for parameters
    of the system HD~73526 where two planets of about 2.5 $M\sub{Jup}$ each orbit a central star. 
    Shown is the gas surface density where yellow denotes higher density and blue lower values. The central red dot
    denotes the position of the star while the locations of the planets are given by the two red dots and the
    red lines indicate their Roche lobes. The scale of the $x,y$ axes is in AU.
    Shown is the configuration 1400 yrs into the simulations after capture in a 2:1 resonance has occurred, 
     see Fig.~\ref{fig:HD73526x}. 
     After \cite{2007A&A...472..981S}
     }
    \label{fig:HD73526}
\end{figure}

\begin{figure}[t]
    \begin{center}
        \includegraphics[width=0.70\textwidth]{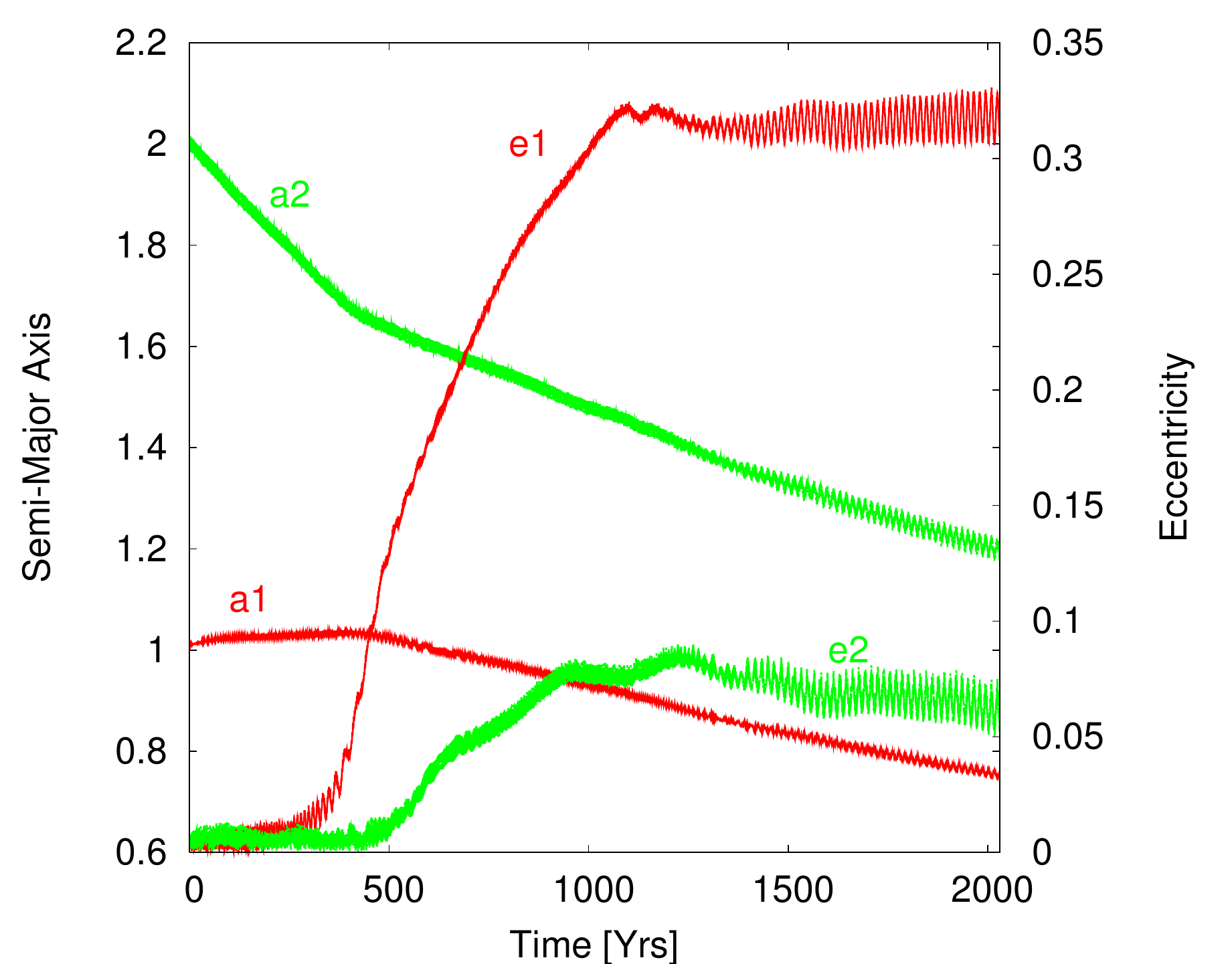}
    \end{center}
    \caption{
         The evolution of the semi-major axes and eccentricities of
         two embedded planets in a disk (those shown in Fig.~\ref{fig:HD73526}) 
         for a full hydrodynamical model. The semimajor axis and eccentricity of the inner
         planet ($a_1$ , $e_1$) are indicated by the red curves and of the outer planet
        ($a_2$ , $e_2$) by the green curves. 
    Starting from their initial positions ($a_1 = 1.0$ and $a_2 = 2.0$ AU) the outer planet migrates
    inward (driven by the outer disk) and the inner one very slowly outward (due to the inner disk).
       After about 400 yrs into the simulations they
         are captured in a 2:1 mean motion resonance and they
         remain coupled during their inward migration. The eccentricities increase
         rapidly after resonant capture and settle to equilibrium values for longer
         times.
     }
    \label{fig:HD73526x}
\end{figure}

A resonant configuration between two planets is a special dynamical state
and it is not very likely that the two planets formed directly in situ at their observed locations,
or were captured into such a state. Instead, the formation of resonant planetary systems is a natural outcome of
a differential migration process. Here, it is assumed that several planets form in a disk and migrate according to the 
torques acting on them. 
Typically, the radial drift speeds are not identical and a differential migration process ensues.
In case the inner planet migrates slower than the outer one, the radial distance between the two planets 
becomes smaller and they approach each other. Whenever the location of a resonance is crossed, i.e. the current orbital periods
of the planets have an integer ratio, the mutual interaction becomes stronger due to the periodic perturbations.
As a consequence the planets can be trapped in a resonance that is maintained during the subsequent migration process.
A typical situation of such a process is shown in Fig.~\ref{fig:HD73526}, which shows the
surface density distribution of a disk with two massive embedded planets, obtained from a numerical
model that simulates the formation of the system HD~73526 \cite{2007A&A...472..981S}.
Because the planets are relatively close, they
orbit within a joint deep gap where the outer planet feels only the torque by the outer disk (which is negative) and
the inner planet feels only the torque of the inner disk (which is positive). Consequently, the two planets approach
each other and a convergent migration process sets in, and eventually the planets are captured in a resonance, here 2:1.
In the subsequent migration process the planets maintain the resonance and move jointly together towards the central star.
The resonant capture is accompanied by an increase in the orbital eccentricities of the planets. The evolution of the 
semimajor axis and eccentricity for such a resonant capture process is displayed in Fig.~\ref{fig:HD73526x}.

%
%

%

The modeling of the resonant capture process is very time consuming because multi-dimensional hydrodynamic simulations have to
be performed, and a full evolution takes typically several weeks to run due to the long evolutionary times to be simulated.
Hence, a simple model for this type of process, using only 3-body simulations consisting of one star and two planets has been
suggested where the action of the disk is taken care of by adding 
additional damping forces using a parameterization of the migration and eccentricity damping.
This type of a modeling takes only a few minutes.
Specifically, the following prescription for the time evolution of the semi-major axis, $a$, and eccentricity, $e$,
of embedded planets has been used in simulations
performed to explain the observed state of GJ 876 \cite{2002ApJ...567..596L}
\beq
\label{eq:tau-ae}
    \frac{\dot{a}}{a} = \frac{1}{\tau\sub{a}}, \quad  \frac{\dot{e}}{e} = \frac{1}{\tau\sub{e}} 
     \equiv  K \,  \frac{\dot{a}}{a} \,.
\eeq
Here $\tau\sub{a}$ and $\tau\sub{e}$ 
denote the damping time scales for the semi-major axis and eccentricity of the planets, respectively,
and $K$ is a constant that describes the ratio between these two.
This prescription for $\dot{a}$ and $\dot{e}$ can be used in principle for both planets because they are both in contact with
the disk, but often it is applied only to the outer one assuming that the inner disk 
has been accreted already onto the star. 
The parameter $K$ specifies the ratio of migration and eccentricity damping. In a study to model GJ~876 Lee \& Peale 
\cite{2002ApJ...567..596L} have applied exactly this model (damping of the outer planet only)
and showed that a configuration similar to the observed one
can be obtained by choosing a value $K=100$ for the eccentricity damping. Smaller values for $K$ typically result
in too small damping and instability of the system.
In follow-up studies, using full hydrodynamic simulations it was found that the disk damping for these systems
with massive planets will produce values of $K \sim 10$ only \cite{2004A&A...414..735K}.
Later it was shown that by taking the inner disk into account, as displayed in Fig.~\ref{fig:HD73526}, it is possible to obtain
full hydrodynamic results in agreement with the observed state of GJ~876 \cite{2008A&A...483..325C}.
The system ends up in apsidal corotation, with correct eccentricities.

This type of differential migration process is a very natural outcome of the dynamical evolution of planets embedded
in protoplanetary disks, and very many resonant configurations should be expected, the only question remaining, in which
resonance are the planets captured.
To become captured in a resonance the migration has to be sufficiently slow such that the excitation mechanism can
operate long enough for them to be captured. If the migration is too fast, the planets cross the resonance and
the system becomes more unstable. In addition the interaction between the two planets has to be strong enough for capture. 
As a consequence, massive planets in the Jupiter mass range and larger are typically captured into a 3:1 or 2:1 
resonance \cite{2004A&A...414..735K}. In fact observationally there is indeed a clustering of massive planets
within the 2:1 resonance \cite{2015ARA&A..53..409W}. 
On the other hand, for smaller mass planets, where the outer planet has up to 
a Saturn mass are typically captured into a 3:2 resonance \cite{2008A&A...482..333P}. 

%
\begin{figure}[t]
    \begin{center}
        \includegraphics[width=0.70\textwidth]{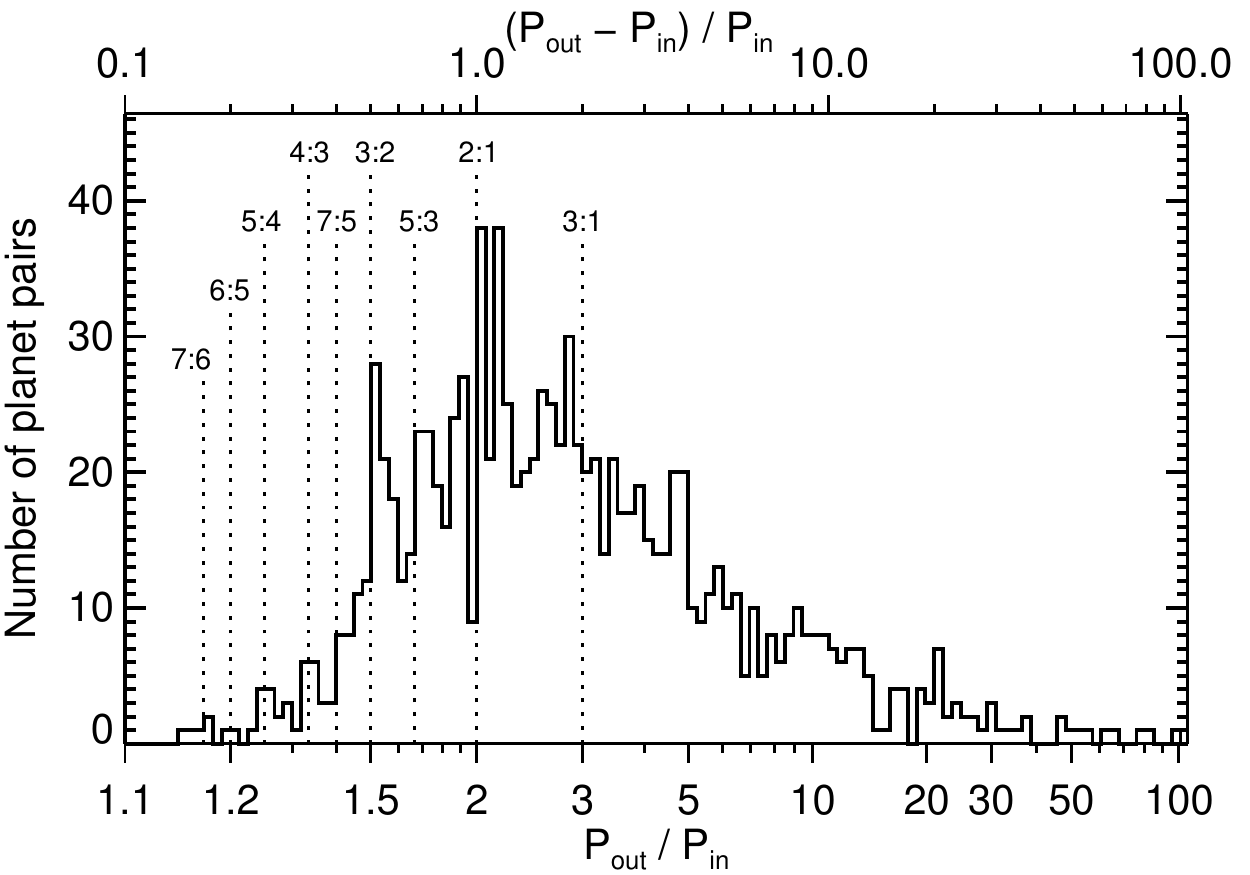}
    \end{center}
    \caption{Histogram of the number of two adjacent planets with a given period ratio for multi-planet system.
     The overall distribution is smooth but enhancements are visible near the 3:2, 2:1 and 3:1 resonances.
    After \cite{2015ARA&A..53..409W}
     }
    \label{fig:ratios}
\end{figure}

Given that embedded planets have to migrate through the disk due to the disk torques and given that resonant capture is
a very probable outcome during the evolution, then the small number of extrasolar planet pairs actually being in resonance
is quite surprising. Indeed, the observed distribution of period ratios between two adjacent extrasolar planets is
relatively smooth and does show only a mild enhancement of planet pairs near the mostly expected 3:2 and 2:1 resonances.
Considering the turbulent nature of the protoplanetary disk the migration process is not smooth anymore
but has a random component due to the gravitational disturbances of individual turbulent eddies.
Through a large set of $N$-body simulations taking disk turbulence into account it was shown that
the overall distribution of the period ratios (see Fig.~\ref{fig:ratios} for the observational data) seems to be in rough
agreement with standard migration including stochastic forces \cite{2012MNRAS.427L..21R}.
This is in agreement with results showing that stochastic perturbations
can break the resonances between two planets \cite{2008ApJ...683.1117A}. This breaking of resonances allows for the possibility
of bringing two planets very close together and producing systems of closely spaced low-mass planets.
For the system Kepler-36 it was shown that this mechanism may be responsible to explain the
unusually close configuration near the high degree first order 7:6 mean motion resonance \cite{2013MNRAS.434.3018P}.

Another feature apparent in Fig.~\ref{fig:ratios} is the fact that the over abundances of period ratios 
near the 2:1 and 3:2 resonances do not occur at exact resonance position but lie just outside of it,
a fact noticed earlier \cite{2013AJ....145....1B}.
It has been argued that this small shift of period ratios just outside of the nominal values may be the
result of resonant repulsion or dissipative divergence in the presence of tidal dissipation within the inner planet
\cite{2012ApJ...756L..11L, 2013AJ....145....1B}. Recently, it has been argued however, that tides alone cannot be held
responsible for this enhancement near the 2:1 resonance, because this would imply too large initial eccentricities
\cite{2015MNRAS.453.4089S}. Another recent suggestion is the interaction of the planets with the remaining planetesimal disk.
Here, it has been found that for the case of a sufficiently massive planetesimal disk the resonances can be
disrupted and planet system remains just outside of the nominal values during the subsequent evolution
\cite{2015ApJ...803...33C}.

\subsection{Dynamics}
\label{subsec:multi02-dynamics}
Already after the first detections of extrasolar planets it was noted that they display a large variation of
orbital eccentricities similar to spectroscopic binaries \cite{2007ARA&A..45..397U}. 
Both $e$-distributions cover the whole range from $0 \leq e \leq 1$ and have
high mean values (with $e \sim 0.3$) in obvious contrast to the Solar System where the majority of the bigger planets (despite
Mercury and Mars) have a rather small eccentricity.
For the overall $e$-distribution the following form has been suggested \cite{2008ApJ...685..553S}
\beq
         \frac{d N}{d e} \propto \left( \frac{1}{(1+e)^b} - \frac{e}{2^b} \right) \,,
\eeq
where $d N$ is the number of planets in an eccentricity interval $[e, e+de]$. According to \cite{2008ApJ...685..553S}
$b =4$ gives a good match to the observed distribution which gives a smooth distribution which peaks at $e=0$.
There is a tendency for lower mass planets to have a lower eccentricity,
in particular for $m\sub{p} < 50 M_{\rm Earth}$ the maximum eccentricity is around $e \approx 0.4$ while larger mass
planets reach over $e > 0.9$. This finding is in agreement with the fact that lower mass planets are preferentially
found in multi-planet systems that have on average lower eccentricities. For an overview of the observational properties
and more references see \cite{2015ARA&A..53..409W}.

In the previous section~\ref{lect:05} we showed that planet-disk interaction will typically lead to eccentricity and inclination
damping on a timescale shorter than the migration time. Consequently, planet-disk interaction cannot be responsible
for the observed eccentricity distribution and two other processes (planet-planet scattering or the Kozai-mechanism)
have been suggested that both require additional objects.
In the first scenario several planets have formed in the disk and undergo initially a convergent migration process
which leads to a compact configuration either in resonance or close to it.
Upon disk dissipation its stabilizing effect disappears and the planetary system my become dynamically unstable.
This will increase the eccentricity and inclination of the objects leading eventually to a scattering processes
\cite{2003Icar..163..290A}.
Some bodies will be scattered towards the star and some outwards, or be thrown out altogether. 
As shown by several simulations using many realizations of such models, these type of planet-planet
scattering processes can lead to eccentricity distributions very similar to the one observed for the ensemble of
extrasolar planets \cite{2008ApJ...686..621F,2008ApJ...686..603J}.

\begin{figure}[t]
    \begin{center}
        \includegraphics[width=0.60\textwidth]{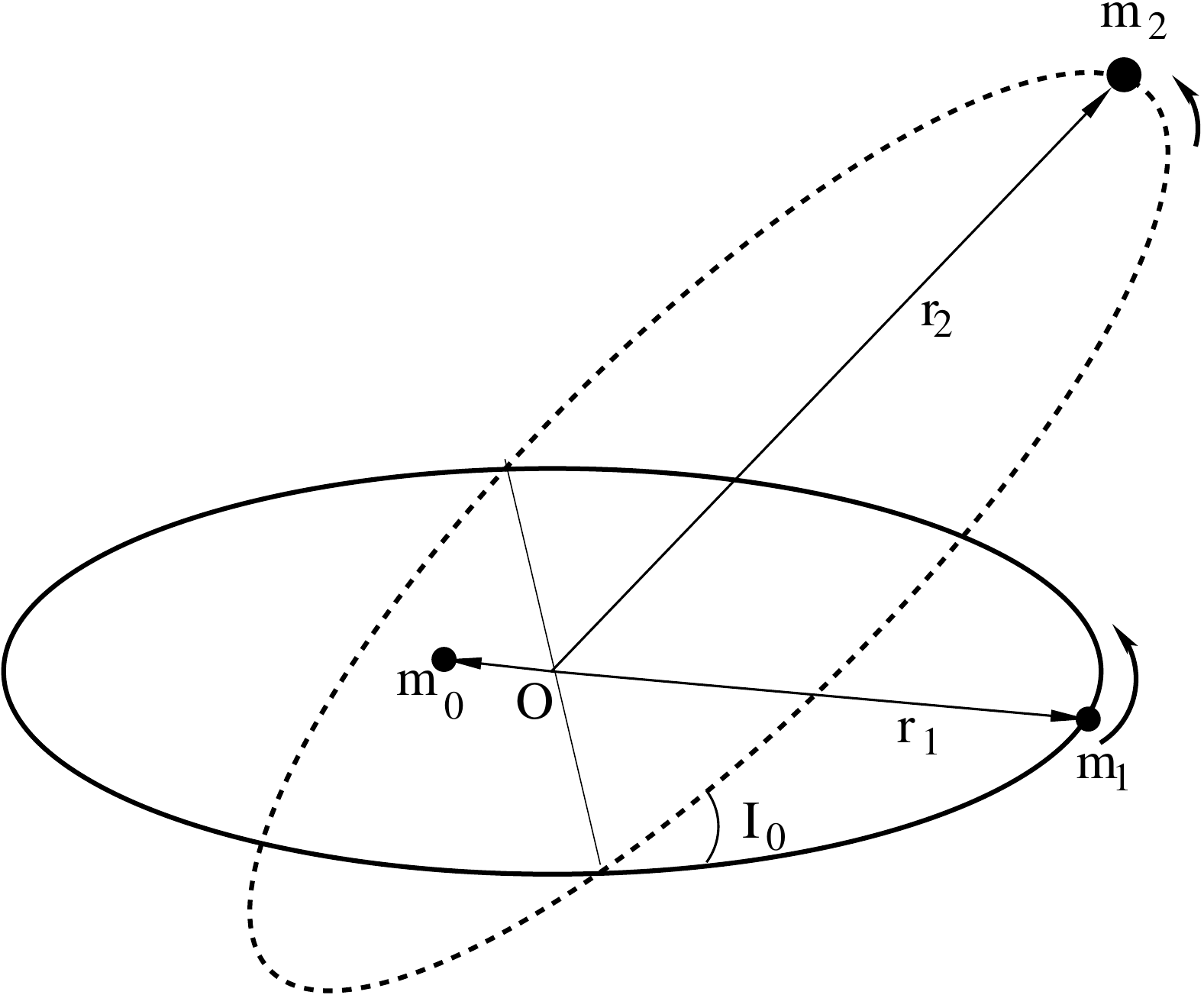}
    \end{center}
    \caption{The geometry of a hierarchical 3-body system susceptible to the Kozai-mechanism.
     A binary system with masses $m_0$ and $m_1$ is orbited by a third object $m_2$ that is on an inclined orbit.
     The inclination of the two orbital planes is initially $I_0$. Please note that the distance (semimajor axis)
     of the third object ($m_2$) is much larger than that of the secondary object ($m_1$) so the real orbits do not
     intersect.
     }
    \label{fig:Kozai-scetch}
\end{figure}
\subsubsection{The Kozai mechanism}
\label{subsubsec:multi02x-kozai}
In the past 20 years or so a dynamical process operating in a hierarchical 3-body system
has gained much of attention to explain certain properties of extrasolar planetary systems.
The mechanism was originally proposed to explain the dynamics of
main-belt asteroids due to the perturbations of Jupiter by Y.~Kozai \cite{1962AJ.....67..591K}. 
The geometry of a hierarchical system is sketched in Fig.~\ref{fig:Kozai-scetch}. A central 2-body system with masses $m_0$
and $m_1$ is orbited by a third object ($m_2$) at a larger distance, $r_2 > r_1$.
The two orbital planes are inclined by a certain angle $I$ with 
each other. Such a configuration describes for example the asteroid case in the Solar System, 
where: $m_0$ = Sun, $m_1$ = asteroid and $m_2$ = Jupiter, or
a protoplanetary case where: $m_0$ = host star, $m_1$ = planet and $m_2$ = secondary star.
The longterm evolution of such systems can result in a periodic exchange of angular momentum between the inner
binary and the third, distant object, given that the initial inclination exceeds a certain critical value.
As Kozai showed, for the satellite (the object $m_1$), the orbit averaged equations of motion have a conserved quantity,
the $z$-component of the angular momentum $J_z = \cos(I) \sqrt{1 - e^2}$,
where $z$ is the direction perpendicular to the orbit of the binary, $m_2$. 
If the initial inclination of the body's orbital plane with that of the perturber, $I_0$, is higher than a critical value $I\sub{crit}$
then the orbit of the secondary ($m_1$) will become oscillatory with phases of high eccentricity (with a reduced inclination $I$)
and low eccentricity (with high inclination), always maintaining a constant $J_z$.
The value of $I\sub{crit}$ depends on the separation of $m_2$ and is about $39^\circ$ for very distant objects with $r_2 \gg r_1$. 
Even though presented by Kozai in 1962, for the first 35 years his work drew little attention with about 60 citations.
However, after the discovery of the first highly eccentric exoplanet orbiting 16 Cyg B \cite{1997ApJ...483..457C} in 1997, immediately 
the importance of the mechanism in the context of exoplanet research was noted \cite{1997Natur.386..254H}, and the number of citations
increased significantly to about 15 per year. Then, in 2003 it was noticed by Wu \& Murray \cite{2003ApJ...589..605W}  
that this mechanism could be used for a new type of migration process, namely eccentric or tidally driven migration.
After that, the recognition of Kozai's work has increased even further to now over 70 citations per year (according to the ADS).
Sometimes the mechanism is also referred to as {\it Kozai-Lidov} mechanism attributing the work by M.~Lidov in the same year of
1962 \cite{1962AJ.....67..591K}.

We will now briefly explain how this mechanism operates in shrinking the orbit of the planet, i.e. how it reduces the semimajor axis,
and produces planets on highly inclined orbits. If a planetary system consisting of a star and a planet is accompanied by a distant
companion star with an orbit that is inclined with respect to the planet's orbit, then the perturbations will give rise
to Kozai oscillations with periodic phases of very high planet eccentricity, reaching nearly $e \approx 1$. In this case, the planet
comes very close to the star and energy is dissipated inside the central star which is taken from the planetary orbit.
Hence, at each pericenter passage the planetary orbits shrinks by a small amount. Over the course of possibly several Gyrs the
orbit can shrink significantly producing a large scale migration of the planet. Because at the same time, the inclination oscillates between
large and small values this mechanism can produce compact systems with close-in planets on possibly highly eccentric orbits.
Wu \& Murray \cite{2003ApJ...589..605W} and later Fabrycky \& Tremaine \cite{2007ApJ...669.1298F} 
applied this model to the planetary system HD 80606 which today has
the following observed orbital parameter: planet mass $m\sub{p} =3.94 M\sub{Jup}$, $a\sub{p} = 0.449$ AU, $e\sub{p} = 0.933$. 
Up to now this is the highest planet eccentricity
observed among the sample of extrasolar planets, and as a consequence the distance from the planet to the stars varies
between 0.03 to 0.88 AU. The central star of about $m_0 = 0.9 M_\odot$ is orbited by a lower mass companion star at a distance
of about $1200$ AU. 
To model the longterm evolution of this system Wu \& Murray used as starting parameter for the planet a mass of $m\sub{p} =3.94 M\sub{Jup}$,
a semi-major axis $a\sub{p} = 5 $AU and an eccentricity $e\sub{p} = 0.1$. They assumed an initial inclination of the planet and binary star orbit of
$I_0 = 85.6^\circ$. Using these initial parameter and a tidal $Q$-value for the star of about $10^6$
they find an initial period of about 0.022 Gyr for the Kozai-oscillations and maximum eccentricities
close too unity. Continuing the simulations until several Gyrs they show that indeed the system approaches today's parameter at an age of about
3~Gyrs, ending up with the observed eccentricity and an inclination of $I= 50^\circ$.

This working example of tidally driven migration was used to suggest that a large fraction of the close-in Jupiter
has migrated not by standard planet-disk interaction but rather by this process \cite{2007ApJ...669.1298F}.
It was shown that the circularization timescale of the planet is comparable to the migration time which implies
that aligned planets are most likely formed coplanar \cite{2009MNRAS.395.2268B}.
Noting that the sky projected spin-orbit angle (stellar obliquity, $\beta$) is
higher for hotter central stars (with $T\sub{eff} >  6250$\,K) than for cooler stars, 
it has been suggested that tidally driven migration may even be the dominant mode of migration \cite{2010ApJ...718L.145W}.
Assuming that most of the planets start with large obliquities lower mass stars  with lower $T\sub{eff}$ 
will tend to damp this misalignment more rapidly due to an active convection zone while more massive stars have smaller
damping and the initial misalignment is maintained.
However, using a sample of 61 transiting hot Jupiters with measured projected spin-orbit angle $\beta$
it was shown that the observed distribution of the angles $\beta$ 
is compatible with the assumption that most hot Jupiters were transported by smooth migration inside a
proto-planetary disk \cite{2014A&A...567A..42C}.

\subsection{Mulit-planet systems}
\label{subsec:multi03-systems}
The large sample of planets discovered by the Kepler space mission has shown that the main group of extrasolar
planets are smaller planets within a radius range between 1.25 and about 5 $R_{\rm Earth}$ that we call here altogether
Super-Earths \cite{2014ApJS..210...19B}. Amongst those, true planetary systems with 2 and more planets are the rule
as Kepler has discovered hundreds of transiting multi-planet systems, showing that statistically about half of all
solar-type stars host at least one planet of this type where additionally for the multi-planet systems
the inclinations between the individual planetary orbits are very small \cite{2015ARA&A..53..409W}.
Mostly, the radial spacing between the planets is very small and they are close to their host stars.
As a consequence these compact systems would fit well into
the orbit of Mercury, i.e. they resemble a Solar System scaled down by a factor of about 10 with somewhat larger
planets, though. Typically, the systems are not in resonance but show separations similar to the Solar System
planets, in terms of their Hill radii \cite{2014prpl.conf..595R}.
The ubiquity of these compact systems of lower mass planets is probably related to their longterm stability
because closely packed systems of larger mass planets would be dynamically unstable with scattering
processes and possible ejections (as shown above), while lower mass planets
might collide and stabilize (see also \cite{2008ApJ...686..621F,2015ARA&A..53..409W}).
The flatness of the complete system is again an indication for a disk-driven evolution.

Concerning the formation of these compact, tightly packed planetary systems different scenarios have been
considered such as 1) in situ formation in a massive disk,
2) accretion during inward type I migration,
3) shepherding by interior mean-motion resonances of inwardly migrating massive planets,
4) shepherding by interior secular resonances of inwardly migrating massive planets,
5) circularization of high-eccentric planets by tidal interactions with the star, and
6) photo-evaporation of close-in giant planets.
In \cite{2014prpl.conf..595R} it is argued that the possibilities 3) to 6) are in conflict with some observational facts
because 3) and 4) require a giant planet to push the planets inward and an additional disk to damp 
orbital eccentricities. In both scenarios a giant planet should be present just outside of the outermost planet
which is not observed.
For scenario 5) circularization is in principle possible (as shown above for tidal migration scenario)
but it requires large initial eccentricities that would result in scattering events and a remaining single-planet system,
which is in conflict with the observations.
For 6) evaporation is possible only for very close-in planets and it requires several Gyrs to operate.
The strong dependence on distance from the central star would only allow the innermost planet to
be evaporated. Hence, it is concluded that only scenarios 1) (in situ) and 2) (inward type I migration)
are leading contestors \cite{2013AREPS..41..469H,2014prpl.conf..595R}.

The in-situ formation scenario requires a lot of material ($\approx 20-40 M_\oplus$) within a fraction of an AU in the inner
protoplanetary disk,
unless radial transport of material is considered \cite{2012ApJ...751..158H}, but this might
also lead to an enhancement of the mass of the outermost planet. 
Within the MMSN context a very steep surface density power law, $\Sigma = \Sigma_0 (r/AU)^{-x}$,
with $x \approx 1.6-1.7$ and a 10 times higher normalization $\Sigma_0$ in comparison to the Solar System 
is required to form the planets within the observed mass range.
This is problematic because the high mass pushes the disk towards instability with respect to fragmentation.
Additionally, the observations indicate shallower profiles with $x \approx 0.5 - 1.0$
\cite{2011ARA&A..49...67W}. Despite these problems, some simulations of in-situ formation
show that orbital properties (eccentricities, inclinations, separations) of the planets seem to match the observations
\cite{2013MNRAS.431.3444C,2014prpl.conf..595R}. However, the absence of a distinct slope in the radial 
mass profile of extrasolar planetary systems is taken as an indication that a radial rearrangement of solids must
have taken place for the majority of systems \cite{2014MNRAS.440L..11R}.
For scenario 2), simulations of migrating Super-Earths indicate the formation of resonant chains with
ongoing destabilization and subsequent collision and accretion events, forming systems similar to the observed ones
\cite{2014A&A...569A..56C}.
A possible distinction between models might be the occurrence of naked high-density rocks for scenario
1) and possibly lower density material containing ice for scenario 2) but planetary atmospheres could
possibly hide the effect \cite{2014prpl.conf..595R}.

\subsection{Circumbinary Planets}
\label{subsec:multi04-circumbinary}
One of the recent findings of the Kepler space mission has been the discovery of 
circumbinary planets that orbit around two stars very similar to the Tatooine system in the {\itshape Star Wars}
series. The first of such systems has been Kepler 16 where a Saturn mass planet orbits a central binary 
(with orbital period $P_{\rm bin} = 41$ days) with a period of about $P_{\rm p} = 229$ days \cite{2011Sci...333.1602D}.
By now about 10 systems are known and in most cases the planets orbit the central binary star very close to the 
mechanical stability limit, i.e. they have the closest possible distance to the binary star. A tighter orbit would
lead to dynamical instability with a possible subsequent loss of the planet through a scattering event with the central
binary star.
This stability limit can be determined by performing sequences of 3-body integrations consisting of a low mass object 
(test particle) that orbit around a binary star. It is determined by the mass ratio $q\sub{bin} = m_1/m_2$, and the eccentricity $e\sub{bin}$ of the
binary star and lies in the range of 2-5 times the semimajor axis of the binary, $a\sub{bin}$ \cite{1989A&A...226..335D}.
Through extensive 3-body simulations Holman \& Wiegert \cite{1999AJ....117..621H} show that the dependence on  $q\sub{bin}$ is
weak and they give the following approximate formula for the critical radius
\beq
\label{eq:a-crit}
        a\sub{crit} =  ( 2.28 + 3.82 e\sub{bin} - 1.71 e\sub{bin}^2  ) \, a\sub{bin} \,.
\eeq
Beyond $a\sub{crit}$ the orbits are in principle stable but 
newer simulations show in addition that for planetary semi-major axes that are near the mean motion resonances 
4:1, 5:1 and so on with the central binary the orbits are very unstable \cite{2015MNRAS.446.1283C}, such that planets at those locations are easily
scattered out of the system. Consequently, the observed locations of the planets around binary stars are typically between these resonances.
In two systems (Kepler 16, Kepler 38) the planets are located between the 5:1 and 6:1 resonances and in 4 others 
(Kepler 35, 47, 64 and 413) they are between the 6:1 and 7:1 resonances. 
These cases comprise over half of the known
systems and the closeness of the planets to the edge of instability raises the question as to their formation.
Table \ref{tab:cb-planets} gives an overview of the orbital parameter of some well known systems, and shows the closeness of the planets
to the boundary of stability, $a\sub{p}/a\sub{crit}$.

\begin{table}[t]
\normalsize
\begin{tabular}{|l|l|l|l|l|l|l|}
   \multicolumn{7}{c}{Binary Parameter:}   \\
  &  Kepler-16 & Kepler-38 &   Kepler-34 & Kepler-35 & Kepler-64 & Kepler-413\\
\hline
$M_1 [M_\odot]$ & 0.62 & 0.95  & 1.05  & 0.89 & 1.38 & 0.82 \\
$M_2 [M_\odot]$ & 0.20 & 0.25  & 1.02  & 0.81 & 0.38 & 0.54 \\
$P\sub{bin}$ [days] & 41 & 18.6  & 28  &  21  & 20  & 10 \\
$a\sub{bin}$ [AU] & 0.22 & 0.15 & 0.23 & 0.18 & 0.17 & 0.11  \\
  $e\sub{bin}$ &  0.16  &  0.10 &   0.52 & 0.14 & 0.22 & 0.04 \\
\hline
  \multicolumn{7}{c}{Planet Parameter:}  \\
  &  Kepler-16 & Kepler-38 &    Kepler-34 & Kepler-35  & Kepler-64 & Kepler-413  \\
\hline
$m\sub{p} [M\sub{Jup}]$ & 0.33 &   0.36  & 0.21 & 0.13 & $<0.1$ &  0.21 \\
$P\sub{p}$ [days]  & 228  &  105.6  & 288  & 131 & 138 &  66 \\
$a\sub{p}$ [AU] & 0.70 & 0.46  & 1.09  &  0.6 & 0.63 & 0.36 \\
$e\sub{p}$ & 0.07 &  0.03 &  0.18  &  0.04 & 0.05 & 0.12  \\
$a\sub{p}/a\sub{crit}$  &   1.09 &   1.13  &   1.28 & 1.13 & 1.15 & 1.36 \\
\hline
\end{tabular}
\caption{Parameters of selected circumbinary planets as listed in \cite{2015ARA&A..53..409W}.
The upper table shows the parameter of the central binary star, and
the bottom table the properties of the circumbinary planet. The last row contains the ratio of the observed
semi-major axis to the critical value $a\sub{crit}$ (eq.~\ref{eq:a-crit})
and indicates clearly the closeness to instability of the circumbinary planets.}
\label{tab:cb-planets}
\end{table}

Typical formation scenarios for planets in general require the growth from smaller particles through
a sequence of collisions as outlined in detail in sections \ref{lect:01} and \ref{lect:02}.
If the relative speed 
between two objects is too large the growth is seriously handicapped. As mentioned above, even
under the conditions prevalent in the early Solar System growth is very difficult to achieve.
Not surprisingly, the additional perturbations generated by the binary star in the center lead to increased
particles velocities that make growth even more difficult, if not impossible \cite{2007MNRAS.380.1119S,
2012ApJ...752...71M,2012ApJ...754L..16P}.
Hence, it is typically assumed that the planets have formed further away from the binary star in environments that
are dynamically much calmer, and then migrated inward to their present location through a planet-disk interaction process
\cite{2007A&A...472..993P,2013A&A...553A..71M}. This migration of the planet through the circumbinary disk
would also naturally explain the flatness of these systems if one assumes that the disk is completely
aligned with the orbital plane of the binary. In the following we shall concentrate on this migration scenario.

\begin{figure}[t]
    \begin{center}
        \includegraphics[width=0.85\textwidth]{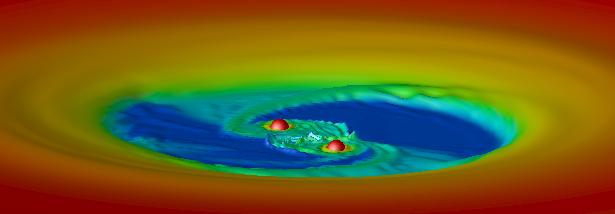}
    \end{center}
    \caption{The structure of a circumbinary disk around a central binary star 
     (indicated by the two red spheres in the middle). Color coded is the surface density of the disk and the vertical
     extension indicates the thickness (temperature) of the disk. (by Richard G\"unther, University of T\"ubingen, 
     based on simulations in \cite{2004A&A...423..559G})
     }
    \label{fig:CB-disk-richie}
\end{figure}

As mentioned above, the presence of a binary star inside a disk creates a large inner cavity cleared from the gas as shown
in Fig.~\ref{fig:CB-disk-richie}.
This clearing is due to the transfer of angular momentum from the binary to the disk through tidal forces.
The extent of the inner gap depends on the binary and disk parameter such as the binary's mass ratio, eccentricity 
and disk viscosity and temperature \cite{1994ApJ...421..651A}.
Numerical simulations of hydrodynamic disks around binary stars show that the inner disk becomes eccentric  
and slowly precesses around the binary star where the precession rate is typically several
100 binary orbits, $P\sub{bin}$ \cite{2015MNRAS.448.3545D,2015A&A...581A..20K}. 
However the full dynamical evolution of circumbinary disks around their central binary stars is complex and 
not fully understood yet \cite{2013A&A...556A.134P}.

\begin{figure}[t]
    \begin{center}
        \includegraphics[width=0.75\textwidth]{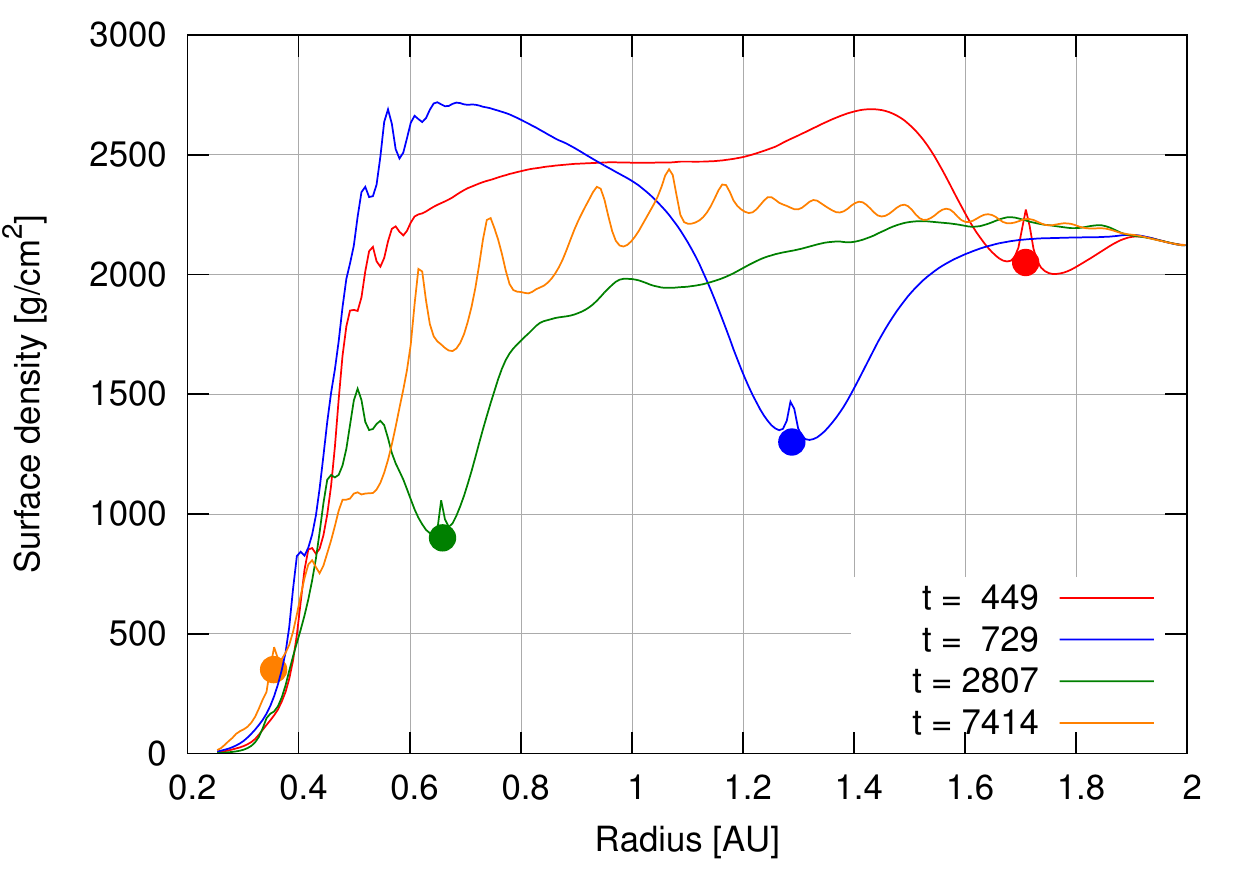}
    \end{center}
    \caption{The migration of a planet in the Kepler-38 system. 
     Shown is the surface density at different times (in yrs) after insertion of the planet. 
    The time levels are indicated by the labels.
     The dots indicate the radial position of the planet at those times, for illustration they have been moved to the
    corresponding surface density distribution. Clearly visible is the (partial) gap formed by the planet during its 
    inward migration. The orange dot marks the final equilibrium position.
    Adapted from \cite{2014A&A...564A..72K}.
     }
    \label{fig:kepler-38}
\end{figure}

Coming back to the evolution of planets embedded in disks we note that Saturn mass planets open a partial gap in the disk.
Hence, the corotation torques are reduced and they are driven primarily by the Lindblad torques that will induce 
inward migration. This inward drift can be stopped at special locations, the so-called planet traps \cite{2006ApJ...642..478M},
where the total torque acting on the planet vanishes, such that the inward migration is terminated.
Regions with a positive density slope near the inner edge of the disk which will act as a natural
location for a planet trap. For single stars a positive density gradient can be generated for example 
by the magnetosphere of the young star while for binary
stars the inner disk cavity is generated naturally by angular momentum transferal from the binary 
to the disk (see Fig.~\ref{fig:CB-disk-richie}). 
The extent of the hole, i.e. the location of the disk's inner edge will then determine the final location of the planet.
For typical disk parameter with $\alpha \approx 0.01$, $H/r \approx 0.05$ and binaries with moderate eccentricity,
$e\sub{bin} \approx 0.1$, such as Kepler-16 and Kepler-38 (see table~\ref{tab:cb-planets}) it was shown 
that in the simulations the final parking position of planets is indeed near the 
the observed locations \cite{2013A&A...556A.134P,2014A&A...564A..72K}. In some simulations for Kepler 38 it was found that
the planet was captured in the 5:1 resonance and remained at this location. This capture into resonance was accompanied by an
increase of the eccentricity of the planetary orbit to about $e\sub{bin} \approx 0.2$ \cite{2014A&A...564A..72K}. 
As explained above, planets in such resonant orbits will typically be unstable in the longterm evolution and 
they can only be stable in the presence of a disk that tends to damp planetary eccentricities as shown in the previous section.
So, after disk dispersal such a planet is expected to become unstable and be scattered out of the system.
It is possible that several planets that initially formed in circumbinary disks experienced such a fate and are now
a member of the free-floating planet population. In this case we are just observing those planets that luckily happened to end up
in a final stable location in between the unstable resonances.
As was shown, different disk parameter with a lower density and viscosity can lead to final parking positions near the
observed location for the Kepler-38 system \cite{2014A&A...564A..72K}. In Fig.~\ref{fig:kepler-38} the final
parking position is indicated by the innermost orange dot. It is right at the inner edge of the disk.

While the final position of planets of planets around binaries with moderate eccentricities (Kepler-16, Kepler-35 and Kepler-38)
can be understood in terms of a migration process of a planet in a disk \cite{2013A&A...556A.134P,2014A&A...564A..72K}, 
it is still difficult to explain the location of the observed planet in the Kepler-34 systems where the central binary has
as large eccentricity of $e\sub{bin} = 0.52$. This high eccentricity of the binary generates a large inner hole of
the disk that is highly eccentric and precesses around the central binary with a period of about 120\,yrs equivalent to 
nearly $2000\,P\sub{bin}$.
The final position of the binary is determined by the size of the inner hole. The planet settles in an elliptic orbit
that precesses around the binary with exactly the same period as the disk, such that the pericenters of the disk and planet
are always aligned. This final orbit has a semi-major axis that is substantially larger than the observed value indicated
in Table~\ref{tab:cb-planets}. Hence, for the eccentric binary star Kepler-34, the observed position of the planet is still
difficult to understand with a migration process but may be that better disk models with more realistic physics
will lead to improved results.

\begin{svgraybox}
In planetary systems with multiple objects interesting dynamics can take place. Disk-planet interaction will often lead to a
convergence of orbits with capture in mean-motion resonances (MMR) between two adjacent planets. However, 
observationally there is only a slight overabundance of planets near the 3:2 and 2:1 MMRs, but 
stochastic and tidal processes can take planets out of exact resonance. 
The dynamical interaction between multiple massive planets can lead to eccentricity and inclination excitation.
Another pathway to generate the high observed eccentricity and inclinations in some systems is through the Kozai-Lidov
mechanism which requires the exchange of angular momentum with a third, distant object in the system. 
Concerning the observed abundance of planetary systems that contain several Super-Earths in a compact configuration, the
most likely pathway appears to be the convergent inward migration process accompanied by stochastic forces with the disk.
A very exciting discovery of the Kepler space mission has been the detection of circumbinary planets.
The observed locations of the planets close to the dynamical stability region can be explained for most systems 
through a migration process of these planets through the circumplanetary disk.
\end{svgraybox}

%

%
\begin{acknowledgement}
This text is based on a series of lectures on the topic {\it Planet formation and disk-planet interactions} given
at the 45th ''Saas-Fee Advanced Course'' of the Swiss Society for Astrophysics and Astronomy (SSAA) held
in Les Diablerets in March 2015.
I acknowledge generous support from the SSAA, and would like to thank the organisers 
(Marc Audard, Yann Alibert and, Michael R. Meyer, Martine Logossou) for providing such a nice and stimulating atmosphere.
I thank Giovanni Picogna for a reading of the manuscript.
\end{acknowledgement}
%

\bibliography{kley,kley00-intro,kley01-dust,kley02-terr,kley03-gas,kley04-self,kley05-pladisk,kley06-multi}

\end{document}